\input pictex.tex   %% Dovrebbe essere in path (altrimenti vedi in Doc)

% $Id: dcpic.sty,v 1.24 2002/11/25 13:51:57 pedro Exp $
%% DC-PiCTeX
%% Realizado por Pedro Quaresma de Almeida, Coimbra 
%% 11/1990 (vers{\~a}o 1.0); 10/1991 (vers{\~a}o 1.1);
%%  9/1993 (vers{\~a}o 1.2);  3/1995 (vers{\~a}o 1.3);
%%  7/1996 (vers{\~a}o 2.1);
%%  5/2001 (vers{\~a}o 3.0); 11/2001 (vers{\~a}o 3.1);
%%  1/2002 (vers{\~a}o 3.2)
%%  5/2002 (versão 4.0)
\immediate\write10{Package DCpic 2002/05/16 v4.0}

\catcode`!=11 %  ***** THIS MUST NEVER BE OMITTED (Ver PiCTeX)

\newcount\aux%
\newcount\auxa%
\newcount\auxb%
\newcount\m%
\newcount\n%
\newcount\x%
\newcount\y%
\newcount\xl%
\newcount\yl%
\newcount\d%
\newcount\dnm%
\newcount\xa%
\newcount\xb%
\newcount\xmed%
\newcount\xc%
\newcount\xd%
\newcount\ya%
\newcount\yb%
\newcount\ymed%
\newcount\yc%
\newcount\yd
%% "variáveis globais"
\newcount\expansao%
\newcount\tipografo%       versão 4.0
\newcount\distanciaobjmor% versão 4.0
\newcount\tipoarco%        versão 4.0
%\newif\ifarredondada%        versão 4.0 (valor inicial "false")
\newif\ifpara%
%% version 3.2
\newbox\caixa%
\newbox\caixaaux%
\newif\ifnvazia%
\newif\ifvazia%
\newif\ifcompara%
\newif\ifdiferentes%
\newcount\xaux%
\newcount\yaux%
\newcount\guardaauxa%
\newcount\alt%
\newcount\larg%
\newcount\prof%
%% para os ajustes
\newcount\auxqx
\newcount\auxqy
\newif\ifajusta%
\newif\ifajustadist
\def\objPartida{}%
\def\objChegada{}%
\def\objNulo{}%

%% 
%% Stack specification
%%

%%
%% Emtpy stack
%%
\def\!vazia{:}

%%
%% Is Empty? : Stack -> Bool
%%
%% nvazia - True if Not Empy
%% vazia  - True if Empty
\def\!pilhanvazia#1{\let\arg=#1%
\if:\arg\ \nvaziafalse\vaziatrue \else \nvaziatrue\vaziafalse\fi}

%%
%% Push : Elems x Stack -> Stack
%%
\def\!coloca#1#2{\edef\pilha{#1.#2}}

%%
%% Top : Stack -> Elems
%%
%% the empty stack is not taken care
%% the element is "kept" ("guardado") 
\def\!guarda(#1)(#2,#3)(#4,#5,#6){\def\id{#1}%
\xaux=#2%
\yaux=#3%
\alt=#4%
\larg=#5%
\prof=#6%
}

\def\!topaux#1.#2:{\!guarda#1}
\def\!topo#1{\expandafter\!topaux#1}

%%
%% Pop : Stack -> Stack
%%
%% the empty stack is not taken care
\def\!popaux#1.#2:{\def\pilha{#2:}}
\def\!retira#1{\expandafter\!popaux#1}

%%
%% Compares words : Word x Word -> Bool
%%
%% compara - True if equal
%% diferentes - True if not equal
\def\!comparaaux#1#2{\let\argA=#1\let\argB=#2%
\ifx\argA\argB\comparatrue\diferentesfalse\else\comparafalse\diferentestrue\fi}

\def\!compara#1#2{\!comparaaux{#1}{#2}}

%%Comando Interno
%% Valor absoluto (absolute value)
%% \absoluto{n}{absn}
%% entrada
%%  n - natural
%% sa{\'\i}da
%%  absn - o valor absoluto de n
\def\!absoluto#1#2{\n=#1%
  \ifnum \n > 0
    #2=\n
  \else
    \multiply \n by -1
    #2=\n
  \fi}

%% Name definitions for edge types and directions

%% Name definitions for edge label placement

%% Tip direction for curved edges

%% Type of graph
\def\commdiag{0}

%% Posicionamento da etiquetas nos grafos

%%Comando Interno
%% Ajusta a dist{\^a}ncia entre as setas e os objectos em fun{\c c}{\~a}o das
%% dimens{\~o}es destes {\'u}ltimos
%% \ajusta{x}{xl}{y}{yl}{d}{Objecto}
%% entrada
%%  (x,y) e (xl,yl), coordenadas dos pontos de {\'\i}nicio e fim da seta
%%  d, dist{\^a}ncia especificada pelo utilizador ou 10 (valor por
%%  omiss{\~a}o), Objecto d{\'a}-nos a refer{\^e}ncia do objecto ao qual se est{\'a} a
%%  efectuar o ajuste.
%% sa{\'\i}da
%%  d, dist{\^a}ncia alterada.
%% 
%% A dist{\^a}ncia alterada {\'e} o maior valor entre 10 e as dimens{\~o}es
%% apropriadas da caixa que cont{\^e}m o objecto. 
%% Se o utilizador especificar um valor essa especifica{\c c}{\~a}o
%% n{\~a}o {\'e} alterada.
%%
%% Se a seta {\'e} horizontal usa-se o valor da largura
%% Se a seta {\'e} vertical usa-se:
%%  o valor da altura se a seta est{\'a} no 1o ou 2o quadrante
%%  o valor da profundidade se a seta est{\'a} no 3o ou 4o quadrante
%% Se a seta {\'e} {\'o}bliqua vai-se escolher o valor conforme:
%%  de 315 a  45 graus usa-se a largura
%%  de  45 a 135 graus usa-se a altura
%%  de 135 a 225 graus usa-se a largura
%%  de 225 a 315 graus usa-se a profundidade
\def\!ajusta#1#2#3#4#5#6{\aux=#5%
  \let\auxobj=#6%
  \ifcase \tipografo    % diagramas comutativos
    \ifnum\number\aux=10 
      \ajustadisttrue % se o valor é o valor por omissão ajusta
    \else
      \ajustadistfalse  % caso contrário não ajusta
    \fi
  \else  % grafos (dirigidos, não dirigidos, com molduras)
   \ajustadistfalse
%  \or  % grafos não dirigidos
%   \ajustadistfalse
%  \else % grafos dirigidos com molduras circulares nos objectos
%    \ifnum\number\aux=8 
%      \ajustadisttrue  % se o valor é o valor por omissão ajusta
%    \else
%      \ajustadistfalse % caso contrário não ajusta
%    \fi
  \fi
  \ifajustadist
%  \tiny Vou ajustar %%%%%%%%%%%%%%%%%%%%%%%%%%%%%%
%  \ifnum\number\aux=10% verificar se s{\~a}o os valores por omiss{\~a}o
   \let\pilhaaux=\pilha%
   \loop%
     \!topo{\pilha}%
     \!retira{\pilha}%
     \!compara{\id}{\auxobj}%
     \ifcompara\nvaziafalse \else\!pilhanvazia\pilha \fi%
     \ifnvazia%
   \repeat%
%% rep{\~o}e os valores na pilha
   \let\pilha=\pilhaaux%
   \ifvazia%
    \ifdiferentes%
%%
%% N{\~a}o {\'e} poss{\'\i}vel efectuar o ajuste dado o utilizador n{\~a}o ter
%% especificado uma etiqueta para o objecto em quest{\~a}o. {\'E} dado o
%% valor de 10, igual ao valor por omiss{\~a}o.
%%
     \larg=1310720% n{\~a}o faz o ajuste
     \prof=655360%
     \alt=655360%
    \fi%
   \fi%
   \divide\larg by 131072
   \divide\prof by 65536
   \divide\alt by 65536
   \ifnum\number\y=\number\yl
%% Caso 1 -- seta horizontal
%%
%% divide-se por 131072 para se obter metade da largura da caixa em
%% pontos (pt), isto dado que o texto est{\'a} centrado na caixa. Soma-se
%% mais tr{\^e}s, que constitue um ajuste imp{\'\i}rico.
    \advance\larg by 3
    \ifnum\number\larg>\aux
     #5=\larg
    \fi
   \else
    \ifnum\number\x=\number\xl
     \ifnum\number\yl>\number\y
%% Caso 2.1 -- seta vertical de cima para baixa
%%
      \ifnum\number\alt>\aux
       #5=\alt
      \fi
     \else
%% Caso 2.2 -- seta vertical de baixo para cima
%%
%% divide-se por 65536 para se obter a altura da caixa em pt. O ajuste
%% de 5 foi obtido imp{\'\i}ricamente
      \advance\prof by 5
      \ifnum\number\prof>\aux
       #5=\prof
      \fi
     \fi
    \else
%% Caso 3 -- seta obl{\'\i}qua 
%% Caso 3.1 de 315o a  45o; |x-xl|>|y-yl| e
%% Caso 3.3 de 135o a 225o; |x-xl|>|y-yl|; Largura
     \auxqx=\x
     \advance\auxqx by -\xl
     \!absoluto{\auxqx}{\auxqx}%
     \auxqy=\y
     \advance\auxqy by -\yl
     \!absoluto{\auxqy}{\auxqy}%
     \ifnum\auxqx>\auxqy
      \ifnum\larg<10
       \larg=10
      \fi
      \advance\larg by 3
      #5=\larg
     \else
%% Caso 3.2 de  45o a 135o; |x-xl|<|y-yl| e y>0; Largura
      \ifnum\yl>\y
       \ifnum\larg<10
        \larg=10
       \fi
      \advance\alt by 6
       #5=\alt
      \else
%% Caso 3.4 de 225o a 315o; |x-xl|<|y-yl| e y<0; Profundidade
      \advance\prof by 11
       #5=\prof
      \fi
     \fi
    \fi
   \fi
\fi} % o ramo "else" {\'e} omisso

%%Comando Interno
%% C{\'a}lculo da Raiz Quadrada
%% raiz{n}{m}
%% entrada
%%   n - natural
%% sa{\'\i}da
%%   n - natural
%%   m - maior natural contido na raiz quadrada de n
\def\!raiz#1#2{\n=#1%
  \m=1%
  \loop
    \aux=\m%
    \advance \aux by 1%
    \multiply \aux by \aux%
    \ifnum \aux < \n%
      \advance \m by 1%
      \paratrue%
    \else\ifnum \aux=\n%
      \advance \m by 1%
      \paratrue%
       \else\parafalse%
       \fi
    \fi
  \ifpara%
  \repeat
#2=\m}

%%Comando Interno
%% Calcula os pontos de 
%%       come{\c c}o da "seta"
%%       fim da "seta"
%%   coloca{\c c}{\~a}o do s{\'\i}mbolo
%% 
%% ucoord{x1}{x2}{x3}{x4}{x5}{x6}{+|- 1}
%% entrada
%%   x1,x2,x3,x4,x5
%% sa{\'\i}da
%%   x6
%%  
%%              x2 - x1
%%  x6 = x3 +|- ------- x4
%%                 x5
\def\!ucoord#1#2#3#4#5#6#7{\aux=#2%
  \advance \aux by -#1%
  \multiply \aux by #4%
  \divide \aux by #5%
  \ifnum #7 = -1 \multiply \aux by -1 \fi%
  \advance \aux by #3%
#6=\aux}

%%Comando Interno 
%% C{\'a}lculo do Quadrado da Dist{\^a}ncia Euclidiana entre dois pontos 
%% quadrado{n}{m}{l}
%% entrada
%%   n - natural
%%   m - natural
%% sa{\'\i}da
%%   l = (n-m)*(n-m)
\def\!quadrado#1#2#3{\aux=#1%
  \advance \aux by -#2%
  \multiply \aux by \aux%
#3=\aux}

%%Comando Interno
%% C{\'a}lculo auxiliar para determinar a dist{\^a}ncia entre o nome do
%% morfismo e a seta.
%% entrada
%%     (x,y), (x',y') e o nome do morfismo
%% sa{\'\i}da
%%     dnm - dist{\^a}ncia do nome ao morfismo respectivo devidamente
%%     compensada pelo tamanho do objecto
%% Observa{\c c}{\~o}es
%%     A compensa{\c c}{\~a}o s{\'o} est{\'a} a ser feita para setas
%%     horizontais e verticais. As obl{\'\i}quas s{\~a}o tratadas de
%%     outra forma.
%% algoritmo
%%  caixa0 <- nome do morfismo
%%  se x-xl = 0 entao                   {recta vertical}
%%     aux <- largura da caixa0
%%     dnm <- convers{\~a}o-sp-pt(aux)/2+3
%%  sen{\~a}o                               {recta n{\~a}o vertical}
%%     se y-yl = 0 entao                {recta horizontal}
%%        aux <- altura+profundidade da caixa0
%%        dnm <- convers{\~a}o-sp-pt(aux)/2+3
%%     sen{\~a}o                            {recta obl{\'\i}qua}
%%        dnm <- 3
%%     fimse
%%  fimse
%% fimalgoritmo
\def\!distnomemor#1#2#3#4#5#6{\setbox0=\hbox{#5}%
  \aux=#1
  \advance \aux by -#3
  \ifnum \aux=0
     \aux=\wd0 \divide \aux by 131072
     \advance \aux by 3
     #6=\aux
  \else
     \aux=#2
     \advance \aux by -#4
     \ifnum \aux=0
        \aux=\ht0 \advance \aux by \dp0 \divide \aux by 131072
        \advance \aux by 3
        #6=\aux%
     \else
     #6=3
     \fi
   \fi
}

%%
%% O ambiente "begindc...enddc"
%%
\def\begindc#1{\!ifnextchar[{\!begindc{#1}}{\!begindc{#1}[30]}}
\def\!begindc#1[#2]{\beginpicture 
  \let\pilha=\!vazia
  \setcoordinatesystem units <1pt,1pt>
  \expansao=#2
  \ifcase #1
    \distanciaobjmor=10
    \tipoarco=0         % seta
    \tipografo=0        % diagrama comutativo
  \or
    \distanciaobjmor=2
    \tipoarco=0         % seta 
    \tipografo=1        % grafo orientado
  \or
    \distanciaobjmor=1
    \tipoarco=2         % linha
    \tipografo=2        % grafo não orientado
  \or
    \distanciaobjmor=8
    \tipoarco=0         % seta 
    \tipografo=3        % grafo orientado
%    \arredondadotrue    % objectos com molduras circulares
  \or
    \distanciaobjmor=8
    \tipoarco=2         % linha
    \tipografo=4        % grafo não orientado
%    \arredondadotrue    % objectos com molduras circulares
  \fi}

\def\enddc{\endpicture}

%%
%% Comando para construir a "seta" entre dois objectos
%%
%% Os pontos definidores da seta e da etiqueta respectiva s{\~a}o:
%% 
%%                (xd,yd)
%%                   o
%%                   |
%%  o------o---------o---------o------o
%%(x,y) (xa,ya)   (xm,ym)   (xb,yb)(xl,yl)
%%
\def\mor{%
  \!ifnextchar({\!morxy}{\!morObjA}}
\def\!morxy(#1,#2){%
  \!ifnextchar({\!morxyl{#1}{#2}}{\!morObjB{#1}{#2}}}
\def\!morxyl#1#2(#3,#4){%
  \!ifnextchar[{\!mora{#1}{#2}{#3}{#4}}{\!mora{#1}{#2}{#3}{#4}[\number\distanciaobjmor,\number\distanciaobjmor]}}%
\def\!morObjA#1{%
 \let\pilhaaux=\pilha%
 \def\objPartida{#1}%
 \loop%
    \!topo\pilha%
    \!retira\pilha%
    \!compara{\id}{\objPartida}%
    \ifcompara \nvaziafalse \else \!pilhanvazia\pilha \fi%
   \ifnvazia%
 \repeat%
 \ifvazia%
  \ifdiferentes%
%%
%% Mensagem de erro e atribui{\c c}{\~a}o de valores fict{\'\i}cios aos 
%% argumentos dos comandos que se seguem.
%%
   Error: Incorrect label specification%
   \xaux=1%
   \yaux=1%
  \fi%
 \fi% 
 \let\pilha=\pilhaaux%
 \!ifnextchar({\!morxyl{\number\xaux}{\number\yaux}}{\!morObjB{\number\xaux}{\number\yaux}}}
\def\!morObjB#1#2#3{%
  \x=#1
  \y=#2
 \def\objChegada{#3}%
 \let\pilhaaux=\pilha%
 \loop
    \!topo\pilha %
    \!retira\pilha%
    \!compara{\id}{\objChegada}%
    \ifcompara \nvaziafalse \else \!pilhanvazia\pilha \fi
   \ifnvazia
 \repeat
 \ifvazia
  \ifdiferentes%
%%
%% Mensagem de erro e atribui{\c c}{\~a}o de valores fict{\'\i}cios aos 
%% argumentos dos comandos que se seguem.
%%
   Error: Incorrect label specification
   \xaux=\x%
   \advance\xaux by \x%
   \yaux=\y%
   \advance\yaux by \y%
  \fi
 \fi
 \let\pilha=\pilhaaux
 \!ifnextchar[{\!mora{\number\x}{\number\y}{\number\xaux}{\number\yaux}}{\!mora{\number\x}{\number\y}{\number\xaux}{\number\yaux}[\number\distanciaobjmor,\number\distanciaobjmor]}}
\def\!mora#1#2#3#4[#5,#6]#7{%
  \!ifnextchar[{\!morb{#1}{#2}{#3}{#4}{#5}{#6}{#7}}{\!morb{#1}{#2}{#3}{#4}{#5}{#6}{#7}[1,\number\tipoarco] }}
\def\!morb#1#2#3#4#5#6#7[#8,#9]{\x=#1%
  \y=#2%
  \xl=#3%
  \yl=#4%
  \multiply \x by \expansao%
  \multiply \y by \expansao%
  \multiply \xl by \expansao%
  \multiply \yl by \expansao%
%%
%% calcular a dist{\^a}ncia Euclidiana entre dois pontos
%% d = \sqrt((x-xl)^2+(y-yl)^2)
%%
  \!quadrado{\number\x}{\number\xl}{\auxa}%
  \!quadrado{\number\y}{\number\yl}{\auxb}%
  \d=\auxa%
  \advance \d by \auxb%
  \!raiz{\d}{\d}%
%%
%% o ponto (xa,ya) est{\'a} {\`a} dist{\^a}ncia #5 (valor por omiss{\~a}o 10) do ponto
%% (x,y)
%%
%% como existem dois pontos em considera{\c c}{\~a}o, o ponto de partida e o
%% ponto de chegada, vai sei necess{\'a}rio recuperar de novo os seus
%% valores por pesquisa na pilha
  \auxa=#5
  \!compara{\objNulo}{\objPartida}%
  \ifdiferentes% S{\'o} vai fazer o ajuste quando {\'e} necess{\'a}rio
   \!ajusta{\x}{\xl}{\y}{\yl}{\auxa}{\objPartida}%
   \ajustatrue
   \def\objPartida{}% re-inicializar o valor do Objecto de Partida
  \fi
%% vai guardar o valor de auxa (ap{\'o}s ajuste) para ser usado no caso
%% dos morfismos de injec{\c c}{\~a}o.
  \guardaauxa=\auxa
  \!ucoord{\number\x}{\number\xl}{\number\x}{\auxa}{\number\d}{\xa}{1}%
  \!ucoord{\number\y}{\number\yl}{\number\y}{\auxa}{\number\d}{\ya}{1}%
%% auxa vai ter o valor da dist{\^a}ncia entre os objectos menos a
%% dist{\^a}ncia da seta ao objecto (10 por omiss{\~a}o)
  \auxa=\d%
%%
%% o ponto (xb,yb) est{\'a} {\`a} dist{\^a}ncia #6 (valor por omiss{\~a}o 10) do ponto
%% (xl,yl)
%%
  \auxb=#6
  \!compara{\objNulo}{\objChegada}%
  \ifdiferentes% S{\'o} vai fazer o ajuste quando {\'e} necess{\'a}rio
%   Vou ajustar
   \!ajusta{\x}{\xl}{\y}{\yl}{\auxb}{\objChegada}%
   \def\objChegada{}% re-inicializar o valor do Objecto de Chegada
  \fi
  \advance \auxa by -\auxb%
  \!ucoord{\number\x}{\number\xl}{\number\x}{\number\auxa}{\number\d}{\xb}{1}%
  \!ucoord{\number\y}{\number\yl}{\number\y}{\number\auxa}{\number\d}{\yb}{1}%
  \xmed=\xa%
  \advance \xmed by \xb%
  \divide \xmed by 2
  \ymed=\ya%
  \advance \ymed by \yb%
  \divide \ymed by 2
  \!distnomemor{\number\x}{\number\y}{\number\xl}{\number\yl}{#7}{\dnm}%
  \!ucoord{\number\y}{\number\yl}{\number\xmed}{\number\dnm}{\number\d}{\xc}{-#8}% 
  \!ucoord{\number\x}{\number\xl}{\number\ymed}{\number\dnm}{\number\d}{\yc}{#8}%
\ifcase #9  % seta s{\'o}lida
  \arrow <4pt> [.2,1.1] from {\xa} {\ya} to {\xb} {\yb}
\or  % seta a tracejado
  \setdashes
  \arrow <4pt> [.2,1.1] from {\xa} {\ya} to {\xb} {\yb}
  \setsolid
\or  % linha s{\'o}lida
  \setlinear
  \plot {\xa} {\ya}  {\xb} {\yb} /
\or  % seta de injec{\c c}{\~a}o
%% C{\'a}lculos auxiliares
%%
%% 3 valor para o raio do "rabo" da "seta"
%%
%% repor o valor de auxa
  \auxa=\guardaauxa
%% dar a compensa{\c c}{\~a}o para o "rabo"
  \advance \auxa by 3%
%%
%% IMPORTANTE os valores de xa e ya v{\~a}o ser alterados
%%
 \!ucoord{\number\x}{\number\xl}{\number\x}{\number\auxa}{\number\d}{\xa}{1}%
 \!ucoord{\number\y}{\number\yl}{\number\y}{\number\auxa}{\number\d}{\ya}{1}%
 \!ucoord{\number\y}{\number\yl}{\number\xa}{3}{\number\d}{\xd}{-1}%
 \!ucoord{\number\x}{\number\xl}{\number\ya}{3}{\number\d}{\yd}{1}%
%% Constru{\c c}{\~a}o da "seta"
  \arrow <4pt> [.2,1.1] from {\xa} {\ya} to {\xb} {\yb}
%% e do seu "rabo"
  \circulararc -180 degrees from {\xa} {\ya} center at {\xd} {\yd}
\or  % seta de aplica{\c c}{\~a}o ("|-->")
  \auxa=3% valor para o meio-segmento do "rabo" da "seta"
%% c{\'a}lculo dos pontos (xmed,ymed) e (xd,yd) para o segmento de recta que
%% define o "rabo" da seta
 \!ucoord{\number\y}{\number\yl}{\number\xa}{\number\auxa}{\number\d}{\xmed}{-1}%
 \!ucoord{\number\x}{\number\xl}{\number\ya}{\number\auxa}{\number\d}{\ymed}{1}%
 \!ucoord{\number\y}{\number\yl}{\number\xa}{\number\auxa}{\number\d}{\xd}{1}%
 \!ucoord{\number\x}{\number\xl}{\number\ya}{\number\auxa}{\number\d}{\yd}{-1}%
%% Constru{\c c}{\~a}o da "seta"
  \arrow <4pt> [.2,1.1] from {\xa} {\ya} to {\xb} {\yb}
%% e do seu "rabo"
  \setlinear
  \plot {\xmed} {\ymed}  {\xd} {\yd} /
\fi
%% Coloca{\c c}{\~a}o do nome do morfismo.
%% Se os morfismos s{\~a}o horizontais ou verticais constro{\'\i} uma caixa
%% centrada no ponto pr{\'e}viamente calculado. Se as setas s{\~a}o
%% obl{\'\i}quas coloca a caixa de forma a n{\~a}o colidir com o morfismo 
%% tendo em aten{\c c}{\~a}o o quadrante assim como a posi{\c c}{\~a}o
%% relativa do morfismo e do respectivo nome.
\auxa=\xl
\advance \auxa by -\x%
\ifnum \auxa=0 
  \put {#7} at {\xc} {\yc}
\else
  \auxb=\yl
  \advance \auxb by -\y%
  \ifnum \auxb=0 \put {#7} at {\xc} {\yc}
  \else 
    \ifnum \auxa > 0 
      \ifnum \auxb > 0
        \ifnum #8=1
          \put {#7} [rb] at {\xc} {\yc}
        \else 
          \put {#7} [lt] at {\xc} {\yc}
        \fi
      \else
        \ifnum #8=1
          \put {#7} [lb] at {\xc} {\yc}
        \else 
          \put {#7} [rt] at {\xc} {\yc}
        \fi
      \fi
    \else
      \ifnum \auxb > 0 
        \ifnum #8=1
          \put {#7} [rt] at {\xc} {\yc}
        \else 
          \put {#7} [lb] at {\xc} {\yc}
        \fi
      \else
        \ifnum #8=1
          \put {#7} [lt] at {\xc} {\yc}
        \else 
          \put {#7} [rb] at {\xc} {\yc}
        \fi
      \fi
    \fi
  \fi
\fi
}

%%
%% Comando para construir a "seta" curvilinea entre dois objectos
%%
%% \cmor(<lista de pontos (n. impar)>){<etiqueta>}
%%
%% Em primeiro lugar {\'e} necess{\'a}rio modificar o comando plot de forma a
%% que a sintaxe de utiliza{\c c}{\~a}o do novo comando seja coerente com a
%% sintaxe dos restantes comandos
%%
\def\modifplot(#1{\!modifqcurve #1}
\def\!modifqcurve(#1,#2){\x=#1%
  \y=#2%
  \multiply \x by \expansao%
  \multiply \y by \expansao%
  \!start (\x,\y)
  \!modifQjoin}
\def\!modifQjoin(#1,#2)(#3,#4){\x=#1%
  \y=#2%
  \xl=#3%
  \yl=#4%
  \multiply \x by \expansao%
  \multiply \y by \expansao%
  \multiply \xl by \expansao%
  \multiply \yl by \expansao%
  \!qjoin (\x,\y) (\xl,\yl)             % \!qjoin  is defined in QUADRATIC
  \!ifnextchar){\!fim}{\!modifQjoin}}
\def\!fim){\ignorespaces}

%%
%% O comando para desenhar a seta vai receber a lista de pontos da qual
%% retira o {\'u}ltimo par de pontos, dependente da escolha dada pelo
%% utilizador a seta vai ser desenhada para cima, para baixo, para a
%% direita ou para a esquerda
%%
\def\setaxy(#1{\!pontosxy #1}
\def\!pontosxy(#1,#2){%
  \!maispontosxy}
\def\!maispontosxy(#1,#2)(#3,#4){%
  \!ifnextchar){\!fimxy#3,#4}{\!maispontosxy}}
\def\!fimxy#1,#2){\x=#1%
  \y=#2
  \multiply \x by \expansao
  \multiply \y by \expansao
  \xl=\x%
  \yl=\y%
  \aux=1%
  \multiply \aux by \auxa%
  \advance\xl by \aux%
  \aux=1%
  \multiply \aux by \auxb%
  \advance\yl by \aux%
  \arrow <4pt> [.2,1.1] from {\x} {\y} to {\xl} {\yl}}

%%
%% Temos agora a defini{\c c}{\~a}o do comando "cmor"
%%
\def\cmor#1 #2(#3,#4)#5{%
  \!ifnextchar[{\!cmora{#1}{#2}{#3}{#4}{#5}}{\!cmora{#1}{#2}{#3}{#4}{#5}[0] }}
\def\!cmora#1#2#3#4#5[#6]{%
  \ifcase #2% para cima "\pup" (pointing up)
      \auxa=0% x mant{\^e}m-se
      \auxb=1% o y "sobe" 
    \or% para baixo "\pdown" (pointing down)
      \auxa=0% x mant{\^e}m-se
      \auxb=-1% o y "desce" 
    \or% para a direita "\pright" (pointing right)
      \auxa=1% o x move-se para a direita
      \auxb=0% o y mant{\^e}m-se
    \or% para a esquerda "\pleft" (pointing left)
      \auxa=-1% o x move-se para a esquerda
      \auxb=0% o y mant{\^e}m-se
    \fi  % constru{\c c}{\~a}o do arco
  \ifcase #6  % arco (com seta) s{\'o}lido
    \modifplot#1% Desenhar o arco
    % constru{\c c}{\~a}o da seta
    \setaxy#1
  \or  % arco (com seta) a tracejado
    \setdashes
    \modifplot#1% Desenhar o arco
    \setaxy#1
    \setsolid
  \or  % arco sem seta
    \modifplot#1% Desenhar o arco
  \fi  % seta de injec{\c c}{\~a}o
%% coloca{\c c}{\~a}o da etiqueta do morfismo
  \x=#3%  
  \y=#4%
  \multiply \x by \expansao%
  \multiply \y by \expansao%
  \put {#5} at {\x} {\y}}

%%
%% Comando para construir os Objectos
%%  \obj(x,y){<texto>}[<etiqueta>]
%% 
\def\obj(#1,#2){%
  \!ifnextchar[{\!obja{#1}{#2}}{\!obja{#1}{#2}[Nulo]}}
\def\!obja#1#2[#3]#4{%
  \!ifnextchar[{\!objb{#1}{#2}{#3}{#4}}{\!objb{#1}{#2}{#3}{#4}[1]}}
\def\!objb#1#2#3#4[#5]{%
  \x=#1%
  \y=#2%
  \def\!pinta{\normalsize$\bullet$}% para definir o tamanho normal das pintas
  \def\!nulo{Nulo}%
  \def\!arg{#3}%
  \!compara{\!arg}{\!nulo}%
  \ifcompara\def\!arg{#4}\fi%
  \multiply \x by \expansao%
  \multiply \y by \expansao%
  \setbox\caixa=\hbox{#4}%
  \!coloca{(\!arg)(#1,#2)(\number\ht\caixa,\number\wd\caixa,\number\dp\caixa)}{\pilha}%
  \auxa=\wd\caixa \divide \auxa by 131072 
  \advance \auxa by 5
  \auxb=\ht\caixa
  \advance \auxb by \number\dp\caixa
  \divide \auxb by 131072 
  \advance \auxb by 5
%(\number\auxa,
%\number\auxb)
%  \aux=\ht\caixa \divide \auxa by 131072 
% \advance \auxa by 5 
%  \auxb=\dp\caixa \divide \auxb by 131072 
%  \advance \auxb by 8
  \ifcase \tipografo    % diagramas comutativos
    \put{#4} at {\x} {\y}
  \or                   % grafos dirigidos
    \ifcase #5 % c=0
      \put{#4} at {\x} {\y}
    \or        % n=1
      \put{\!pinta} at {\x} {\y}
      \advance \y by \number\auxb  % height+depth+5
      \put{#4} at {\x} {\y}
    \or        % ne=2
      \put{\!pinta} at {\x} {\y}
      \advance \auxa by -2  % para fazer o ajuste (imperfeito)
      \advance \auxb by -2  % ao raio da circunferência de centro (x,y)
      \advance \x by \number\auxa  % width+5
      \advance \y by \number\auxb  % height+depth+5
      \put{#4} at {\x} {\y}   
    \or        % e=3
      \put{\!pinta} at {\x} {\y}
      \advance \x by \number\auxa  % width+5
      \put{#4} at {\x} {\y}   
    \or        % se=4
      \put{\!pinta} at {\x} {\y}
      \advance \auxa by -2  % para fazer o ajuste (imperfeito)
      \advance \auxb by -2  % ao raio da circunferência de centro (x,y)
      \advance \x by \number\auxa  % width+5
      \advance \y by -\number\auxb  % height+depth+5
      \put{#4} at {\x} {\y}   
    \or        % s=5
      \put{\!pinta} at {\x} {\y}
      \advance \y by -\number\auxb  % height+depth+5
      \put{#4} at {\x} {\y}   
    \or        % sw=6
      \put{\!pinta} at {\x} {\y}
      \advance \auxa by -2  % para fazer o ajuste (imperfeito)
      \advance \auxb by -2  % ao raio da circunferência de centro (x,y)
      \advance \x by -\number\auxa  % width+5
      \advance \y by -\number\auxb  % height+depth+5
      \put{#4} at {\x} {\y}   
    \or        % w=7
      \put{\!pinta} at {\x} {\y}
      \advance \x by -\number\auxa  % width+5
      \put{#4} at {\x} {\y}   
    \or        % nw=8
      \put{\!pinta} at {\x} {\y}
      \advance \auxa by -2  % para fazer o ajuste (imperfeito)
      \advance \auxb by -2  % ao raio da circunferência de centro (x,y)
      \advance \x by -\number\auxa  % width+5
      \advance \y by \number\auxb  % height+depth+5
      \put{#4} at {\x} {\y}   
    \fi
  \or                   % grafos não dirigidos
    \ifcase #5 % c=0
      \put{#4} at {\x} {\y}
    \or        % n=1
      \put{\!pinta} at {\x} {\y}
      \advance \y by \number\auxb  % height+depth+5
      \put{#4} at {\x} {\y}
    \or        % ne=2
      \put{\!pinta} at {\x} {\y}
      \advance \auxa by -2  % para fazer o ajuste (imperfeito)
      \advance \auxb by -2  % ao raio da circunferência de centro (x,y)
      \advance \x by \number\auxa  % width+5
      \advance \y by \number\auxb  % height+depth+5
      \put{#4} at {\x} {\y}   
    \or        % e=3
      \put{\!pinta} at {\x} {\y}
      \advance \x by \number\auxa  % width+5
      \put{#4} at {\x} {\y}   
    \or        % se=4
      \put{\!pinta} at {\x} {\y}
      \advance \auxa by -2  % ver acima
      \advance \auxb by -2
      \advance \x by \number\auxa  % width+5
      \advance \y by -\number\auxb % height+depth+5
      \put{#4} at {\x} {\y}   
    \or        % s=5
      \put{\!pinta} at {\x} {\y}
      \advance \y by -\number\auxb % height+depth+5
      \put{#4} at {\x} {\y}   
    \or        % sw=6
      \put{\!pinta} at {\x} {\y}
      \advance \auxa by -2  % ver acima
      \advance \auxb by -2
      \advance \x by -\number\auxa % width+5
      \advance \y by -\number\auxb % height+depth+5
      \put{#4} at {\x} {\y}   
    \or        % w=7
      \put{\!pinta} at {\x} {\y}
      \advance \x by -\number\auxa % width+5
      \put{#4} at {\x} {\y}   
    \or        % nw=8
      \put{\!pinta} at {\x} {\y}
      \advance \auxa by -2  % ver acima
      \advance \auxb by -2
      \advance \x by -\number\auxa % width+5
      \advance \y by \number\auxb  % height+depth+5
      \put{#4} at {\x} {\y}   
    \fi
%  \or % grafos dirigidos com molduras circulares nos objectos
%    \advance \auxa by 4
%    \put{\circle{\auxa}} [Bl] at {\x} {\y}
%    \put{#4} at {\x} {\y}
%  \or % grafos não dirigidos com molduras circulares nos objectos
   \else % grafos com molduras circulares nos objectos
     \ifnum\auxa<\auxb % determina a maior das dimensões
       \aux=\auxb
     \else
       \aux=\auxa
     \fi
% se a largura da caixa é menor do que 1em então o tamanho 
% tamanho é ajustado para esse valor mínimo
     \ifdim\wd\caixa<1em
       \dimen99 = 1em
       \aux=\dimen99 \divide \aux by 131072 
       \advance \aux by 5
     \fi
     \advance\aux by -2 %folga entre o objecto e a moldura
     \multiply\aux by 2 % 
     \ifnum\aux<30
       \put{\circle{\aux}} [Bl] at {\x} {\y}
     \else
       \multiply\auxa by 2
       \multiply\auxb by 2
       \put{\oval(\auxa,\auxb)} [Bl] at {\x} {\y}
     \fi
     \put{#4} at {\x} {\y}
   \fi   
}

\catcode`!=12 %  *****  THIS MUST NEVER BE OMITTED (Ver PiCTeX)

  \input miniltx
  \def\Gin@driver{pdftex.def}
  \input color.sty
  \input graphicx.sty
  \resetatcatcode

%\input graphicx.tex
%
%    Bundle of my macros.    Version 1.2.0.beta
%    The best use is to paste all of them into the papers
%     1/8/2005
%

%
%    Fonts.    Version 1.2.0.beta
%    The best use is to paste all of them into the papers
%     1/8/2005
%
%
% History:
%	3 Agosto 2005: ChernSimons.tex
%
%%%%%%%%%%%%%%%%%%%%%%%%%%%%

%%%%%%%%%%%%%%%%%%%%%%%%%%%%%%%%%%%%%%%%%%%%%%%%%%%%
%% Font Types	%%%%%%%%%%%%%%%%%%%%%%%%%%%%%%%%%%%%%%%%%%%%
%%%%%%%%%%%%%%%%%%%%%%%%%%%%%%%%%%%%%%%%%%%%%%%%%%%%%
\def\Serif{cmr}
\def\SerifBold{cmbx}
\def\SerifItalics{cmti}
\def\SerifSlanted{cmsl}
\def\SerifBoldItalics{cmbxti}
\def\SansSerif{cmss}
\def\SansSerifBold{cmssbx}
\def\SansSerifItalics{cmssi}
\def\SansSerifSlanted{cmssi}%%
\def\Math{cmmi}
\def\Symbols{cmsy}
\def\MathBold{cmmib}
\def\MoreSymbols{cmex}
\def\Typewriter{cmtt}
\def\Gothic{eufm}
\def\Double{msbm}
\def\Relazioni{msam}

%% Font Declarations	
%\font\tenbg=cmmib10%
%\def\bg{\tenbg}%
%%%%%%%%%%%%%%%%%%%%%%%%%%%%%%%%%%%%%%%%%%%%%%%%%%%%%%
%%%	5		%%%%%%%%%%%%%%%%%%%%%%%%%%%%%%%%%%%%%%%%%%%%
%%%	%%%%%%%%%%%%%%%%%%%%%%
= 			\Serif10 			at 5pt
= 		\SerifBold10 		at 5pt
= 	\SerifItalics10 	at 5pt
=		\SerifSlanted10 	at 5pt
=	\SerifBoldItalics10	at 5pt
= 		\SansSerif10 		at 5pt
=	\SansSerifBold10	at 5pt
=	\SansSerifItalics10	at 5pt
=	\SansSerifSlanted10	at 5pt
=				\Math10				at 5pt
=			\MathBold10			at 5pt
=			\Symbols10			at 5pt
=		\MoreSymbols10		at 5pt
=		\Typewriter10		at 5pt
=			\Gothic10			at 5pt
=			\Double10			at 5pt

%%%	7		%%%%%%%%%%%%%%%%%%%%%%%%%%%%%%%%%%%%%%%%%%%
%%%	%%%%%%%%%%%%%%%%%%%%%%%
= 			\Serif10 			at 7pt
= 		\SerifBold10 		at 7pt
= 	\SerifItalics10 	at 7pt
=	\SerifSlanted10 	at 7pt
=\SerifBoldItalics10	at 7pt
= 		\SansSerif10 		at 7pt
= 	\SansSerifBold10 	at 7pt
=\SansSerifItalics10	at 7pt
=\SansSerifSlanted10	at 7pt
=			\Math10				at 7pt
=		\MathBold10			at 7pt
=			\Symbols10			at 7pt
=		\MoreSymbols10		at 7pt
=		\Typewriter10		at 7pt
=			\Gothic10			at 7pt
=			\Double10			at 7pt

%%%	8		%%%%%%%%%%%%%%%%%%%%%%%%%%%%%%%%%%%%%%%%%
%%%	%%%%%%%%%%%%%%%%%%%%%%%%%
= 			\Serif10 			at 8pt
= 		\SerifBold10 		at 8pt
= 	\SerifItalics10 	at 8pt
=	\SerifSlanted10 	at 8pt
=\SerifBoldItalics10	at 8pt
= 		\SansSerif10 		at 8pt
= 	\SansSerifBold10 	at 8pt
=\SansSerifItalics10 at 8pt
=\SansSerifSlanted10 at 8pt
=			\Math10				at 8pt
=		\MathBold10			at 8pt
=			\Symbols10			at 8pt
=		\MoreSymbols10		at 8pt
=		\Typewriter10		at 8pt
=			\Gothic10			at 8pt
=			\Double10			at 8pt

%%%	10		%%%%%%%%%%%%%%%%%%%%%%%%%%%%%%%%%%%%%
%%%	%%%%%%%%%%%%%%%%%%%%%%%%%%%%%
= 			\Serif10 			at 10pt
= 		\SerifBold10 		at 10pt
= 		\SerifItalics10 	at 10pt
=		\SerifSlanted10 	at 10pt
=	\SerifBoldItalics10	at 10pt
= 		\SansSerif10 		at 10pt
= 	\SansSerifBold10 	at 10pt
= 	\SansSerifItalics10 at 10pt
= 	\SansSerifSlanted10 at 10pt
=				\Math10				at 10pt
=			\MathBold10			at 10pt
=			\Symbols10			at 10pt
=		\MoreSymbols10		at 10pt
=		\Typewriter10		at 10pt
=			\Gothic10			at 10pt
=			\Double10			at 10pt
=			\Relazioni10			at 10pt

%%%	12		%%%%%%%%%%%%%%%%%%%%%%%%%%%%%%%%%%%%%
%%%	%%%%%%%%%%%%%%%%%%%%%%%%%%%%%
= 				\Serif10 			at 12pt
= 			\SerifBold10 		at 12pt
= 		\SerifItalics10 	at 12pt
=		\SerifSlanted10 	at 12pt
=	\SerifBoldItalics10	at 12pt
= 			\SansSerif10 		at 12pt
= 		\SansSerifBold10 	at 12pt
= 	\SansSerifItalics10 at 12pt
= 	\SansSerifSlanted10 at 12pt
=				\Math10				at 12pt
=			\MathBold10			at 12pt
=			\Symbols10			at 12pt
=		\MoreSymbols10		at 12pt
=			\Typewriter10		at 12pt
=				\Gothic10			at 12pt
=				\Double10			at 12pt

%%%	14		%%%%%%%%%%%%%%%%%%%%%%%%%%%%%%%%
= 			\Serif10 			at 14pt
= 		\SerifBold10 		at 14pt
= 	\SerifItalics10 	at 14pt
=		\SerifSlanted10 	at 14pt
=	\SerifBoldItalics10	at 14pt
= 		\SansSerif10 		at 14pt
= 	\SansSerifBold10 	at 14pt
= \SansSerifSlanted10 at 14pt
= \SansSerifItalics10 at 14pt
=				\Math10				at 14pt
=			\MathBold10			at 14pt
=			\Symbols10			at 14pt
=		\MoreSymbols10		at 14pt
=		\Typewriter10		at 14pt
=			\Gothic10			at 14pt
=			\Double10			at 14pt

%% Styles	%%%%%%%%%%%%%%%%%%%%%%%%%%%%%%%%%%%%%%%%%%%%%
%% %%%%%%%%%%%%%%%%%%%%%%%%%%%%%%%%%%%%%%%%%%%%%
\def\NormalStyle{\parindent=5pt\parskip=3pt\normalbaselineskip=14pt%
\def\nt{\tenSerif}%
\def\rm{\fam0\tenSerif}%
\textfont0=\tenSerif\scriptfont0=\sevenSerif\scriptscriptfont0=\fiveSerif%text(\tenrm)
\textfont1=\tenMath\scriptfont1=\sevenMath\scriptscriptfont1=\fiveMath%math(\tenmi)
\textfont2=\tenSymbols\scriptfont2=\sevenSymbols\scriptscriptfont2=\fiveSymbols%symbol(\tensy)
\textfont3=\tenMoreSymbols\scriptfont3=\sevenMoreSymbols\scriptscriptfont3=\fiveMoreSymbols%ex(tenex)
\textfont\itfam=\tenSerifItalics\def\it{\fam\itfam\tenSerifItalics}%
\textfont\slfam=\tenSerifSlanted\def\sl{\fam\slfam\tenSerifSlanted}%
\textfont\ttfam=\tenTypewriter\def\tt{\fam\ttfam\tenTypewriter}%
\textfont\bffam=\tenSerifBold%
\def\bf{\fam\bffam\tenSerifBold}\scriptfont\bffam=\sevenSerifBold\scriptscriptfont\bffam=\fiveSerifBold%
\def\cal{\tenSymbols}%
\def\greekbold{\tenMathBold}%
\def\gothic{\tenGothic}%
\def\Bbb{\tenDouble}%
\def\LieFont{\tenSerifItalics}%
\nt\normalbaselines\baselineskip=14pt%
}

%%%%%%%%%%%%%%%%%%%%%%%%%%%%%%%%%%%%%%%%%%%%%%%%%%%%%%%
%%%%%%%%%%%%%%%%%%%%%%
\def\TitleStyle{\parindent=0pt\parskip=0pt\normalbaselineskip=15pt%
\def\nt{\fourteenSansSerifBold}%
\def\rm{\fam0\fourteenSansSerifBold}%
\textfont0=\fourteenSansSerifBold\scriptfont0=\tenSansSerifBold\scriptscriptfont0=\eightSansSerifBold%text(\fourteenrm)
\textfont1=\fourteenMath\scriptfont1=\tenMath\scriptscriptfont1=\eightMath%math(\fourteenmi)
\textfont2=\fourteenSymbols\scriptfont2=\tenSymbols\scriptscriptfont2=\eightSymbols%symbol(\fourteensy)
\textfont3=\fourteenMoreSymbols\scriptfont3=\tenMoreSymbols\scriptscriptfont3=\eightMoreSymbols%ex(fourteenex)
\textfont\itfam=\fourteenSansSerifItalics\def\it{\fam\itfam\fourteenSansSerifItalics}%
\textfont\slfam=\fourteenSansSerifSlanted\def\sl{\fam\slfam\fourteenSerifSansSlanted}%
\textfont\ttfam=\fourteenTypewriter\def\tt{\fam\ttfam\fourteenTypewriter}%
\textfont\bffam=\fourteenSansSerif%
\def\bf{\fam\bffam\fourteenSansSerif}\scriptfont\bffam=\tenSansSerif\scriptscriptfont\bffam=\eightSansSerif%
\def\cal{\fourteenSymbols}%
\def\greekbold{\fourteenMathBold}%
\def\gothic{\fourteenGothic}%
\def\Bbb{\fourteenDouble}%
\def\LieFont{\fourteenSerifItalics}%
\nt\normalbaselines\baselineskip=15pt%
}

%%%%%%%%%%%%%%%%%%%%%%%%%%%%%%%%%%%%%%%%%%%%%%%%%%%%%%%%
%%%%%%%%%%%%%%%%%%%%%
\def\PartStyle{\parindent=0pt\parskip=0pt\normalbaselineskip=15pt%
\def\nt{\fourteenSansSerifBold}%
\def\rm{\fam0\fourteenSansSerifBold}%
\textfont0=\fourteenSansSerifBold\scriptfont0=\tenSansSerifBold\scriptscriptfont0=\eightSansSerifBold%text(\fourteenrm)
\textfont1=\fourteenMath\scriptfont1=\tenMath\scriptscriptfont1=\eightMath%math(\fourteenmi)
\textfont2=\fourteenSymbols\scriptfont2=\tenSymbols\scriptscriptfont2=\eightSymbols%symbol(\fourteensy)
\textfont3=\fourteenMoreSymbols\scriptfont3=\tenMoreSymbols\scriptscriptfont3=\eightMoreSymbols%ex(fourteenex)
\textfont\itfam=\fourteenSansSerifItalics\def\it{\fam\itfam\fourteenSansSerifItalics}%
\textfont\slfam=\fourteenSansSerifSlanted\def\sl{\fam\slfam\fourteenSerifSansSlanted}%
\textfont\ttfam=\fourteenTypewriter\def\tt{\fam\ttfam\fourteenTypewriter}%
\textfont\bffam=\fourteenSansSerif%
\def\bf{\fam\bffam\fourteenSansSerif}\scriptfont\bffam=\tenSansSerif\scriptscriptfont\bffam=\eightSansSerif%
\def\cal{\fourteenSymbols}%
\def\greekbold{\fourteenMathBold}%
\def\gothic{\fourteenGothic}%
\def\Bbb{\fourteenDouble}%
\def\LieFont{\fourteenSerifItalics}%
\nt\normalbaselines\baselineskip=15pt%
}

%%%%%%%%%%%%%%%%%%%%%%%%%%%%%%%%%%%%%%%%%%%%%%%%%%%%%%%%
%%%%%%%%%%%%%%%%%%%%%
\def\ChapterStyle{\parindent=0pt\parskip=0pt\normalbaselineskip=15pt%
\def\nt{\fourteenSansSerifBold}%
\def\rm{\fam0\fourteenSansSerifBold}%
\textfont0=\fourteenSansSerifBold\scriptfont0=\tenSansSerifBold\scriptscriptfont0=\eightSansSerifBold%text(\fourteenrm)
\textfont1=\fourteenMath\scriptfont1=\tenMath\scriptscriptfont1=\eightMath%math(\fourteenmi)
\textfont2=\fourteenSymbols\scriptfont2=\tenSymbols\scriptscriptfont2=\eightSymbols%symbol(\fourteensy)
\textfont3=\fourteenMoreSymbols\scriptfont3=\tenMoreSymbols\scriptscriptfont3=\eightMoreSymbols%ex(fourteenex)
\textfont\itfam=\fourteenSansSerifItalics\def\it{\fam\itfam\fourteenSansSerifItalics}%
\textfont\slfam=\fourteenSansSerifSlanted\def\sl{\fam\slfam\fourteenSerifSansSlanted}%
\textfont\ttfam=\fourteenTypewriter\def\tt{\fam\ttfam\fourteenTypewriter}%
\textfont\bffam=\fourteenSansSerif%
\def\bf{\fam\bffam\fourteenSansSerif}\scriptfont\bffam=\tenSansSerif\scriptscriptfont\bffam=\eightSansSerif%
\def\cal{\fourteenSymbols}%
\def\greekbold{\fourteenMathBold}%
\def\gothic{\fourteenGothic}%
\def\Bbb{\fourteenDouble}%
\def\LieFont{\fourteenSerifItalics}%
\nt\normalbaselines\baselineskip=15pt%
}

%%%%%%%%%%%%%%%%%%%%%%%%%%%%%%%%%%%%%%%%%%%%%%%%%%%%%%%%
%%%%%%%%%%%%%%%%%%%%%
\def\SectionStyle{\parindent=0pt\parskip=0pt\normalbaselineskip=13pt%
\def\nt{\twelveSansSerifBold}%
\def\rm{\fam0\twelveSansSerifBold}%
\textfont0=\twelveSansSerifBold\scriptfont0=\eightSansSerifBold\scriptscriptfont0=\eightSansSerifBold%text(\fourteenrm)
\textfont1=\twelveMath\scriptfont1=\eightMath\scriptscriptfont1=\eightMath%math(\fourteenmi)
\textfont2=\twelveSymbols\scriptfont2=\eightSymbols\scriptscriptfont2=\eightSymbols%symbol(\fourteensy)
\textfont3=\twelveMoreSymbols\scriptfont3=\eightMoreSymbols\scriptscriptfont3=\eightMoreSymbols%ex(fourteenex)
\textfont\itfam=\twelveSansSerifItalics\def\it{\fam\itfam\twelveSansSerifItalics}%
\textfont\slfam=\twelveSansSerifSlanted\def\sl{\fam\slfam\twelveSerifSansSlanted}%
\textfont\ttfam=\twelveTypewriter\def\tt{\fam\ttfam\twelveTypewriter}%
\textfont\bffam=\twelveSansSerif%
\def\bf{\fam\bffam\twelveSansSerif}\scriptfont\bffam=\eightSansSerif\scriptscriptfont\bffam=\eightSansSerif%
\def\cal{\twelveSymbols}%
\def\bg{\twelveMathBold}%
\def\gothic{\twelveGothic}%
\def\Bbb{\twelveDouble}%
\def\LieFont{\twelveSerifItalics}%
\nt\normalbaselines\baselineskip=13pt%
}

%%%%%%%%%%%%%%%%%%%%%%%%%%%%%%%%%%%%%%%%%%%%%%%
\def\SubSectionStyle{\parindent=0pt\parskip=0pt\normalbaselineskip=13pt%
\def\nt{\twelveSansSerifItalics}%
\def\rm{\fam0\twelveSansSerifItalics}%
\textfont0=\twelveSansSerifItalics\scriptfont0=\eightSansSerifItalics\scriptscriptfont0=\eightSansSerifItalics%
\textfont1=\twelveMath\scriptfont1=\eightMath\scriptscriptfont1=\eightMath%
\textfont2=\twelveSymbols\scriptfont2=\eightSymbols\scriptscriptfont2=\eightSymbols%
\textfont3=\twelveMoreSymbols\scriptfont3=\eightMoreSymbols\scriptscriptfont3=\eightMoreSymbols%
\textfont\itfam=\twelveSansSerif\def\it{\fam\itfam\twelveSansSerif}%
\textfont\slfam=\twelveSansSerifSlanted\def\sl{\fam\slfam\twelveSerifSansSlanted}%
\textfont\ttfam=\twelveTypewriter\def\tt{\fam\ttfam\twelveTypewriter}%
\textfont\bffam=\twelveSansSerifBold%
\def\bf{\fam\bffam\twelveSansSerifBold}\scriptfont\bffam=\eightSansSerifBold\scriptscriptfont\bffam=\eightSansSerifBold%
\def\cal{\twelveSymbols}%
\def\greekbold{\twelveMathBold}%
\def\gothic{\twelveGothic}%
\def\Bbb{\twelveDouble}%
\def\LieFont{\twelveSerifItalics}%
\nt\normalbaselines\baselineskip=13pt%
}

%%%%%%%%%%%%%%%%%%%%%%%%%%%%%%%%%%%%%%%%%%%%%%%%%%%%%%%%%%%
%%%%%%%%%%%%%%%%%%
\def\AuthorStyle{\parindent=0pt\parskip=0pt\normalbaselineskip=14pt%
\def\nt{\tenSerif}%
\def\rm{\fam0\tenSerif}%
\textfont0=\tenSerif\scriptfont0=\sevenSerif\scriptscriptfont0=\fiveSerif%text(\tenrm)
\textfont1=\tenMath\scriptfont1=\sevenMath\scriptscriptfont1=\fiveMath%math(\tenmi)
\textfont2=\tenSymbols\scriptfont2=\sevenSymbols\scriptscriptfont2=\fiveSymbols%symbol(\tensy)
\textfont3=\tenMoreSymbols\scriptfont3=\sevenMoreSymbols\scriptscriptfont3=\fiveMoreSymbols%ex(tenex)
\textfont\itfam=\tenSerifItalics\def\it{\fam\itfam\tenSerifItalics}%
\textfont\slfam=\tenSerifSlanted\def\sl{\fam\slfam\tenSerifSlanted}%
\textfont\ttfam=\tenTypewriter\def\tt{\fam\ttfam\tenTypewriter}%
\textfont\bffam=\tenSerifBold%
\def\bf{\fam\bffam\tenSerifBold}\scriptfont\bffam=\sevenSerifBold\scriptscriptfont\bffam=\fiveSerifBold%
\def\cal{\tenSymbols}%
\def\greekbold{\tenMathBold}%
\def\gothic{\tenGothic}%
\def\Bbb{\tenDouble}%
\def\LieFont{\tenSerifItalics}%
\nt\normalbaselines\baselineskip=14pt%
}

%%%%%%%%%%%%%%%%%%%%%%%%%%%%%%%%%%%%%%%%%%%%%%%%%%%%%%%%%%%
%%%%%%%%%%%%%%%%%%
\def\AddressStyle{\parindent=0pt\parskip=0pt\normalbaselineskip=14pt%
\def\nt{\eightSerif}%
\def\rm{\fam0\eightSerif}%
\textfont0=\eightSerif\scriptfont0=\sevenSerif\scriptscriptfont0=\fiveSerif%text(\tenrm)
\textfont1=\eightMath\scriptfont1=\sevenMath\scriptscriptfont1=\fiveMath%math(\tenmi)
\textfont2=\eightSymbols\scriptfont2=\sevenSymbols\scriptscriptfont2=\fiveSymbols%symbol(\tensy)
\textfont3=\eightMoreSymbols\scriptfont3=\sevenMoreSymbols\scriptscriptfont3=\fiveMoreSymbols%ex(tenex)
\textfont\itfam=\eightSerifItalics\def\it{\fam\itfam\eightSerifItalics}%
\textfont\slfam=\eightSerifSlanted\def\sl{\fam\slfam\eightSerifSlanted}%
\textfont\ttfam=\eightTypewriter\def\tt{\fam\ttfam\eightTypewriter}%
\textfont\bffam=\eightSerifBold%
\def\bf{\fam\bffam\eightSerifBold}\scriptfont\bffam=\sevenSerifBold\scriptscriptfont\bffam=\fiveSerifBold%
\def\cal{\eightSymbols}%
\def\greekbold{\eightMathBold}%
\def\gothic{\eightGothic}%
\def\Bbb{\eightDouble}%
\def\LieFont{\eightSerifItalics}%
\nt\normalbaselines\baselineskip=14pt%
}

%%%%%%%%%%%%%%%%%%%%%%%%%%%%%%%%%%%%%%%%%%%%%%%%%%%%%%%%%%%
%%%%%%%%%%%%%%%%%%
\def\AbstractStyle{\parindent=0pt\parskip=0pt\normalbaselineskip=12pt%
\def\nt{\eightSerif}%
\def\rm{\fam0\eightSerif}%
\textfont0=\eightSerif\scriptfont0=\sevenSerif\scriptscriptfont0=\fiveSerif%text(\tenrm)
\textfont1=\eightMath\scriptfont1=\sevenMath\scriptscriptfont1=\fiveMath%math(\tenmi)
\textfont2=\eightSymbols\scriptfont2=\sevenSymbols\scriptscriptfont2=\fiveSymbols%symbol(\tensy)
\textfont3=\eightMoreSymbols\scriptfont3=\sevenMoreSymbols\scriptscriptfont3=\fiveMoreSymbols%ex(tenex)
\textfont\itfam=\eightSerifItalics\def\it{\fam\itfam\eightSerifItalics}%
\textfont\slfam=\eightSerifSlanted\def\sl{\fam\slfam\eightSerifSlanted}%
\textfont\ttfam=\eightTypewriter\def\tt{\fam\ttfam\eightTypewriter}%
\textfont\bffam=\eightSerifBold%
\def\bf{\fam\bffam\eightSerifBold}\scriptfont\bffam=\sevenSerifBold\scriptscriptfont\bffam=\fiveSerifBold%
\def\cal{\eightSymbols}%
\def\greekbold{\eightMathBold}%
\def\gothic{\eightGothic}%
\def\Bbb{\eightDouble}%
\def\LieFont{\eightSerifItalics}%
\nt\normalbaselines\baselineskip=12pt%
}

%%%%%%%%%%%%%%%%%%%%%%%%%%%%%%%%%%%%%%%%%%%%%
\def\RefsStyle{\parindent=0pt\parskip=0pt%
\def\nt{\eightSerif}%
\def\rm{\fam0\eightSerif}%
\textfont0=\eightSerif\scriptfont0=\sevenSerif\scriptscriptfont0=\fiveSerif%text(\tenrm)
\textfont1=\eightMath\scriptfont1=\sevenMath\scriptscriptfont1=\fiveMath%math(\tenmi)
\textfont2=\eightSymbols\scriptfont2=\sevenSymbols\scriptscriptfont2=\fiveSymbols%symbol(\tensy)
\textfont3=\eightMoreSymbols\scriptfont3=\sevenMoreSymbols\scriptscriptfont3=\fiveMoreSymbols%ex(tenex)
\textfont\itfam=\eightSerifItalics\def\it{\fam\itfam\eightSerifItalics}%
\textfont\slfam=\eightSerifSlanted\def\sl{\fam\slfam\eightSerifSlanted}%
\textfont\ttfam=\eightTypewriter\def\tt{\fam\ttfam\eightTypewriter}%
\textfont\bffam=\eightSerifBold%
\def\bf{\fam\bffam\eightSerifBold}\scriptfont\bffam=\sevenSerifBold\scriptscriptfont\bffam=\fiveSerifBold%
\def\cal{\eightSymbols}%
\def\greekbold{\eightMathBold}%
\def\gothic{\eightGothic}%
\def\Bbb{\eightDouble}%
\def\LieFont{\eightSerifItalics}%
\nt\normalbaselines\baselineskip=10pt%
}

%%%%%%%%%%%%%%%%%%%%%%%%%%%%%%%%%%%%%%%%%%%%%%%%%%%%%%%%%%%
%%%%%%%%%%%%%%%%%%%

%%%%%%%%%%%%%%%%%%%%%%%%%%%%%%%%%%%%%%%%%%%%%%%%%%%%%%%%%%%%
%%%%%%%%%%%%%%%%%%

%
%    Various Libraries.    Version 1.2.0.beta
%    The best use is to paste all of them into the papers
%     1/8/2005
%

%%%%%%%%%%%%%%%%%%%%%%%%%%%%%%
%%%%%%			Utilities		 %%%%%%
%%%%%%%%%%%%%%%%%%%%%%%%%%%%%%

% Definition modes  %
\def\ModeYes{yes}
\def\ModeNo{no}

\def\ModeUndef{undefined}

%%%%%%%%%%%%

\def\nx{\noexpand}
\def\ni{\noindent}
\def\newpage{\vfill\eject}

\def\ss{\vskip 5pt}
\def\ms{\vskip 10pt}
\def\bs{\vskip 20pt}

 \def\,{\mskip\thinmuskip}
 \def\!{\mskip-\thinmuskip}
 \def\>{\mskip\medmuskip}
 \def\;{\mskip\thickmuskip}

%%%%%%%%%%%%%%%%%%%%%%%%%%%%%%
%%%%%%		Bibliography		 %%%%%%
%%%%%%%%%%%%%%%%%%%%%%%%%%%%%%
%
% Usage:
%	[\SetModeAuto]
% ... 
%	\bib{libro1}{L.Fatibene, ...}
%	\bib{libro2}{L.Fatibene, ...}
% ...
%	(see \ref{libro2} and \ref{libro1})
% ...
% 	\ShowBiblio
%

% Definition modes  %
\def\refsModePost{post}
\def\refsModeAuto{auto}

\def\dbRefsSatusModeOk{ok}
\def\dbRefsSatusModeError{error}
\def\dbRefsSatusModeWarning{warning}

%%%%%%%%%%%%

\newcount\BNUM
\BNUM=0

\def\refs{}

\def\SetModePost{\xdef\refsMode{\refsModePost}}			%	Items are numbered by Citation order
		%	Items are numbered by Insertion order
\SetModePost

\def\dbRefsStatusOk{%
	\xdef\dbRefsStatus{\dbRefsSatusModeOk}%
	\xdef\dbRefsError{\ModeNo}%
	\xdef\dbRefsWarning{\ModeNo}%
	\xdef\dbRefsInfo{\ModeNo}%
}

\def\dbRefs{%
}

\def\dbRefsGet#1{%
	\xdef\found{N}\xdef\ikey{#1}\dbRefsStatusOk%
	\xdef\key{\ModeUndef}\xdef\tag{\ModeUndef}\xdef\tail{\ModeUndef}%
	\dbRefs%
}

\def\NextRefsTag{%
	\global\advance\BNUM by 1%
}
\def\ShowTag#1{{\bf [#1]}}

\def\dbRefsInsert#1#2{%
\dbRefsGet{#1}%
\if\found Y %
   \xdef\dbRefsStatus{\dbRefsSatusModeWarning}%
   \xdef\dbRefsWarning{record is already there}%
   \xdef\dbRefsInfo{record not inserted}%
\else%
   \toks2=\expandafter{\dbRefs}%
   \ifx\refsMode\refsModeAuto \NextRefsTag
    \xdef\dbRefs{%
   	\the\toks2 \nx\xdef\nx\dbx{#1}%
	\nx\ifx\nx\ikey %
		\nx\dbx\nx\xdef\nx\found{Y}%
		\nx\xdef\nx\key{#1}%
		\nx\xdef\nx\tag{\the\BNUM}%
		\nx\xdef\nx\tail{#2}%
	\nx\fi}%
	\global\xdef\refs{\refs \ss\ni[\the\BNUM]\ #2\par}%%%%
   \fi%   	
   \ifx\refsMode\refsModePost 
    \xdef\dbRefs{%
   	\the\toks2 \nx\xdef\nx\dbx{#1}%
	\nx\ifx\nx\ikey %
		\nx\dbx\nx\xdef\nx\found{Y}%
		\nx\xdef\nx\key{#1}%
		\nx\xdef\nx\tag{\ModeUndef}%
		\nx\xdef\nx\tail{#2}%
	\nx\fi}%
   \fi%
\fi%
}

\def\dbRefsEdit#1#2#3{\dbRefsGet{#1}%
\if\found N 
   \xdef\dbRefsStatus{\dbRefsSatusModeError}%
   \xdef\dbRefsError{record is not there}%
   \xdef\dbRefsInfo{record not edited}%
\else%
   \toks2=\expandafter{\dbRefs}%
   \xdef\dbRefs{\the\toks2%
   \nx\xdef\nx\dbx{#1}%
   \nx\ifx\nx\ikey\nx\dbx %
	\nx\xdef\nx\found{Y}%
	\nx\xdef\nx\key{#1}%
	\nx\xdef\nx\tag{#2}%
	\nx\xdef\nx\tail{#3}%
   \nx\fi}%
\fi%
}

\def\bib#1#2{\RefsStyle\dbRefsInsert{#1}{#2}%
	\ifx\dbRefsStatus\dbRefsSatusModeWarning %
		\message{^^J}%
		\message{WARNING: Reference [#1] is doubled.^^J}%
	\fi%
}

\def\ref#1{\dbRefsGet{#1}%
\ifx\found N %
  \message{^^J}%
  \message{ERROR: Reference [#1] unknown.^^J}%
  \ShowTag{??}%
\else%
	\ifx\tag\ModeUndef \NextRefsTag%
		\dbRefsEdit{#1}{\the\BNUM}{\tail}%
		\dbRefsGet{#1}%
		\global\xdef\refs{\refs \ss\ni [\tag]\ \tail\par}%%%%
	\fi
	\ShowTag{\tag}%
\fi%
}

\def\ShowBiblio{\ms\Ensure{\SectionEnsure}%
{\SectionStyle\ni References}%
{\RefsStyle\refs}%
}

%%%%%%%%%%%%%%%%%%%%%%%%%%%%%%
%%%%%%		Label DB			 %%%%%%
%%%%%%%%%%%%%%%%%%%%%%%%%%%%%%
\newcount\CHANGES
\CHANGES=0
\def\AuxFile{7}
\def\PreventDoubleOn{\xdef\PreventDoubleLabel{\ModeYes}}

\PreventDoubleOn

\def\StoreLabel#1#2{\xdef\itag{#2}% Mantiene FileAux e ritorna #2 in \itag
 \ifx\PreModeStatus\ModeNo %
   \message{^^J}%
   \errmessage{You can't use Check without starting with OpenPreMode (and finishing with ClosePreMode)^^J}%
 \else%
   \immediate\write\AuxFile{\nx\dbLabelPreInsert{#1}{\itag}}%     
   \dbLabelGet{#1}%
   \ifx\itag\tag %
   \else%
	\global\advance\CHANGES by 1%
 	\xdef\itag{(?.??)}%
    \fi%
   \fi%
}

\def\PreModeStatus{\ModeNo}

\def\ClosePreMode{\immediate\closeout\AuxFile%
  \ifnum\CHANGES=0%
	\message{^^J}%
	\message{**********************************^^J}%
	\message{**  NO CHANGES TO THE AuxFile  **^^J}%
	\message{**********************************^^J}%
 \else%
	\message{^^J}%
	\message{**************************************************^^J}%
	\message{**  PLAEASE TYPESET IT AGAIN (\the\CHANGES)  **^^J}%
    \errmessage{**************************************************^^ J}%
  \fi%
  \edef\PreModeStatus{\ModeNo}%
}

\def\dbLabelSatusModeOk{ok}

\def\dbLabelSatusModeWarning{warning}

\def\dbLabelStatusOk{%
	\xdef\dbLabelStatus{\dbLabelSatusModeOk}%
	\xdef\dbLabelError{\ModeNo}%
	\xdef\dbLabelWarning{\ModeNo}%
	\xdef\dbLabelInfo{\ModeNo}%
}

\def\dbLabel{%
}

\def\dbLabelGet#1{%
	\xdef\found{N}\xdef\ikey{#1}\dbLabelStatusOk%
	\xdef\key{\ModeUndef}\xdef\tag{\ModeUndef}\xdef\pre{\ModeUndef}%
	\dbLabel%
}

\def\ShowLabel#1{%
 \dbLabelGet{#1}%
 \ifx\tag \ModeUndef %
 	\global\advance\CHANGES by 1%
 	(?.??)%
 \else%
 	\tag%
 \fi%
}

\def\dbLabelPreInsert#1#2{\dbLabelGet{#1}%
\if\found Y %
  \xdef\dbLabelStatus{\dbLabelSatusModeWarning}%
   \xdef\dbLabelWarning{Label is already there}%
   \xdef\dbLabelInfo{Label not inserted}%
   \message{^^J}%
   \errmessage{Double pre definition of label [#1]^^J}%
\else%
   \toks2=\expandafter{\dbLabel}%
    \xdef\dbLabel{%
   	\the\toks2 \nx\xdef\nx\dbx{#1}%
	\nx\ifx\nx\ikey %
		\nx\dbx\nx\xdef\nx\found{Y}%
		\nx\xdef\nx\key{#1}%
		\nx\xdef\nx\tag{#2}%
		\nx\xdef\nx\pre{\ModeYes}%
	\nx\fi}%
\fi%
}

\def\dbLabelInsert#1#2{\dbLabelGet{#1}%
\xdef\itag{#2}%
\dbLabelGet{#1}%
\if\found Y %
	\ifx\tag\itag %
	\else%
	   \ifx\PreventDoubleLabel\ModeYes %
		\message{^^J}%
		\errmessage{Double definition of label [#1]^^J}%
	   \else%
		\message{^^J}%
		\message{Double definition of label [#1]^^J}%
	   \fi%	
	\fi%
   \xdef\dbLabelStatus{\dbLabelSatusModeWarning}%
   \xdef\dbLabelWarning{Label is already there}%
   \xdef\dbLabelInfo{Label not inserted}%
\else%
   \toks2=\expandafter{\dbLabel}%
    \xdef\dbLabel{%
   	\the\toks2 \nx\xdef\nx\dbx{#1}%
	\nx\ifx\nx\ikey %
		\nx\dbx\nx\xdef\nx\found{Y}%
		\nx\xdef\nx\key{#1}%
		\nx\xdef\nx\tag{#2}%
		\nx\xdef\nx\pre{\ModeNo}%
	\nx\fi}%
\fi%
}

%%%%%%%%%%%%%%%%%%%%%%%%%%%%%%
%%%%%%		Numbering			 %%%%%%
%%%%%%%%%%%%%%%%%%%%%%%%%%%%%%

\newcount\PART
\newcount\CHAPTER
\newcount\SECTION
\newcount\SUBSECTION
\newcount\FNUMBER
%\newdimen\TOBOTTOM
%\newdimen\LIMIT

\PART=0
\CHAPTER=0
\SECTION=0
\SUBSECTION=0	
\FNUMBER=0

\def\LastPart{\ModeUndef}
\def\LastChapter{\ModeUndef}
\def\LastSection{\ModeUndef}
\def\LastSubSection{\ModeUndef}
\def\LastClaim{\ModeUndef}
\def\Last{\ModeUndef}

\newdimen\TOBOTTOM
\newdimen\LIMIT

\def\Ensure#1{\ \par\ \immediate\LIMIT=#1\immediate\TOBOTTOM=\the\pagegoal\advance\TOBOTTOM by -\pagetotal%
\ifdim\TOBOTTOM<\LIMIT\newpage \else%
\vskip-\parskip\vskip-\parskip\vskip-\baselineskip\fi}

%%%%%%%%%%%%%%%%%%%%%%%%%%%%%
\def\PartLabel{\the\PART}
\def\NewPart#1{\global\advance\PART by 1%
         \bs\ni{\PartStyle  Part \PartLabel:}
         \bs\ni{\PartStyle #1}\newpage%
         \CHAPTER=0\SECTION=0\SUBSECTION=0\FNUMBER=0%
         \gdef\Left{#1}%
         \global\edef\Last{\PartLabel}%
         \global\edef\LastPart{\PartLabel}%
         \global\edef\LastChapter{\ModeUndef}%
         \global\edef\LastSection{\ModeUndef}%
         \global\edef\LastSubSection{\ModeUndef}%
         \global\edef\LastClaim{\ModeUndef}}
%%%%%%%%%%%%%%%%%%%%%%%%%%%%%
\def\ChapterLabel{\the\CHAPTER}
\def\NewChapter#1{\global\advance\CHAPTER by 1%
         \bs\ni{\ChapterStyle  Chapter \ChapterLabel: #1}\ms%
         \SECTION=0\SUBSECTION=0\FNUMBER=0%
         \gdef\Left{#1}%
         \global\edef\Last{\ChapterLabel}%
         \global\edef\LastChapter{\ChapterLabel}%
         \global\edef\LastSection{\ModeUndef}%
         \global\edef\LastSubSection{\ModeUndef}%
         \global\edef\LastClaim{\ModeUndef}}
%%%%%%%%%%%%%%%%%%%%%%%%%%%%%
\def\SectionEnsure{3cm}
\def\NewSection#1{\Ensure{\SectionEnsure}\gdef\SectionLabel{\the\SECTION}\global\advance\SECTION by 1%
         \ms\ni{\SectionStyle  \SectionLabel.\ #1}\ss%
         \SUBSECTION=0\FNUMBER=0%
         \gdef\Left{#1}%
         \global\edef\Last{\SectionLabel}%
         \global\edef\LastSection{\SectionLabel}%
         \global\edef\LastSubSection{\ModeUndef}%
         \global\edef\LastClaim{\ModeUndef}}
%%%%%%%%%%%%%%%%%%%%%%%%%%%%%
\def\NewAppendix#1#2{\Ensure{\SectionEnsure}\gdef\SectionLabel{#1}\global\advance\SECTION by 1%
         \bs\ni{\SectionStyle  Appendix \SectionLabel.\ #2}\ss%
         \SUBSECTION=0\FNUMBER=0%
         \gdef\Left{#2}%
         \global\edef\Last{\SectionLabel}%
         \global\edef\LastSection{\SectionLabel}%
         \global\edef\LastSubSection{\ModeUndef}%
         \global\edef\LastClaim{\ModeUndef}}
%%%%%%%%%%%%%%%%%%%%%%%%%%%%%
\def\Acknowledgements{\Ensure{\SectionEnsure}\gdef\SectionLabel{}%
         \ms\ni{\SectionStyle  Acknowledgments}\ss%
         \SECTION=0\SUBSECTION=0\FNUMBER=0%
         \gdef\Left{}%
         \global\edef\Last{\ModeUndef}%
         \global\edef\LastSection{\ModeUndef}%
         \global\edef\LastSubSection{\ModeUndef}%
         \global\edef\LastClaim{\ModeUndef}}
%%%%%%%%%%%%%%%%%%%%%%%%%%%%%
\def\SubSectionEnsure{2cm}
\def\SubSectionLabel{\ifnum\SECTION>0 \the\SECTION.\fi\the\SUBSECTION}
\def\NewSubSection#1{\Ensure{\SubSectionEnsure}\global\advance\SUBSECTION by 1%
         \ms\ni{\SubSectionStyle #1}\ss%
         \global\edef\Last{\SubSectionLabel}%
         \global\edef\LastSubSection{\SubSectionLabel}}
%%%%%%%%%%%%%%%%%%%%%%%%%%%%%
\def\SetNumberingModeN{\def\ClaimLabel{(\the\FNUMBER)}}
\def\SetNumberingModeSN{\def\ClaimLabel{(\ifnum\SECTION>0 \SectionLabel.\fi%
      \the\FNUMBER)}}
\def\SetNumberingModeCSN{\def\ClaimLabel{(\ifnum\CHAPTER>0 \the\CHAPTER.\fi%
      \ifnum\SECTION>0 \SectionLabel.\fi%
      \the\FNUMBER)}}

\def\NewClaim{\global\advance\FNUMBER by 1%
    \ClaimLabel%
    \global\edef\LastClaim{\ClaimLabel}%
    \global\edef\Last{\ClaimLabel}}
%%%%%%%%%%%%%%%%%%%%%%%%%%%%%

\def\HideLabels{\xdef\ShowLabelsMode{\ModeNo}}
\HideLabels

\def\fn{\eqno{\NewClaim}} 
\def\fl#1{%
\ifx\ShowLabelsMode\ModeYes%
%\eqno{\relax\hbox to 1cm{\NewClaim\hbox{[#1]}}}%
 \eqno{{\buildrel{\hbox{\AbstractStyle[#1]}}\over{\hfill\NewClaim}}}%
\else%
 \eqno{\NewClaim}%
\fi% 
\dbLabelInsert{#1}{\ClaimLabel}}
\def\fprel#1{\global\advance\FNUMBER by 1\StoreLabel{#1}{\ClaimLabel}%
\ifx\ShowLabelsMode\ModeYes%
%\eqno{\relax\hbox to 1cm{ .\itag\hbox{[#1]}}}%
\eqno{{\buildrel{\hbox{\AbstractStyle[#1]}}\over{\hfill.\itag}}}%
\else%
 \eqno{\itag}%
\fi% 
}

\def\cl#1{\global\advance\FNUMBER by 1\dbLabelInsert{#1}{\ClaimLabel}%
\ifx\ShowLabelsMode\ModeYes%
${\buildrel{\hbox{\AbstractStyle[#1]}}\over{\hfill\ClaimLabel}}$%
\else%
  $\ClaimLabel$%
\fi% 
}
\def\cprel#1{\global\advance\FNUMBER by 1\StoreLabel{#1}{\ClaimLabel}%
\ifx\ShowLabelsMode\ModeYes%
${\buildrel{\hbox{\AbstractStyle[#1]}}\over{\hfill.\itag}}$%
\else%
  $\itag$%
\fi% 
}
%%%%%%%%%%%%%%%%%%%%%%%%%%%%%

\def\Note{\ms\leftskip 3cm\rightskip 1.5cm\AbstractStyle}
\def\endNote{\par\leftskip 2cm\rightskip 0cm\NormalStyle\ss}

%%%%%%%%%%%%%%%%%%%%%%%%%%%%%%
%%%%%%		   Sidebars		          %%%%%%
%%%%%%%%%%%%%%%%%%%%%%%%%%%%%%

\parindent=7pt
\leftskip=2cm
\newcount\SideIndent
\newcount\SideIndentTemp
\SideIndent=0
\newdimen\SectionIndent
\SectionIndent=-8pt

\def\sidebar{\vrule height15pt width.2pt }
\def\endcorner{\hbox{\hbox{\vrule height6pt width.2pt}\vbox to6pt{\vfill\hbox
to4pt{\leaders\hrule height0.2pt\hfill}}}}
\def\begincorner{\hbox{\hbox{\vrule height6pt width.2pt}\vbox to6pt{\hbox
to4pt{\leaders\hrule height0.2pt\hfill}}}}
\def\endbegincorner{\hbox{\vbox to15pt{\endcorner\vskip-6pt\begincorner\vfill}}}
\def\SideShow{\SideIndentTemp=\SideIndent \ifnum \SideIndentTemp>0 
\loop\sidebar\hskip 2pt \advance\SideIndentTemp by-1\ifnum \SideIndentTemp>1 \repeat\fi}

\def\BeginSection{{\vbadness 100000 \par\ni\hskip\SectionIndent%
\SideShow\vbox to 15pt{\vfill\begincorner}}\global\advance\SideIndent by1\vskip-10pt}

\def\EndSection{{\vbadness 100000 \par\ni\global\advance\SideIndent by-1%
\hskip\SectionIndent\SideShow\vbox to15pt{\endcorner\vfill}\vskip-10pt}}

\def\EndBeginSection{{\vbadness 100000\par\ni%
\global\advance\SideIndent by-1\hskip\SectionIndent\SideShow
\vbox to15pt{\vfill\endbegincorner}}%
\global\advance\SideIndent by1\vskip-10pt}

\def\ShowBeginCorners#1{%
\SideIndentTemp =#1 \advance\SideIndentTemp by-1%
\ifnum \SideIndentTemp>0 %
\vskip-15truept\hbox{\kern 2truept\vbox{\hbox{\begincorner}%
\ShowBeginCorners{\SideIndentTemp}\vskip-3truept}}%				
\fi%
}

\def\ShowEndCorners#1{%
\SideIndentTemp =#1 \advance\SideIndentTemp by-1%
\ifnum \SideIndentTemp>0 %
\vskip-15truept\hbox{\kern 2truept\vbox{\hbox{\endcorner}%
\ShowEndCorners{\SideIndentTemp}\vskip 2truept}}%				
\fi%
}

\def\BeginSections#1{{\vbadness 100000 \par\ni\hskip\SectionIndent%
\SideShow\vbox to 15pt{\vfill\ShowBeginCorners{#1}}}\global\advance\SideIndent by#1\vskip-10pt}

\def\EndSections#1{{\vbadness 100000 \par\ni\global\advance\SideIndent by-#1%
\hskip\SectionIndent\SideShow\vbox to15pt{\vskip15pt\ShowEndCorners{#1}\vfill}\vskip-10pt}}

\def\EndBeginSections#1#2{{\vbadness 100000\par\ni%
\global\advance\SideIndent by-#1%
\hbox{\hskip\SectionIndent\SideShow\kern-2pt%
\vbox to15pt{\vskip15pt\ShowEndCorners{#1}\vskip4pt\ShowBeginCorners{#2}}}}%
\global\advance\SideIndent by#2\vskip-10pt}

%%%%%%%%%%%%%%%%%%%%%%%%%%%%%%
%%%%%%		Margin notes		 %%%%%%
%%%%%%%%%%%%%%%%%%%%%%%%%%%%%%

%%%%%%%%%%%%%%%%%%%%%%%%%%%%%

%%%%%%%%%%%%%%%%%%%%%%%%%%%%%

%
%    Macros.    Version 1.2.0.beta
%    The best use is to paste all of them into the papers
%     1/8/2005
%

%%%%%%%%%%%%%%%%%%%%%%%%%%%%%%
%%%%%%			Greek		 %%%%%%
%%%%%%%%%%%%%%%%%%%%%%%%%%%%%%

\def\al{\alpha}
\def\be{\beta}
\def\de{\delta}
\def\ga{\gamma}

\def\ep{\epsilon}

\def\te{\theta}
\def\la{\lambda}

\def\om{\omega}
\def\si{\sigma}

\def\ka{\kappa}

\def\De{\Delta}
\def\Ga{\Gamma}

\def\Om{\Omega}

%%%%%%%%%%%%%%%%%%%%%%%%%%%%%%
%%%%%%			Cal			 %%%%%%
%%%%%%%%%%%%%%%%%%%%%%%%%%%%%%

 \def\calC{{\hbox{\cal C}}}
 \def\calQ{{\hbox{\cal Q}}}

 \def\calE{{\hbox{\cal E}}}

%%%%%%%%%%%%%%%%%%%%%%%%%%%%%%
%%%%%%			gothic		 %%%%%%
%%%%%%%%%%%%%%%%%%%%%%%%%%%%%%

 		% to prevent \sl redefinition

%%%%%%%%%%%%%%%%%%%%%%%%%%%%%%
%%%%%%			Bbb			 %%%%%%
%%%%%%%%%%%%%%%%%%%%%%%%%%%%%%

 \def\R{{\hbox{\Bbb R}}}

 \def\R{{\hbox{\Bbb R}}}

%%%%%%%%%%%%%%%%%%%%%%%%%%%%%%
%%%%%%		MathRoman		 %%%%%%
%%%%%%%%%%%%%%%%%%%%%%%%%%%%%%

\def\det{{\hbox{det}}}

\def\id{{\hbox{\rm id}}}

%%%%%%%%%%%%%%%%%%%%%%%%%%%%%%
%%%%%%		OtherSymbols		 %%%%%%
%%%%%%%%%%%%%%%%%%%%%%%%%%%%%%
\def\ip{\hbox to4pt{\leaders\hrule height0.3pt\hfill}\vbox to8pt{\leaders\vrule width0.3pt\vfill}\kern 2pt}
% inner product
 
\def\del{\partial}
\def\na{\nabla}

\def\Lie{\hbox{\LieFont \$}}

\def\arr{\rightarrow}

\def\then{\Rightarrow}

%
%    Format.    Version 1.2.0.beta
%    The best use is to paste all of them into the papers
%     1/8/2005
%

\def\cases#1{\left\{\eqalign{#1}\right.}
%%%%%%%%%%%%%%%%%%%%
\NormalStyle
\SetNumberingModeSN
\PreventDoubleOn

\long\def\title#1{\centerline{\TitleStyle\ni#1}}
\long\def\moretitle#1{\baselineskip18pt\centerline{\TitleStyle\ni#1}}
\long\def\author#1{\ms\centerline{\AuthorStyle by {\it #1}}}

\long\def\address#1{\ss\centerline{\AddressStyle #1}\par}
\long\def\moreaddress#1{\centerline{\AddressStyle #1}\par}
\def\abstract{\ms\leftskip 3cm\rightskip .5cm\AbstractStyle{\bf \ni Abstract:}\ }
\def\endabstract{\par\leftskip 2cm\rightskip 0cm\NormalStyle\ss}

%%%%%%%%%%%%%%%%%%%%%%%%%%%%%
\SetNumberingModeSN
%\ShowLabels

\def\calH{{\hbox{\cal H}}}

\def\frac[#1/#2]{\hbox{$#1\over#2$}}
\def\Frac[#1/#2]{{#1\over#2}}
\def\({\left(}
\def\){\right)}
\def\[{\left[}
\def\]{\right]}
\def\^#1{{}^{#1}_{\>\cdot}}
\def\_#1{{}_{#1}^{\>\cdot}}
\def\Label=#1{{\buildrel {\hbox{\fiveSerif \ShowLabel{#1}}}\over =}}
\def\<{\kern -1pt}

\def\Dal{\hbox{\tenRelazioni  \char003}}

%%%%%%%%  		Collapsable Notes		%%%%%%%%%%%%%%%%%%%%%%%%

\def\ExpandAllCNotes{\long\def\CNote##1{%
\BeginSection%\Margine{{\AbstractStyle To be collapsed}}%
	\Note%
 		##1%
	\endNote% 
\EndSection%
}}
\ExpandAllCNotes
%
% If you want to collapse classes of CNotes independently one of the other, just clone the definition as
%
%	\def\CollapseAllCNotesClassA{\long\def\CNoteClassA##1{}}
%	\def\ExpandAllCNotesClassA{\long\def\CNoteClassA##1{\BeginSection\Note ##1 \endNote\EndSection}}
%	\ExpandAllCNotesClassA
%
%	\def\CollapseAllCNotesClassB{\long\def\CNoteClassB##1{}}
%	\def\ExpandAllCNotesClassB{\long\def\CNoteClassB##1{\BeginSection\Note ##1 \endNote\EndSection}}
%	\ExpandAllCNotesClassB
%
%%%%%%%%%%%%%%%%%%%%%%%%%%%%%%%%%%%%%%%%%%%%%%%%%

%%%%%%%%%%%%			frames 				%%%%%%%%%%%%%%%%%%%

\long\def\red#1{\textcolor{red}{#1}}
\long\def\blue#1{\textcolor{blue}{#1}}
\long\def\green#1{\textcolor{green}{#1}}

\def\frame#1{\vbox{\hrule\hbox{\vrule\vbox{\kern2pt\hbox{\kern2pt#1\kern2pt}\kern2pt}\vrule}\hrule\kern-4pt}} 
\def\redframe#1{\red{\frame{#1}}} 
\def\greenframe#1{\green{\frame{#1}}}

\def\uline#1{\underline{#1}}
\def\uuline#1{\underline{\underline{#1}}}
\def\Box to #1#2#3{\frame{\vtop{\hbox to #1{\hfill #2 \hfill}\hbox to #1{\hfill #3 \hfill}}}}

%%%%%%%%%%%%%%%%%%%%%%%%%%%%%%%%%%%%%%%%%%%%%%%%%

\bib{Nester}{C.\ M.\ Chen, J.\ M.\ Nester,  
{\it Quasilocal quantities for GR and other gravity theories},
Class.Quant.Grav. 16 (1999) 1279-1304;\goodbreak
C.-M. Chen, J.M. Nester,
Gravitation \& Cosmology {\bf 6}, (2000), 257  (gr--qc/0001088)
}

\bib{Augmented}{L.~Fatibene, M.~Ferraris, M.~Francaviglia,
{\it Augmented Variational Principles and Relative Conservation Laws in Classical Field Theory},
{  Int. J. Geom. Methods Mod. Phys.}, {\bf 2}(3), (2005), pp. 373-392; [math-ph/0411029v1]
}

\bib{Sinicco}{M.\ Ferraris, M.\ Francaviglia and I.\ Sinicco, Il Nuovo Cimento, {\bf 107B},(11), 1992, 1303
}

\bib{ADM}{R.  Arnowitt, S.  Deser and C.  W.  Misner, in: 
{\it Gravitation: An Introduction to Current Research}, 
L. Witten ed. Wyley,  227, (New York, 1962)}

\bib{RT}{T.\ Regge, C.\ Teitelboim, Annals of Physics {\bf 88}, 286  (1974).
}

\bib{Wald}{L.ÊFatibene, M. Ferraris, M.Francaviglia, M. Raiteri, 
{\it Remarks on Noether charges and black holes entropy.}  
Ann. Physics 275(1) (1999)
}

\bib{Kerr2}{{L. Fatibene, M. Ferraris, M. Francaviglia, S.Mercadante,
{\it About the Energy of AdS-Kerr Solutions},
Int. J. Geom. Methods Mod. Phys. 5(7), 2008.}
}

\bib{Kerr}{L. Fatibene, M.Ferraris, M. Francaviglia,
{\it The Energy of a Solution from Different Lagrangians},
{ Int. J. Geom. Methods Mod. Phys.}, {\bf 3}(7), (2006), pp. 1341-1347
}

\bib{Book}{L. Fatibene, M. Francaviglia,
{\it Natural and Gauge Natural Formalism for Classical Field Theories},
Kluwer Academic Publishers, (Dordrecht, 2003), xxii
}

\bib{OurKosman}{L. Fatibene, M. Ferraris, M. Francaviglia, M. Godina,
{\it A geometric definition of Lie derivative for Spinor Fields},
in: Proceedings of 
{\it ``6th International Conference on Differential Geometry
and its Applications, August 28--September 1, 1995"}, (Brno, Czech Republic),
Editor: I. Kol{\'a}{\v r}, MU University, Brno, Czech Republic (1996)}

\bib{Kosman}{Kosmann, Y., (1972), Ann. di Matematica Pura et Appl. {\bf 91} 317--395.\goodbreak
Kosmann, Y., (1966), Comptes Rendus Acad. Sc. Paris, s\'erie A, {\bf 262} 289--292.\goodbreak
Kosmann, Y., (1966), Comptes Rendus Acad. Sc. Paris, s\'erie A, {\bf 262} 394--397.\goodbreak
Kosmann, Y., (1967), Comptes Rendus Acad. Sc. Paris, s\'erie A, {\bf 264} 355--358.
}

\bib{Cavalese}{M.~Ferraris, M.~Francaviglia, 
in: {\sl 8th Italian Conference on General Relativity and    Gravitational Physics}, 
Cavalese (Trento), August 30--September 3, World Scientific, Singapore, 1988, 183}

\bib{Maple}{K.Chu, C.Farel, G.Fee, R.McLenaghan, Fields Inst. Comm. {\bf 15}, (1997)
}

\bib{Julia}{B. Julia and S. Silva, 
{\it Currents and superpotentials in classical gauge theories}, 
Classical Quantum Gravity , vol. 17, No. 22 (2000), 4733-4743}

\bib{Torre}{I.M. Anderson, C.G. Torre, Phys. Rev. Lett. 77 (1996) 4109 (hepÐ th/9608008); \goodbreak
C.G. Torre, hepÐth/9706092, Lectures given at 2nd Mexican School on Gravitation and Mathematical Physics, Tlaxcala, Mexico (1996)}

\bib{BY}{L.\ Fatibene, M.\ Ferraris, M.\ Francaviglia, M.\ Raiteri, J.  Math. Phys., {\bf 42}, No. 3, 1173 (2001)  (gr--qc/0003019). }

\bib{BTZ}{L.~Fatibene, M.~Ferraris, M.~Francaviglia, M.~Raiteri, 
{\sl Remarks on conserved      quantities and entropy of BTZ black hole solutions.  I. The general setting.} 
Phys.\ Rev.\ D (3) 60 (1999), n.12, 124012
}

\bib{Taub}{L. Fatibene, M. Ferraris, M. Francaviglia, M. Raiteri,
{\it The entropy of the Taub-bolt solution},
Ann. Physics 284 (2000), no. 2, 197Ð214.}

\bib{Taub2}{R. Clarkson, L. Fatibene, R.B. Mann,
{\it Thermodynamics of $(d+1)$-dimensional NUT-charged AdS spacetimes},
Nuclear Phys. B 652 (2003), no. 1-3, 348Ð382.}

\bib{Weinberg}{{S. Weinberg,
{\it Gravitation and Cosmology: Principles and Applications of the General Theory of Relativity}}
Wiley, New York (a.o.) (1972). XXVIII, 657 S. : graph. Darst.. ISBN: 0-471-92567-5.}

\bib{BL}{J. Katz, J. Bicak, D. Lynden-Bell.
{\it Relativistic conservation laws and integral constraints for large  cosmological perturbations},
{Phys. Rev. D}, 55:5957--5969, 1997.}

\bib{Landau}{L. Landau, E. Lifchitz,
{\it Th\'eorie du champ},
Moscou, ƒditions Mir. 1966.
}

\bib{Petrov}{A.N.Petrov,
{\it Nonlinear Perturbations and Conservation Laws on Curved Backgrounds in GR and Other Metric Theories},
; arXiv:0705.0019v1
}

\bib{Murchadha}{R.Beig, N. \'o Murchadha,
{\it The Poincar\'e Group as the Symmetry Group of Canonical General Relativity},
Annals Phys.  {\bf 174} (1987) 463-498.
}

\bib{Freud}{Von Ph. Freud,
{\it \"Uber die Ausdr\"ucke der Gesamtenergie und des Gesamt-Impuses eines materiellen Systems in der allgemeinen Relativit\"atstheorie},
Annals of Mathematics, {\bf 40}(2) (1939)}

\bib{Geroch}{S. Persides,
{\it A definition of asymptotically Minkowskian space--times},
J. Math. Phys. {\bf 20}, (1979) 1731;  %doi:10.1063/1.524258
}

\bib{L1}{D.Alba and L.Lusanna,  
{\it Charged Particles and the Electro-Magnetic Field in Non-Inertial Frames: I.  Admissible $3+1$ Splittings of
Minkowski Spacetime and the Non-Inertial Rest Frames},
(in press on Int.J.Geom.Methods in Physics); arXhive: 0908.0213}

\bib{L2}{D.Alba and L.Lusanna,  
{\it II. Applications: Rotating Frames, Sagnac Effect, Faraday Rotation, Wrap-up Effect}
(in press on Int.J.Geom.Methods in Physics); arXhive: 0908.0215}

\bib{L3}{D.Alba, H.W.Crater and L.Lusanna,  
{\it Towards Relativistic Atom Physics. I. The Rest-Frame Instant Form of Dynamics and a
Canonical Transformation for a system of Charged Particles plus the Electro-Magnetic Field},
(in press on Canad.J.Phys.); arXhive: 0806.2383}

\bib{L4}{D.Alba, H.W.Crater and L.Lusanna,  
{\it II. Collective and Relative Relativistic Variables for a System of Charged Particles plus the Electro-Magnetic Field},
(in press on Canad.J.Phys.); arXhive: 0811.0715.}

%2) relativita' generale

\bib{L5}{D.Alba and L.Lusanna, 
{\it The York Map as a Shanmugadhasan Canonical Transformation in Tetrad Gravity and the Role of Non-Inertial Frames in the Geometrical View of the Gravitational Field}, 
Gen.Rel.Grav. {\bf 39}, 2149 (2007) (gr-qc/0604086, v2)}

\bib{L6}{D.Alba and L.Lusanna,  
{\it The Einstein-Maxwell-Particle System in the York Canonical Basis of ADM Tetrad Gravity: I) The Equations of Motion in Arbitrary Schwinger Time Gauges.}, 
arXive: (0907.4087)}

%3) review generale

\bib{L7}{L.Lusanna, 
{\it Post-Minkowskian Gravity as a Relativistic Inertial Effect?},
talk at the  1st Mediterranean Conference in Classical and
Quantum Gravity, held in the Orthodox Academy of Crete in Kolymbari
(Greece) from Monday, September 14th to Friday, September 18th,
2009; arXive: 0912.2935)}.

\bib{Szaba}{L.B Szabados,
{\it Towards the quasi-localization of canonical GR},
Class. Quantum Grav. {\bf 26} (2009) 125013; arXiv: 0902.3199v2
}

%%%%%%%%%%%%%%%%%%%%%%%%%%%%%%%%%%%%%%%%%%%%%%%%%%%
\def\AppA{A}
\def\AppB{B}
\def\AppC{C}

\def\ubal{\underline{\al}\kern1pt}
\def\obal{\overline{\al}\kern1pt}

\def\ubR{\underline{R}\kern1pt}
\def\obR{\overline{R}\kern1pt}
\def\ubom{\underline{\om}\kern1pt}
\def\obxi{\overline{\xi}\kern1pt}
\def\ubu{\underline{u}\kern1pt}
\def\ube{\underline{e}\kern1pt}
\def\obe{\overline{e}\kern1pt}
\def\Limit{\>{\buildrel{r\arr\infty}\over \longrightarrow}\,}

%%%%%%%%%%%%%%%%%%%%%%%%%%%%%%%%%%%%%%%%%%%%%%%%%%%
\NormalStyle
%\ShowLabels
%\CollapseAllCNotes

\title{ADM Pseudotensors, Conserved Quantities and }
\moretitle{Covariant Conservation Laws in General Relativity\footnote{$^*$}{{\AbstractStyle
	This paper is published despite the effects of the Italian law 133/08 ({\tt http://groups.google.it/group/scienceaction}). 
        This law drastically reduces public funds to public Italian universities, which is particularly dangerous for free scientific research, 
        and it will prevent young researchers from getting a position, either temporary or tenured, in Italy.
        The authors are protesting against this law to obtain its cancellation.\goodbreak}}}

\author{L.Fatibene$^{a, b}$, M.Ferraris$^a$, M.Francaviglia$^{a,c}$, L.Lusanna$^d$\footnote{}{\AbstractStyle eMail: {\tt lorenzo.fatibene@unito.it}, {\tt marco.farraris@unito.it}, {\tt mauro.francaviglia@unito.it}, {\tt luca.lusanna@fi.infn.it}}}

\address{$^a$ Department of Mathematics, University of Torino (Italy)}

\moreaddress{$^b$ INFN - Sezione Torino (Italy). In. Spec. Na12}

\moreaddress{$^c$ LCS, University of Calabria}

\moreaddress{$^d$ INFN - Sezione Firenze}

\abstract
The ADM formalism is reviewed and techniques for decomposing generic components of metric, connection and curvature are obtained.
These techniques will turn out to be enough to decompose not only Einstein equations but also covariant conservation laws.

Then a  number of independent sets of hypotheses that are sufficient (though non--necessary) to obtain standard ADM quantities (and Hamiltonian) from covariant conservation laws are considered.
This determines explicitely the range in which standard techniques are equivalent to covariant conserved quantities.

The Schwarzschild metric in different coordinates is then considered, showing how the standard ADM quantities fail dramatically in non-Cartesian coordinates or even worse when asymptotically flatness is not manifest;
while, in view of their covariance, covariant conservation laws give the correct result in all cases.

\endabstract

%%%%%%%%%%%%%%%%%%%%%%%%%%%%%%%%%%%%%%%%%%%%%%%%%%%
\NewSection{Introduction}

Almost a century after the birth of GR there is yet no universal consensus on how energy, momentum and other conserved quantities should be defined  in it 
from a fundamental viewpoint.
There is a number of simple situations in which everybody agrees on the expected result, but such results can be obtained 
from a great number of, quite different, prescriptions that usually differ when used in a generic situation.

From a fundamental viewpoint the situation could not be worse: 
some people use prescriptions based on pseudotensors (here this term refers to coordinate expressions with non-tensorial, sometimes undetermined, transformation properties) argueing that generally covariant prescriptions cannot catch physical properties
of conserved quantities; 
some use covariant prescriptions argueing that non-covariant objects have no meaning in GR.
The main reason to defend covariant conserved quantities in GR is that, according to the general covariance principle, if conserved quantities were intrinsically non-covariant they would be irrelevant to the description of Nature. 
Here we are using {\it general covariance principle} in the naive form: {\it one can describe Physics in an observer independent way}, that is just the definition of what one should understand for a (classical) Nature which exists out there independently of the observer.
This is enough for us to accept at least the challenge to understand conserved quantities on a covariant stance.

Let us stress here once for all that obeying general covariance principle does not mean to use always intrinsic quantities or covariant expressions.
On the contrary, one necessarily has to break the coordinate gauge at some stage to compare results with experiments; see \ref{L7}. 
Experiments are by their own nature performed in some coordinate system set by some observer which  (or who) has set its own conventions for determining position and time of events.

To be precise, the general covariance principle claims that the description of Physics {\it can} be done independently of any {\it a priori} coordinate fixing.
It does not exclude that in particular situations one has {\it a posteriori} preferred  coordinates, preferred splittings between space and time, or preferred observers;
see \ref{L1}, \ref{L2}, \ref{L3}, \ref{L5}.
One very well--known example of such a situation is Cosmology: in Friedmann-Robertson-Walker solutions one has canonical clocks (e.g.~the temperature of the cosmic background radiation) that not only break Lorentz invariance defining a cosmic (global) time but break the Galilei invariance defining observers which are at rest with respect to the cosmic background radiation. In fact, there are special observers which see the CBR to be isotropic, while observers in motion with respect to them see part of the sky blue shifted and part red shifted. These observers are {\it universal}, by this meaning that one should not be surprised to learn that a civilization on the other side of the universe has some special name for them.

{This situation in Cosmology is not a violation of the general covariance principle. In fact, these observers can be defined only once a particular class of solutions
(FRW) has been given. They are given {\it a posteriori} with respect to solving the dynamics of the system, not {\it a priori} (in which case they would be given in any solution of Einstein theory).}

{Another issue in which these sort of {\it a posteriori} gauge fixings could be important is quantum physics. It is quite well understood that quantizing a gauge system and then fixing the gauge at quantum level gives in general a different system with respect to fixing the gauge at classical level and then quantizing the physical
degrees of freedom (see \ref{L4}, \ref{L5}, \ref{L6}). 
In particular  it is not yet clear whether in GR there exists a quantum description independent of the observer or quantum gravity necessarily describes the point of view of a fixed quantum observer.}

{In any event, at a classical level, if one accepts GR as a fundamental description of Physics, whenever coordinate fixings are used one should explain
in detail which gauge fixing has been done, why or whether this gauge fixing is needed and why or whether it defines {\it preferred observers} in the description of the world.}

Another ground of misunderstanding is the physical fundamental meaning of {\it the} energy (as well as other conserved quantities) of a system, 
opposed to {\it an} energy for the system. Nester noted (see \ref{Nester}) that in many areas of Physics it is well known that many different energies can be associated to the same system. 
Besides the obvious dependence on the observer's reference frame, which is well known also in Mechanics, Nester pointed out the dependence
of energy on control variables (or boundary conditions)  well known in Thermodynamics, where besides internal energy one can associate other {\it energies} to a system. We completely agree with this position. We believe that many prescriptions define {\it an} energy for the system and ambiguities in the prescriptions just 
mathematically reflect the physical ambiguity in the notion of energy.

As it is well known at least two communities have grown and become incompatible about the issue of the energy of the gravitational field.
We believe it is time to make some effort to reconsider the whole issue in view of decades of research in both directions, at least trying to contribute in
addressing the issue once and for all.
It is our opinion that part of the misunderstanding comes from the overestimated role given to special relativity (SR) in GR.
People have often tried to mimick what is done in SR and extend it directly to GR, while it is clear that GR demands instead a deep re-thinking about what are to be understood 
as legitimate techniques and prescriptions. We collect some motivations of this position in Appendix \AppA.
 There are also a number of claims, also from valuable reseachers in the area (not excluding Einstein himself), that have been shown to
be flat wrong. We collect a number of them, together with their counterarguments, in Appendix \AppB.
In Appendix \AppC\ we will finally present a summary of relations among the different quantities defined in this paper.

This paper is a logical review of ADM tecniques and a contribution to the discussion about pseudotensors and covariant prescriptions. 
In particular we shall investigate in detail the relation between one of the covariant prescriptions for conserved quantities (based on the so-called {\it augmented variational principle} and Noether theorem) and one of the non-covariant ones (namely, ADM conserved quantities).

ADM prescription (see \ref{ADM}, \ref{RT}, \ref{Sinicco}) is very well--established and deeply connected to certain issues concerning the Hamiltonian structure of GR theory.
It is known to hold for asymptotically flat systems in quasi-Cartesian coordinates (i.e., Cartesian coordinates for the underlying Minkowski reference background).

We shall obtain ADM prescriptions as the ADM decomposition of augmented conserved quantities and then discuss in detail what happens when the hypotheses required for standard ADM do not hold true. 
To investigate in  this direction we shall start performing ADM decomposition in a quite general setting without assuming too much about the solution under consideration, adding hypotheses once the decomposition has been obtained.
We shall find out that augmented covariant conservation laws keep providing the correct result, while the standard pseudotensorial ADM quantities fail, sometimes introducing dramatic unphysical divergencies, out of their original scope.

However, the aim of this paper is not to kill pseudotensorial conservation laws ultimately and without appeal. 
Pseudotensors can in fact be legitemately used with due attention, provided that it is clearly stated which preferred observers they assume.
Deriving pseudotensors from covariant conservation laws is one effective way, though probably not the only one, of keeping under control which hypotheses have been used and which gauge fixings have been done.

\ms
In Section 2 we shall introduce ADM foliations and a systematic decomposition of objects along the foliation.
The main goal of this Section is to obtain some Lie derivatives of the connection which then enter conservation laws.
The main computational resource here is to write all results in a frame adapted to the foliation to obtain easier expressions to be dealt with.

In Section 3 we shall review the main results about augmented covariant conservation laws. 
For an extended introduction and motivations we refer to \ref{Augmented}.

In Section 4 we shall apply ADM decomposition to the covariant conservation laws. 
This in particular contains the standard canonical analysis of Hamiltonian formalism which was already 
discussed in \ref{Sinicco}.
Here we shall rely on systematic decomposition to obtain directly the ADM mass and the ADM momentum from ADM decomposition of augmented covariant conservation laws.
Moreover, we shall discuss various independent set of hypotheses which allow to obtain ADM conserved quantities with no extra correction.
We believe this will definitely set the long standing discussion about which hypotheses are needed to obtain a ``meaningful'' notion of energy in GR.

As an example, in Section 5 we shall apply the results of the previous Section to various forms of Schwarzschild solution and with different hypotheses to obtain its ADM conserved quantities. 
In what follows we shall use homogeneous units, in which $c=1$ and $G=1$.

\CNote{
Hoping it could help readers who are approaching ADM techniques for the first time we add a detailed guideline to a systematic
derivation of decomposition of spacetime objects with respect to a given ADM foliation.
In literature it is relatively easy to find decompositions of some relevant object. 
Here we decided to establish techiques to decompose {\it anything}.
If the reader wishes to skip these details, recompile this \TeX\ sourcefile uncommenting (just above the title) the command
{\tt $\backslash$CollapseAllCNotes}. 

We shall use below the following typographic conventions: paired color terms cancel out,
underlined terms are similar (or to be collected together), framed terms are zero.
}

\NewSection{ADM Foliations}

Let us consider a spacetime manifold $M$, here for simplicity assumed of dimension $m=4$, even though one can easily generalize 
the discussion to a generic dimension.
We shall denote coordinates on spacetime $M$ by $x^\mu$, with $\mu=0,\dots ,3$.

A (global) {\it ADM foliation} is a bundle structure $\pi:M\arr \R$; the fibers $\pi^{-1}(t)= S_t\subset M$ are identified with the leaves of the foliation.
From a  physical viewpoint, the fibers $S_t\subset M$ are defined as the set of simultaneous events for a given observer.
Of course, different observers might have different synchronization protocols and define different ADM foliations on the same spacetime $M$. 
In fact, fixing an ADM foliation is part of the observer's specification.

Here, instead of looking for a preferred class of observers (which of course could lead to simplifications in special situations) we shall work with an arbitrary 
but fixed observer. ADM formalism imposes the choice of an observer, but at least working with an {\it arbitrary} one is as close  as possible to the principle 
of general covariance. Moreover, special classes of observers are often depending on the class of solutions under consideration and, as such, they are not suitable for discussing {\it from a fundamental viewpoint} how one should define conserved quantities.

The standard fiber $S$ of an ADM foliation represents an abstract model for space. Coordinates on $S$ are denoted by $k^A$,
with $A=1,2 ,3$.
Fibered coordinates $(x^0, x^i)$, with $i=1,2 ,3$, on $M$ are called {\it adapted coordinates} with respect to a given ADM foliation.
In adapted coordinates the fibers (i.e.~space submanifolds) are given by $S_t=\{x^0=t\}$.
The fiber coordinates  provide a canonical parametrization of the fiber $S_t$  in the form 
$$
i_t: S \arr M: k^A\mapsto (x^0=t, x^i=\al^i_A k^A)
\qquad\qquad
\al^i_A\equiv \de^i_A
\quad
(\hbox{and } \al_i^A\equiv \de_i^A)
\fl{AlphaAdapted}$$
These parametrizations are called the {\it adapted  parametrizations}. They depend of course on the ADM foliation, a trivialization and the fiber coordinates chosen.
For notational convenience we shall also define the (maximal rank) matrix $\al^\mu_A=\del_A x^\mu$, by setting $\al^0_A=0$ and $\al^i_A=\de^i_A$.

\CNote{
Notice that the fiber $S_t$, being vertical, can be covered within a single trivialization domain. Therefore, we shall here work in a fixed trivialization.

In general, one could use different parametrizations of the fibers; the generic parametrization is
$$
i_t: F\arr M: k^A \mapsto (x^0=f(t), x^i=\phi^i(t, k))
\fl{Embeddings}$$
In this more general case, let us set $\obal=d_t f(t)$, $\al^i= d_t\phi^i(t,k)$, $\al^i_A= \del_A \phi^i(t, k)$.
The family of embeddings $i_t$ is a diffeomorphism $i:\R\times S\arr M$ which can be inverted to give
$$
\cases{
&t= f^{-1}(x^0)\cr
&k^A= \phi^A(x^0, x^i)\cr
}
\fl{InverseEmbeddings}$$
Let us define $\ubal=d_0 f^{-1}(x^0)$, $\al^A= d_0 \phi^A(x)$ and $\al^A_i= d_i \phi^A(x)$.
Each object can be expressed as a function of $x^\mu$ or $(t, k^A)$. 
The fact that \ShowLabel{InverseEmbeddings} are the inverse of \ShowLabel{Embeddings} implies
$x^0=f(f^{-1}(x^0))$ and $x^i=\phi^i(f^{-1}(x^0), \phi^A(x^0, x^i))$. 
By differentiating we obtain
$$
\cases{
&\ubal\cdot \obal=1\cr
&0= \al^i \ubal + \al^i_A \al^A\cr
&\al_A^i \al^A_j =\de^i_j\cr
}
\fn$$
For later convenience let us prove that $d_0\al^i_A=0$.
In fact, one has
$$
d_0\al^i_A=d_0 d_A \phi^i= d_A d_0 \phi^i=  d_A \(\al^i \ubal + \al^i_B \al^B\)\equiv 0
\fl{Lemmad0al}$$

Moreover, let us remark that
$$
d_A \al^i_B= d_{AB} \phi^i=d_B \al^i_A
\fn$$
is symmetric in the indices $(AB)$; it will be denoted by $\al^i_{AB}$.

Generic changes of adapted coordinates are in the form
$$
\cases{
&x'{}^0= x'{} ^0(x^0)\cr
&x'{}^i= x'{}^i(x^0, x^i)\cr
}
\fl{ChangeOfADMCoordinates}$$
These changes of coordinates preserve the ADM foliation. There always exists an atlas of $M$ made only of adapted coordinates.
Let us now consider two trivializations, two charts on $S$  with transition functions $k'^A=k'^A(k)$ (the Jacobian being denoted by $J_A^B$) and two parametrizations
$$
\eqalign{
i_t:& F\arr M: k^A \mapsto (x^0=f(t), x^i=\phi^i(t, k))
\cr
i'_{t'}:& F\arr M: k'^A \mapsto (x'^0=f'(t'), x'^i=\phi'^i(t', k'))
}
\fn$$
Then $i_t = i'_{t'}$ glue together to define a single embedding if and only if
$$
\cases{
&f'(t') = x'^0(f(t))\cr
&\phi'^i(t', k') = x'^i(t, \phi^i(t, k))\cr
}
\fl{glueEmbeddings}$$

From this it follows that changing trivialization on the ADM foliation does not preserve adapted parametrizations;
if one starts with an adapted parametrization $\phi^i(t, k)=\de^i_A k^A$, the transformed parametrization
$\phi'^i(t', k') = x'^i(t, \de^i_A k^A)$ is not adapted unless in the special cases when $x'^i(x^0, x^l)\equiv x'^i( \de^i_A k^A)=  \de^i_A k'^A$ 
(i.e.~one is keeping the trivialization fixed and one is just changing coodinates on $S$).
However, this is enough to prove that in a fixed trivialization one can cover the whole fibers by means of local adapted parametrizations.

Below we shall need to keep under control global properties of objects defined using these ADM foliation and fiber parametrizations.
Hence we need to trace how quantities transform under changes of trivialization. 
In particular, by taking the derivative of \ShowLabel{glueEmbeddings}, one has:
$$
\al'^i_A=\del'_A \phi'^i(t', k')= \bar J_A^B J^i_j \del_B \phi^j= J^j_i \> \al_B^j\>   \bar J_A^B
\qquad\then
\al'^A_i=J^A_B \> \al^B_j\>    \bar  J_j^i
\fl{eTranformationRules}$$
}

Once the parametrizations of fibers have been fixed we can define a canonical covector:
$$
\ubu = u_\al \>d x^\al=\frac[1/3!]\ep_{0ijk}\> \al_A^i \al_B^j \al_C^k \>\ep^{ABC}\> dx^\al
= \frac[1/3!] \al \> \ep_{\al ABC}\>\ep^{ABC}\> dx^\al= \al  dx^0
\fn$$
where we set $\al=\det(\al_A^i)$.
In view of the definition  \ShowLabel{AlphaAdapted}
of $\al_A^i$ one has  $\ubu = dx^0$ on adapted parametrizations.

Let us also define a basis of vectors tangent to $T_xS_t$ (also known as {\it vertical vectors}) by:
$$
e_A = \del_A x^\mu(k)\>\del_\mu=\al_A^i \>\del_i
\fl{ek}$$

This basis of vertical vectors transforms as $e'_A=\bar J_A^B e_B$, where $J^B_A$ denotes the Jacobian of change of space coordinates.
\CNote{
In view of \ShowLabel{eTranformationRules} this is also true when changes of trivializations are allowed.
Notice that $\del_i$ are vertical vectors and hence they transform as $\del'_i= \bar J^j_i \del_j$.
}
We stress that no metric structure has been yet fixed on $M$.
Let us now consider a Lorentzian metric $g$ on $M$ and assume that
$S_t$ are spacelike submanifolds on the Lorentzian manifold $(M, g)$, i.e.~that
the metric $\ga$ induced on $S$ given by
$$
\ga=\ga_{AB} \> dk^A\otimes dk^B= \al_A^\mu \> g_{\mu\nu} \>\al_B^\nu\> dk^A\otimes dk^B
\fn$$
is positive definite.
We can define a future directed unit vector normal to the leaves
$$
\vec n= n^\al\>\del_\al=\pm \Frac[1/|u|] g^{\al\be} u_\be\>\del_\al
\fn$$
The global vector field $\vec n$ (as well as any global vector field everywhere transverse to the leafs of the ADM foliation) 
defines a connection on the bundle $\pi:M\arr \R$ (depending on the 
spacetime metric and on the observer conventions). 
Integral trajectories of $\vec n$ (as well as the horizontal trajectories of any connection) 
define what it means to {\it stand still} for an observer associated to the fixed ADM foliation.
Notice that two parallel vector fields (as $\vec n$ and $\vec m=N\vec n$ which will be defined below)
define, under this viewpoint, the same observer, but different standards of time.

We can hence define a (not-necessarily-orthonormal) frame $e_a=\ube_a^\mu \>\del_\mu=(e_0= \vec n, e_A)$ adapted to the foliation.
In adapted coordinates the vector $\del_0$ associated to the coordinate $x^0$ can be expanded along the basis $e_a$ as follows
$$
\del_0= N \vec n + N^A e_A
\fl{del0}$$
and defines the {\it lapse function} $N$ and the {\it shift vector} $\vec N= N^A \>e_A$.

\CNote{
In view of transformation laws \ShowLabel{eTranformationRules}, $N$ is a scalar on $S$ and $N^A$ transforms as a vector on $S$.
Of course, both are time-dependent as well.
}

The dual basis  $e^a= \obe^a_\mu \>dx^\mu$ of covectors can be defined accordingly.
Here $ \obe^a_\mu $ denotes the inverse matrix of $\ube_a^\mu$.
In view of \ShowLabel{ek}  and \ShowLabel{del0} in adapted coordinates one has
$$
\(\matrix{
\ube_0^0= N^{-1} \hfill			&\ube_A^0=0			\cr
\ube_0^i=-N^{-1} N^i		&\ube_A^i=\al^i_A		\cr
}\)
\qquad\qquad
\(\matrix{
\obe_0^0=N					& \obe_j^0=0 \cr
\obe_0^A= N^A				&\obe_j^A=\al_j^A\cr
}\)
\fn$$
where we set $N^i:=N^A \al_A^i$.
Of course, one can check that $\ube_a^\mu \obe^a_\nu=\de^\mu_\nu$, $\ube_a^\mu \obe^b_\mu=\de^b_a$.
By using the induced metric $\ga_{AB}$ (and its inverse $\ga^{AB}$) we define
$$
\eta_{ab}=g(e_a, e_b)=\(\matrix{
-1& 0\cr
0 & \ga_{AB}\cr
}\)
\qquad\qquad
\eta^{ab}= g(e^a, e^b)=\(\matrix{
-1& 0\cr
0 & \ga^{AB}\cr
}\)
\fn$$
The ADM expression of the spacetime metric
$g_{\mu\nu}= \obe^a_\mu \eta_{ab}\obe^b_\nu$ can be easily obtained as
$$
\matrix{
g_{00}=-\obe^0_0 \obe^0_0 + \obe^A_0 \ga_{AB}\obe^B_0=	-N^2+ |\vec N|^2			\hfill&\hskip 2cm&
g_{0j}= \obe^A_0 \ga_{AB}\obe^B_j=	 N\_j{}			\hfill\cr
g_{i0}= \obe^A_i \ga_{AB}\obe^B_0=	 N\_i{}			\hfill&&
g_{ij}= \obe^A_i \ga_{AB}\obe^B_j= \ga_{ij}			\hfill\cr
}
\fl{MetricADMDec}$$
where we set $\ga_{ij}:=  \al^A_i \ga_{AB}\al^B_j$ and $N\_i{} := N_B \al^B_i= N^j\ga_{ji}$.

Analogously, for the contravariant metric $g^{\mu\nu}= \ube_a^\mu \eta^{ab}\ube_b^\nu$ we obtain
$$
\matrix{
g^{00}=- \ube_0^0 \ube_0^0=	-N^{-2}			\hfill& \hskip 2cm&
g^{0j}= -\ube_0^0 \ube_0^j=	N^{-2} N^j			\hfill\cr
g^{i0}=-\ube_0^i \ube_0^0=	N^{-2} N^i			\hfill&&
g^{ij}= -\ube_0^i \ube_0^j	+ \ube_A^i \ga^{AB}\ube_B^j=	 \ga^{ij} - N^{-2} N^iN^j				\hfill\cr
}
\fn$$
where we set  $\ga^{ij}:=\al^i_A \ga^{AB} \al_B^j$ (which is of course the inverse of $\ga_{ij}$).

Here and hereafter spatial indices $A,B,C,\dots$ will be raised and lowered by the induced metric $\ga_{AB}$,
the triad indices  $i, j, k, \dots$ by the metric $\ga_{ij}$,
spacetime indices $\mu, \nu, \la, \dots$ by the metric $g_{\mu\nu}$ and tetrad indices $a, b, c, \dots$ by the
frame metric $\eta_{ab}$. Moreover, spatial indices can be transmuted into triad indices (and vice versa) by using the matrix $\al_A^i$. 

The element $g^{00}=-N^{-2}$ can be also expressed as $g^{00}=- \frac[1/ |g|]\ga $. Hence one has
$\sqrt{g}= N\sqrt{\ga}$.

The spacetime covariant derivative (with respect to the Levi-Civita connection of $g_{\mu\nu}$) will be denoted as usual by $\na_\mu$, while the covariant derivative of space objects
(with respect to the Levi-Civita connection of $\ga_{AB}$)  will be denoted by $D_A$.

\CNote{
Equation \ShowLabel{glueEmbeddings} can be regarded as a (time--dependent) change of coordinates on the fibers given by
$x'^i= \phi^i(t, k)$.
Accordingly, we can consider the metric $\ga_{ij}$ as the pull--back metric obtained by dragging $\ga_{AB}$ along this transformation on the fibers.
Hence one can define $^3\Ga^i_{jk}$ to be the Christoffell symbols of $\ga_{ij}$ and denote by $D_i$ its covariant derivative.
The Christoffell symbols $^3\Ga^i_{jk}$ and $^3\Ga^A_{BC}$ are related as follows
$$
\eqalign{
^3\Ga^A_{BC}=&\frac[1/2] \ga^{AD}\(-d_D \ga_{BC} + d_B \ga_{CD} + d_C\ga_{DB}\)=\cr
=&\frac[1/2] \ga^{AD}\Big(-d_D \al_B^j\ga_{jk}\al^k_C - \al_B^jd_D\ga_{jk}\al^k_C - \al_B^j\ga_{jk}d_D\al^k_C+\cr 
  &\hskip1cm+ d_B \al_C^k\ga_{ki}\al^i_D +  \al_C^kd_B\ga_{ki}\al^i_D +  \al_C^k\ga_{ki}d_B \al^i_D +\cr
  &\hskip1cm+ d_C\al_D^i\ga_{ij}\al^j_B + \al_D^i d_C\ga_{ij}\al^j_B+ \al_D^i\ga_{ij}d_C\al^j_B \Big)=\cr
=&\frac[1/2] \ga^{AD}\Big(- \red{\al_{DB}^j\ga_{jk}\al^k_C} - \uline{\al_B^jd_i\ga_{jk}\al^i_D\al^k_C} - \green{\al_B^j\ga_{jk}\al^k_{DC}}+\cr 
  		    &\hskip1cm+ \uuline{\al_{BC}^k\ga_{ki}\al^i_D} +  \uline{\al_C^kd_j\ga_{ki}\al^j_B\al^i_D} +  \red{\al_C^k\ga_{ki} \al^i_{BD} }+\cr
  		    &\hskip1cm+ \green{\al_{CD}^i\ga_{ij}\al^j_B} + \uline{\al_D^i d_k\ga_{ij}\al^j_B\al^k_C}+ \uuline{\al_D^i\ga_{ij}\al^j_{CB}} \Big)=\cr
=&\frac[1/2] \ga^{AD}\al^i_D\( -d_i\ga_{jk} + d_j\ga_{ki}+ d_k\ga_{ij} \)\al^j_B\al^k_C +  \ga^{AD}\al_D^i\ga_{ij}\al^j_{CB}=\cr
=&\frac[1/2] 	\al^A_l \ga^{li}\( -d_i\ga_{jk} + d_j\ga_{ki}+ d_k\ga_{ij} \)\al^j_B\al^k_C +  \al^A_jd_k\al^j_{B}\al^k_C=\cr
=&	\al^A_l \({}^3\Ga^l_{jk}\al^j_B + d_k\al^l_{B}\)\al^k_C\cr
}
\fn$$
as well as vice versa
$$
{}^3\Ga^l_{jk} =\al^l_A\({}^3\Ga^A_{BC} \al_j^B\al_k^C + d_k\al_j^A\)
\fn$$

This means that we can transform $D_A$ into $D_i$, and vice versa, by suitable multiplication for $\al^i_A$. 
For example % we define $\hat\na_i N^k = \al_i^A d_A N^k + {}^3\Ga^k_{ji} N^j$ and 
we have
$$
\eqalign{
 \al_i^A D_A N^B \al_B^k=& \al_i^A \(d_A N^B  + {}^3\Ga^B_{CA} N^C\) \al_B^k=\cr
=&\al_i^A \(d_A (N^B \al_B^k) -  N^B d_A\al_B^k + {}^3\Ga^B_{CA} N^C\al_B^k\) =\cr
=& \al_i^A d_A N^k + N^l \(\al_i^A  d_A\al_l^B\al_B^k +\al_i^A {}^3\Ga^B_{CA} \al_l^C\al_B^k\) =\cr
=& \al_i^A d_A N^k + N^l \al_B^k \(   {}^3\Ga^B_{CA} \al_l^C +  d_A\al_l^B\) \al_i^A=
%\cr=&
d_i N^k +    {}^3\Ga^k{}_{li} N^l  = D_i N^k\cr
}
\fn$$

We can prove the following Lemma:
$$
\eqalign{
D_k \al^i_A :=& \al^C_k d_C \al^i_A + {}^3\Ga^i{}_{jk} \al^j_A -   \al^C_k{}^3\Ga^B_{AC} \al^i_B=\cr
=& \({}^3\Ga^i{}_{jk} -   \al^i_B\({}^3\Ga^B_{EC}\al^E_j  -  \al^B_l d_C \al^l_E \al^E_j\)  \al^C_k \)  \al^j_A=\cr
=& \({}^3\Ga^i{}_{jk} -   \al^i_B\({}^3\Ga^B_{EC}\al^E_j  + d_C \al^B_j\) \al^C_k \)  \al^j_A \equiv 0\cr
}
\fn$$
}

Moreover, the fibers $S_t$ are submanifolds embedded into spacetime $M$ and one can define the extrinsic curvature
$$
K_{ij}:=\Frac[1/N] \( -D_{(i} N_{j)}+\frac[1/2] d_0 \ga_{ij}\) = \Frac[1/2N] \de_0 \ga_{ij}
\fl{ExtrinsicCurvature}$$
where we introduced the operator $ \de_0 := d_0 -\Lie_{\vec N}$ and $\Lie_{\vec N}$ denotes the Lie derivative with respect to the shift vector.
The extrinsic curvature $K_{ij}$ is, as usual, a symmetric space  tensor on $S_t$. 

\CNote{
We shall later prove that this is actually the extrinsic curvature defined in differential geometry and how it relates to the derivative of the normal $\vec n$ of the submanifold $S_t$. 
Until then \ShowLabel{ExtrinsicCurvature} is to be regarded as a notation.
One can also show that 
$$
\eqalign{
\al_A^i K_{ij} \al^j_B=&\Frac[1/2N]  \al_A^i \(  -D_{i} N_{j} -D_{j} N_{i}+d_0 \ga_{ij}\) \al^j_B=\cr
=&\Frac[1/2N]   \(  -D_{A} N_{B} -D_{B} N_{A} + d_0 \ga_{AB} - \redframe{$d_0 \al_A^i  \ga_{ij} \al^j_B$}-  \greenframe{$\al_A^i  \ga_{ij} d_0\al^j_B$}\)= 
K_{AB}
}
\fn$$
where Lemma \ShowLabel{Lemmad0al} has been used.

In the meanwhile, for later convenience let us prove that we are able to express  the ``time'' derivative of the spatial connection $^3\Ga^A_{BC}$
as a quantity on the fibers. In fact
$$
\eqalign{
\del_0 {}^3\Ga^A_{BC}=&\frac[1/2] \ga^{AD} \(-D_D \del_0\ga_{BC} + D_B \del_0\ga_{CD} + D_C \del_0\ga_{DB}\)=\cr
=&-D^A \(NK_{BC}\) + D_B \(NK_{C}{}^A\) + D_C \(NK^A{}_{B}\)+\cr
&-\frac[1/2]  [D^A ,D_{B}] N_{C}
 -\frac[1/2] [D^A, D_{C}] N_{B}
 + D_{(B} D_{C)} N^A=\cr
 =&-D^A \(NK_{BC}\) + D_B \(NK_{C}\^A\) + D_C \(NK\^A{}_{}{B}\)
 -\> {}^3R^A{}_{(BC)D} N^D 
 + D_{(B} D_{C)} N^A\cr
}
\fl{SpaceConnectionEvolution}$$
}
\CNote{
The ADM splitting of the spacetime metric induces the ADM splitting of its Christoffel symbols.

Before starting let us consider some special combinations used herafter;
in particular we have 
$$
\del_i g_{00}=D_i \(-N^2+|\vec N|^2 \)= -2N D_i N+2N^j D_iN_j
\fn$$
and
$$
\del_i g_{0j}= D_iN_j + {}^3\Ga^k_{ji} N_k
\fn$$

Then we obtain:
$$
\eqalign{
\Ga^0_{00}=&\frac[1/2] g^{00}\del_0 g_{00} -\frac[1/2] g^{0i}\del_i g_{00} +g^{0i}\del_0 g_{0i}
= \cr
=&-\frac[1/N^2] \(-N\del_0 N +\red{ N^k\del_0 N_k} + \frac[1/2]N_i N_j \del_0 \ga^{ij}  \) +\cr
&-\frac[N^i/N^2] \(-N D_i N + N^k D_iN_k \)  +\red{\frac[N^k/N^2] \del_0 N_k}
= \cr
=& \frac[1/N]\del_0 N+\underline{\frac[N^i N^j/2N^2]  \del_0 \ga_{ij} }
+\frac[N^i/N] D_i N -\underline{\frac[N^iN^k/N^2]  D_{(i}N_{k)}  }
= \cr
=&\frac[1/N]\( \del_0 N+ N^i D_iN+N^iN^j\>K_{ij}\)
}
\fn$$

$$
\eqalign{
\Ga^0_{0i}=&\frac[1/2] g^{00}\del_i g_{00} 
+\frac[1/2] g^{0k}\(-\del_k g_{0i} + \del_0 g_{ik}+ \del_i g_{k0} \)
=\cr
=&-\frac[1/2N^2]\(-2N D_iN + 2N^k D_iN_k\) 
+\frac[N^k/2N^2]\( D_iN_k -D_kN_i + \del_0 \ga_{ik} \)
=\cr
=&\frac[1/N]D_iN -\frac[N^k/N^2] D_iN_k
+\frac[N^k/2N^2]\( D_iN_k -D_kN_i + \del_0 \ga_{ik}\)
=\cr
=&\frac[1/N] D_iN 
+\frac[N^k/2N^2]\( -D_iN_k -D_kN_i + \del_0 \ga_{ik}\)
=\cr
=&\frac[1/N]\(D_iN +N^kK_{ik}\)
\cr
}
\fn$$

$$
\eqalign{
\Ga^0_{ij}=&\frac[1/2] g^{00}\(-\del_0 g_{ij} + \del_i g_{j0}+\del_j g_{i0}\)
+g^{0k}\>{}^3\Ga\_k{}_{ij}
=\cr
=&-\frac[1/2N^2] \(-\del_0 \ga_{ij} + D_i N_j+D_j N_i +\red{2\ {}^3\Ga^k_{ij} N_k}\) +\red{\frac[N_k/N^2]\>{}^3\Ga^k_{ij}}
= \frac[1/N] K_{ij} \cr
}
\fn$$

$$
\eqalign{
\Ga^i_{00}=&\frac[1/2] g^{i0}\del_0g_{00} 
+\frac[1/2] g^{ij}\>\(-\del_j g_{00} + 2\del_0 g_{j0}\)
=\cr
=&-\uline{\frac[1/2] N^ig^{00}\del_0g_{00} }
+\frac[1/2] \ga^{ij}\>\(-\del_j g_{00} + 2\del_0 g_{j0}\)
-\uline{\frac[1/2] N^ig^{0j}\>\(-\del_j g_{00} + 2\del_0 g_{j0}\)}
=\cr
=&-N^i\Ga^0_{00}
+\frac[1/2] D^i\(N^2-|\vec N|^2\) +\ga^{ij} \del_0 N_j
\cr
}
\fn$$

$$
\eqalign{
\Ga^i_{0j}=&\frac[1/2] g^{i0}\del_j g_{00} 
+\frac[1/2] g^{ik}\>\(-\del_k g_{0j} + \del_0 g_{jk}+\del_j g_{0k}\)
=\cr
=&-\frac[1/2] N^ig^{00}\del_j g_{00} 
+\frac[1/2] \ga^{ik}\>\(-\del_k g_{0j} + \del_0 g_{jk}+\del_j g_{0k}\)+\cr
&-\frac[1/2] N^i g^{0k}\>\(-\del_k g_{0j} + \del_0 g_{jk}+\del_j g_{0k}\)
=\cr
=&-N^i\Ga^0_{0j}
+\frac[1/2] \ga^{ik}\>\(-D_k N_j + \del_0 \ga_{jk}+D_j N_k\)
=\cr
=& -N^i \Ga^0_{0j}
+N K\^i{}_{j} 
+D_j N^i\cr
}
\fl{Gai0j}$$

$$
\eqalign{
\Ga^i_{jk}=&\frac[1/2] g^{i0}\(-\del_0 g_{jk}+ \del_j g_{k0}+ \del_k g_{0j}\) 
+g^{il}\>{}^3\Ga_{ljk}
=\cr
=&-N^i\frac[1/2] g^{00}\(-\del_0 g_{jk}+ \del_j g_{k0}+ \del_k g_{0j}\) 
+\ga^{il}\>{}^3\Ga_{ljk}- N^i g^{0l}\>{}^3\Ga_{ljk}
=\cr
=&-N^i\Ga^0_{jk}+{}^3\Ga^i_{jk}
}
\fl{Gaijk}$$

}

\CNote{
The {\it extrinsic curvature} of a time-dependent embedded hypersurface $i_t:S\arr M$ is a measure of the change in time $t$ of the normal unit vector.
As the name suggests that is a quantity depending on the embedding maps $i_t:S\mapsto M$.
Let $v= v^A \>\del_A$ a vector tangent to $S$, also 
interpreted as a vector $v=(i_t)_\ast(v)= v^A \> e_A$ tangent to $S_t$.

Let us define the endomorphism $\chi:TS_t\mapsto TS_t: v\mapsto \na_v \vec n$.
In adapted coordinates one has:
$$
\chi(v)= \na_v \vec n = v^\mu\( \del_\mu n^\al + \Ga^\al_{\be\mu} n^\be\)\del_\al
 = v^B \al_B^\mu \( \del_\mu n^\al + \Ga^\al_{\be\mu} n^\be\)\del_\al
= \chi^A_B v^B \> e_A 
\fn$$
The last equality is a consequence of the fact that the vector $\chi(v)$ is again tangent to $S_t$; in fact 
$$
\vec n\cdot \na_v \vec n=\frac[1/2] \na_v(\vec n\cdot \vec n)= 0
\fn$$
Hence one has
$$
\eqalign{
 \chi^A_B=&\al_B^j\(	\del_j n^i + \Ga^i_{0 j} n^0 + \Ga^i_{k j} n^k\) \al^A_i 
=\al_B^j \chi^i_j\al^A_i 
 }
\fn$$
$$
\eqalign{
 \chi^i_j=&	\del_j n^i + \Ga^i_{0 j} n^0 + \Ga^i_{k j} n^k = \blue{-D_j \(N^{-1}N^i\)}   -\blue{N^{-2}N^i D_jN}+\cr
 & -\red{N^{-2}N^iN^kK_{jk}}+K\^i{}_{j} +\blue{N^{-1} D_j N^i}   +\red{N^{-2}K_{kj} N^i N^k}=  K\^i{}_{j} 
 \cr
 }
\fn$$
The normal vector is $\vec n= N^{-1} (\del_0 - N^i\del_i)$ --- see \ShowLabel{del0}; the Christoffel symbols are expressed in terms of the metric tensor
which is in turn expressed as \ShowLabel{Gai0j} and \ShowLabel{Gaijk}.
By expanding all terms one easily obtains $K_{ij}\equiv \ga_{il}\chi^l_j$.
}

Despite the frame timelike vector $e_0$ is a unit vector and orthogonal to each $e_A$, the frame $e_a$ is not orthonormal in spacetime since the space vectors $e_A$ are not necessarily orthonormal in space. 
However, it is covenient to express geometric objects in this frame. For example, the Christoffel symbols $\Ga^\al_{\be\mu}$ of the metric $g$, 
namely the coefficients of its Levi-Civita connection, define the spin coefficients:
$$
\om^a{}_{b\la}=\obe^a_\al \( \Ga^\al_{\be\la} \ube^\be_b + d_\la \ube^\al_b\)
\qquad
\ubom^a{}_{bc}= \om^a{}_{b\la}\ube^\la_c
\fn$$
that have particularly simple expressions
$$
\matrix{
\ubom^0{}_{00}=0						\hfill&\hskip 1cm&
\ubom^A{}_{00}=N^{-1} D^A N				\hfill\cr
\ubom^0{}_{0C}=0						\hfill&&
\ubom^A{}_{0C}=K\^A{}_C				\hfill\cr
\ubom^0{}_{B0}= N^{-1} D_B N 				\hfill&&
\ubom^A{}_{B0}=	K\^A{}_B +  N^{-1}\(D_B N^A -{}^3\Ga^A_{BC} N^C\) 						\hfill\cr
\ubom^0{}_{BC}= K_{BC}					\hfill&&
\ubom^A{}_{BC}=	{}^3\Ga^A_{BC}		\hfill\cr
}
\fn$$

\CNote{
Let us prove here  the previous expressions for spin coefficients. One has:

$$
\eqalign{
\ubom^0{}_{00}=&
\obe^0_\al \( \Ga^\al_{\be\la} \ube^\be_0 + d_\la \ube^\al_0\)\ube_0^\la=\cr
=&\(N^{-2}\( \red{\del_0 N}+\green{ N^i D_iN}+\blue{N^iN^j\>K_{ij}}\) -   N^{-2} N^j  \(\green{D_jN} +\blue{N^kK_{jk}}\)  + \red{d_0 N^{-1}}\)+\cr
&-   \( N^{-2}N^j\(\green{\uline{D_jN}} +\red{\uline{N^kK_{jk}}}\) - \red{\uline{N^{-2} K_{jk} N^k N^j}} +  \green{\uline{N^kd_k N^{-1}}}\)\equiv 0
}
\fn$$

$$
\eqalign{
\ubom^0{}_{0C}=&
\obe^0_\al \( \Ga^\al_{\be\la} \ube^\be_0 + d_\la \ube^\al_0\)\ube_C^\la=\cr
=& \( \Ga^0_{0 k} - \Ga^0_{j k} N^j +N d_k N^{-1}\)\al_C^k=
  \( N^{-1}\(\red{D_kN} +\green{N^lK_{kl}}\)- \green{N^{-1} K_{j k} N^j} -\red{N^{-1} d_k N}\)\al_C^k\equiv 0
}
\fn$$

$$
\eqalign{
\ubom^0{}_{B0}=&
 \obe^0_\al \( \Ga^\al_{\be\la} \ube^\be_B + d_\la \ube^\al_B\)\ube_0^\la=\cr
=&\Ga^0_{j0} \al^j_B -  \Ga^0_{jk} \al^j_BN^k=
 \( N^{-1}\(D_jN +\red{N^kK_{jk}}\) - \red{ N^{-1}K_{jk}N^k } \)\al^j_B
 = N^{-1} D_jN \al^j_B=\cr
 =&N^{-1} D_B N 
}
\fn$$

$$
\eqalign{
\ubom^0{}_{BC}=&
\obe^0_\al \( \Ga^\al_{\be\la} \ube^\be_B + d_\la \ube^\al_B\)\ube_C^\la
=N \Ga^0_{jk} \ube^j_B \ube_C^k=K_{jk} \al^j_B \al_C^k
=: K_{BC} 
}
\fn$$

$$
\eqalign{
\ubom^A{}_{00}=&
\obe^A_\al \( \Ga^\al_{\be\la} \ube^\be_0 + d_\la \ube^\al_0\)\ube_0^\la=\cr
=&N^{-2} N^A \( \Ga^0_{00}   	-\Ga^0_{j0} N^j 	+ Nd_0 N^{-1} \)
+N^{-2} \al^A_i \( \Ga^i_{00} 	-\Ga^i_{j0} N^j		 - Nd_0 \(N^{-1} N^i\)\)+\cr
-&N^{-2} N^A \(\Ga^0_{0k}   	- \Ga^0_{jk} N^j 	+ Nd_k N^{-1} \)N^k
-N^{-2} \al^A_i \(\Ga^i_{0k}	 - \Ga^i_{jk} N^j 	- Nd_k \(N^{-1} N^i\)\)N^k
=\cr
=&N^{-2} N^A \( \Ga^0_{00}   	- \Ga^0_{j0} N^j 	+ \red{Nd_0 N^{-1}} \)
+N^{-2} \al^A_i \( \Ga^i_{00} 	-\Ga^i_{j0} N^j 		- \red{N d_0N^{-1}  N^i}		- d_0  N^i\)+\cr
-&N^{-2} N^A \(\Ga^0_{0k}   	- \blue{\Ga^0_{jk} N^j }	+ \green{Nd_k N^{-1}} \)N^k
-N^{-2} \al^A_i \(\Ga^i_{0k}	+\blue{N^i\Ga^0_{jk} N^j} 	- \green{N d_k N^{-1} N^i	}	- D_k N^i\)N^k=\cr
=&N^{-2} N^A \( \red{\Ga^0_{00} }  	- \Ga^0_{j0} N^j 	\)
+N^{-2} \al^A_i \(
-\red{N^i\Ga^0_{00} }		+N D^i N		-\uline{N^j D_l N_j \ga^{il}} 			+ \green{d_0 N^i}		+\uline{\ga^{il}d_0 \ga_{lj}  N^j}
					 	-\Ga^i_{j0} N^j 		- \green{d_0  N^i}\)+\cr
-&N^{-2} N^A \blue{\Ga^0_{0k} }  N^k
-N^{-2} \al^A_i \(
-\blue{N^i \Ga^0_{0k}}		+N K\^i{}_{k} 		+\red{\uline{D_k N^i	}}	-\red{\uline{ D_k N^i}}\)N^k=\cr
=&-\red{N^{-2} N^A  \Ga^0_{j0} N^j }
+N^{-2} \al^A_i \(
N D^i N		+\green{N^j D_j N^i }			+\red{\uline{2NK\^i{}_{j}  N^j}}
+\red{N^i \Ga^0_{0j} N^j}		-\red{\uline{N K\^i{}_{j}N^j} }			-\green{ D_j N^iN^j}\)+\cr
-&\red{\uline{N^{-1} \al^A_i K\^i{}_{k} N^k}}=
 N^{-1} \al^A_i D^i N= N^{-1} \al^A_i \ga^{ik}D_k N= N^{-1} \ga^{AB} \al_B^k D_k N=:  N^{-1}D^A N \cr
}
\fn$$

$$
\eqalign{
\ubom^A{}_{0C}=&
\obe^A_\al \( \Ga^\al_{0 k} \ube^0_0  +  \Ga^\al_{j k} \ube^j_0 + d_k \ube^\al_0\)\ube_C^k=\cr
=&N^A \(N^{-1}  \Ga^0_{0 k}   - N^{-1}  N^j \Ga^0_{j k}  + \red{d_k N^{-1}} \)\al_C^k
+\al^A_i \( N^{-1} \Ga^i_{0 k}    - N^{-1} N^j  \Ga^i_{j k}  - \red{d_k N^{-1} N^i}-  N^{-1} d_kN^i\)\al_C^k=\cr
=&N^A \( \green{N^{-1} \Ga^0_{0 k}}  - \red{N^{-1}  N^j \Ga^0_{j k} }  \)\al_C^k
+\al^A_i \(-\green{N^{-1}  N^i\Ga^0_{0 k}}    + N^{-1} NK\^i{}_k + \red{\uline{N^{-1} D_k N^i}}
+ \red{N^{-1} N^j  N^i \Ga^0_{j k}} -  \red{\uline{N^{-1}D_kN^i}}\)\al_C^k=\cr
=&\al^A_i    K\^i{}_k \al_C^k=: K\^A{}_C\cr
}
\fn$$

$$
\eqalign{
\ubom^A{}_{BC}=&
\obe^A_\al \( \Ga^\al_{j k} \ube^j_B + d_k \ube^\al_B\)\ube_C^k= \cr
=& \red{N^A  \Ga^0_{j k} \al^j_B \al_C^k }
+ \al^A_i \(-\red{N^i \Ga^0_{j k} \al^j_B} + D_k \al^i_B\)\al_C^k=
 \al^A_i  \(  {}^3\Ga^i_{jk} \al^j_B +d_k \al^i_B  \) \al_C^k
 =:{}^3\Ga^A_{BC}\cr
}
\fn$$

$$
\eqalign{
\ubom^A{}_{B0}=&
\obe^A_\al \( \Ga^\al_{j\la} \ube^j_B + d_\la \ube^\al_B\)\ube_0^\la=\cr
=& N^{-1}\( \(N^A  \Ga^0_{j0}  - N^A  \Ga^0_{jk}  N^k\)\al^j_B
+ \al^A_i  \( \Ga^i_{j0} \al^j_B + \redframe{$d_0 \al^i_B$}\)
- \al^A_i \( \Ga^i_{jk} \al^j_B + d_k \al^i_B\)N^k\)=\cr
=& N^{-1}\Big( \(\green{N^A  \Ga^0_{j0}}  -\red{ N^A  \Ga^0_{jk}  N^k}\)\al^j_B
+ \al^A_i  \( -\green{N^i\Ga^0_{j0} \al^j_B}  +N K\^i{}_j \al^j_B +D_j N^i \al^j_B \)+ \cr
&+ \al^A_i \(\red{N^i \Ga^0_{jk} \al^j_B} - D_k \al^i_B\)N^k\Big)=
%\cr=& 
 N^{-1}\al^A_i\( 
   N K\^i{}_j \al^j_B +D_j N^i \al^j_B 
-   D_k \al^i_B N^k\)=\cr
=&  N^{-1}\(  N \al^A_i\ K\^i{}_j \al^j_B
+  \al^A_i D_j N^i  \al^j_B
 -   \al^A_i D_k \al^i_CN^k  \al^C_j\al^j_B \) =\cr
=&   K\^A{}_B+N^{-1}\( 
D_B N^A 
 -   \al^A_iD_k \al^i_B \al_C^k N^C  \) =\cr
=&  K\^A{}_B+N^{-1}\(  D_B N^A  -  {}^3\Ga^A{}_{BC}  N^C  \) \cr
}
\fn$$
where Lemma \ShowLabel{Lemmad0al} has been used.

}

We can now define the curvature tensor of the spin coefficients
$$
\ubR^a{}_{bcd}= \( d_\mu \ubom^a{}_{b\nu} - d_\nu \ubom^a{}_{b\mu} + \ubom^a{}_{e\mu}\ubom^e{}_{b\nu} - \ubom^a{}_{e\nu}\ubom^e{}_{b\mu}\) \ube^\mu_c  \ube^\nu_d
\fn$$
This can be proven to be the tedrad expression of the Riemann tensor of the spacetime metric $g$, i.e.
$$
\ubR^a{}_{bcd}= \obe^a_\al \>R^\al{}_{\be\mu\nu}\> \ube^\be_b  \ube^\mu_c  \ube^\nu_d
\fn$$
\CNote{
This is trivial in tetrad formalism but here we are generalizing to non-orthonormal frames. 
For example, in our case the spin coefficients $\ubom^{ab}{}_c:= \eta^{bd}\ubom^a{}_{dc}$ are not antisymmetric in $[ab]$. Hence let us prove the claim
from scratch.
In fact:
$$
\eqalign{
\ubR^a{}_{b\mu\nu}=& d_\mu \om^a{}_{b\nu} + \om^a{}_{e\mu}\om^e{}_{b\nu} - [\mu\nu]=\cr
=& d_\mu  \obe^a_\al\( \Ga^\al_{\be\nu} \ube^\be_b + d_\nu \ube^\al_b\) 
+   \obe^a_\al\( d_\mu \Ga^\al_{\be\nu} \ube^\be_b +  \Ga^\al_{\be\nu} d_\mu \ube^\be_b  + \redframe{$d_{\mu\nu} \ube^\al_b$}\) +\cr
 &+   \obe^a_\al  \obe^e_\la\( \Ga^\al_{\be\mu} \ube^\be_e + d_\mu \ube^\al_e\) \( \Ga^\la_{\si\nu} \ube^\si_b + d_\nu \ube^\la_b\) - [\mu\nu]=\cr
=& d_\mu  \obe^a_\al \Ga^\al_{\be\nu} \ube^\be_b + d_\mu  \obe^a_\al d_\nu \ube^\al_b
+    \obe^a_\al d_\mu \Ga^\al_{\be\nu} \ube^\be_b +  \Ga^\al_{\be\nu} \obe^a_\al d_\mu \ube^\be_b +\cr
 &+   \obe^a_\al  \obe^e_\la\( 
 \Ga^\al_{\be\mu} \ube^\be_e \Ga^\la_{\si\nu} \ube^\si_b + \Ga^\la_{\be\nu} \ube^\be_b d_\mu \ube^\al_e
+  d_\nu \ube^\la_b \Ga^\al_{\be\mu} \ube^\be_e + d_\nu \ube^\la_b d_\mu \ube^\al_e 
\) - [\mu\nu]=\cr
=& \blue{ \Ga^\al_{\be\nu} d_\mu  \obe^a_\al \ube^\be_b} + \red{d_\mu  \obe^a_\al d_\nu \ube^\al_b}
+   d_\mu \Ga^\al_{\be\nu}  \obe^a_\al \ube^\be_b +  \green{\Ga^\al_{\be\nu} \obe^a_\al d_\mu \ube^\be_b }+\cr
 &+  
 \Ga^\al_{\be\mu}    \Ga^\be_{\si\nu} \obe^a_\al\ube^\si_b - \blue{\Ga^\la_{\be\nu} \ube^\be_b d_\mu \obe^a_\la }
-  \green{ \Ga^\al_{\be\nu}\obe^a_\al d_\mu \ube^\be_b} -    \red{d_\nu \ube^\la_b d_\mu\obe^a_\la  }
 - [\mu\nu]=\cr
=&   \obe^a_\al  \(d_\mu \Ga^\al_{\si\nu}  + \Ga^\al_{\be\mu}    \Ga^\be_{\si\nu} - [\mu\nu] \) \ube^\si_b=
  \obe^a_\al  R^\al{}_{\be\mu\nu} \ube^\be_b\cr
}
\fn$$
Here $[\mu\nu]$ means ``the same expession with the indices $[\mu\nu]$ exchanged''.
}

The frame components of the Riemann tensor $\ubR^{a}{}_{bcd}$ can be obtained as
$$
\eqalign{
&\ubR^{0}{}_{00D}=0								\cr
&\ubR^{0}{}_{B0D}	\equiv \ga_{BE}  \ubR^E{}_{00D}		\cr
&\ubR^{0}{}_{BCD}=	2D_{[C} K_{D] B}					\cr
&\ubR^{0}{}_{0CD}=0				\cr
}
\hskip 1cm
\eqalign{
&\ubR^{A}{}_{B0D}=2\ga^{AE} D_{[B} K_{E]D}					\cr
&\ubR^{A}{}_{00D}=	N^{-1}  \( \de_0K\^A{}_{D}   -D_D D^A N \) + K^{EA}K_{E D}  
				\cr
&\ubR^{A}{}_{0CD}=	2 D_{[C} K_{D]}\^A					\cr
&\ubR^{A}{}_{BCD}={}^3R^A{}_{BCD}  +2 K^A{}_{[C} K_{D]B}	\cr
}$$

\CNote{
Let us here prove these expressions.

$$
\eqalign{
\ubR^{0}{}_{00D}=&
 \( \redframe{$d_\mu \ubom^0{}_{0\nu}$} - \greenframe{$d_\nu \ubom^0{}_{0\mu}$} + \ubom^0{}_{e\mu}\ubom^e{}_{0\nu} - \ubom^0{}_{e\nu}\ubom^e{}_{0\mu}\)\ube^\mu_0  \ube^\nu_D=
   \ubom^0{}_{E0}\ubom^E{}_{0D} - \ubom^0{}_{ED}\ubom^E{}_{00}=\cr
=& N^{-1} D_E N  K\^E{}_D- K_{ED}N^{-1} D^E N	\equiv 0
}
\fn$$

$$
\eqalign{
\ubR^{0}{}_{B0D}=&
\( d_\mu \(\ubom^0{}_{BE}\obe^E_k \)- d_k \(\ubom^0{}_{B0} \obe^0_\mu + \ubom^0{}_{BE} \obe^E_\mu\)
+ \ubom^0{}_{E\mu}\ubom^E{}_{Bk} - \ubom^0{}_{Ek}\ubom^E{}_{B\mu}\) \ube^\mu_0  \al^k_D=\cr
=& d_\mu \(K_{BE}\al^E_k \)\ube^\mu_0  \al^k_D - d_D \(\ubom^0{}_{B0} \obe^0_0\) \ube^0_0 
- d_D \(K_{BE} \obe^E_\mu\)\ube^\mu_0 
+N^{-1} D_E N\>^3\Ga^E_{BD} +\cr
&- K_{ED}K\^E{}_B - N^{-1}K_{ED} D_B N^E + N^{-1}K_{ED}{}^3\Ga^E_{BC} N^C  =\cr
=& N^{-1}d_0 \(K_{BE}\al^E_k \)  \al^k_D  - \uline{N^{-1} d_D D_BN }
-N^{-1} d_i \(K_{BE}\al^E_k \)N^i  \al^k_D
- N^{-1}d_D \(K_{BE} N^E\) + N^{-1} d_D \(K_{BE} \al^E_i\)N^i 
 +\cr
&+\uline{N^{-1} D_E N\>^3\Ga^E_{BD}}- K_{ED}K\^E{}_B - N^{-1}K_{ED} D_B N^E + N^{-1}K_{ED}{}^3\Ga^E_{BC} N^C  =\cr
=& N^{-1}\Big(d_0 K_{BD}    -  D_D D_BN  - D_i \(K_{BE}\al^E_k \)N^i  \al^k_D
- D_D \(K_{BE} N^E\) + D_D \(K_{BE} \al^E_i\)N^i +\cr
&-\blue{\>^3\Ga^A_{BC} K_{AD} N^C}   -\green{ \>^3\Ga^l_{ki} K_{BE}\al^E_l N^i  \al^k_D}
- \red{\>^3\Ga^C_{BD} K_{CE} N^E }
+ \red{\>^3\Ga^C_{BD} K_{CE} N^E} + \green{\>^3\Ga^l_{ik}\al^k_D K_{BE} \al^E_lN^i }
+\cr
&- NK_{ED}K\^E{}_B - K_{ED} D_B N^E \Big)  =\cr
=& N^{-1}\Big(d_0 K_{BD}   -  D_D D_BN  - \uline{N^A  D_A K_{BD} }
- \red{D_D K_{BE} N^E}-  \uline{K_{BE} D_DN^E} + \red{D_D K_{BE} N^E} +\cr
&- NK_{ED}K\^E{}_B - \uline{K_{ED} D_B N^E} \Big)  
=
N^{-1}\(\de_0 K_{BD}  - D_D D_BN   - NK_{ED}K\^E{}_B \)  =\cr
=&
N^{-1}\(\uline{2NK_{BA}K\^A{}_{D} } +  \ga_{BA}\de_0 K\^A{}_{D}  - D_D D_BN   - \uline{NK_{ED}K\^E{}_B} \)  =\cr
=&N^{-1} \ga_{BA} \( \de_0 K\^A{}_{D}  - D_D D^AN   +NK_{ED}K^{EA} \)  \cr
}
\fn$$

$$
\eqalign{
\ubR^{0}{}_{BCD}=&
 d_C \(\ubom^0{}_{BA}\al^A_i\)   \al^i_D - d_D \(\ubom^0{}_{BA}\al^A_i \) \al^i_C  
+ \ubom^0{}_{EC}\ubom^E{}_{BD} - \ubom^0{}_{ED}\ubom^E{}_{BC}=\cr
=&
 D_C \(K_{BA}\al^A_i\)   \al^i_D - D_D \(K_{BA}\al^A_i \) \al^i_C 
 +\red{ \>^3\Ga^E_{BC} K_{ED}} + \uline{\>^3\Ga^k_{ij}\al^j_C K_{BA}\al^A_k  \al^i_D} +\cr
& - \green{\>^3{}\Ga^E_{BD} K_{EC}} - \uline{\>^3{}\Ga^k_{ij}\al_D^j K_{BA}\al^A_k  \al^i_C }+ \green{K_{EC}\>^3\Ga^E_{BD}} - \red{K_{ED}\>^3\Ga^E_{BC}}=\cr
=&
 D_C K_{BD} - D_D K_{BC} = 2 D_{[C} K_{D]B}\cr
}
\fn$$

$$
\eqalign{
\ubR^{0}{}_{0CD}=&
\( \redframe{$d_\mu \ubom^0{}_{0\nu}$} - \greenframe{$d_\nu \ubom^0{}_{0\mu}$} + \ubom^0{}_{e\mu}\ubom^e{}_{0\nu} - \ubom^0{}_{e\nu}\ubom^e{}_{0\mu}\) \ube^\mu_C  \ube^\nu_D=
\ubom^0{}_{EC}\ubom^E{}_{0D} - \ubom^0{}_{ED}\ubom^E{}_{0C}=\cr
=& K_{EC} K\^E{}_D -K_{ED}K\^E{}_C \equiv 0
}
\fn$$

$$
\eqalign{
\ubR^{A}{}_{00D}=&
N^{-1} d_0\(\ubom^A{}_{0C}\al^C_i\)\al^i_D -N^{-1} d_l\(\ubom^A{}_{0C}\al^C_i\)\al^i_D N^l
 -d_D\(\ubom^A{}_{0a}\obe^a_\mu\) \ube^\mu_0 
 + \ubom^A{}_{E0}\ubom^E{}_{0D} - \ubom^A{}_{ED}\ubom^E{}_{00}=\cr
 =&
N^{-1} d_0\ubom^A{}_{0D} -N^{-1} d_l\(K\^A{}_{C}\al^C_i\)\al^i_D N^l
-N^{-1}d_D\(\ubom^A{}_{00}N + \ubom^A{}_{0B} N^B\)  +N^{-1}d_D\(\ubom^A{}_{0B}\al^B_i\) N^i+\cr
& + \ubom^A{}_{E0}\ubom^E{}_{0D} - \>^3\Ga^A_{ED}\ubom^E{}_{00}
= 
N^{-1} d_0K\^A{}_{D} -N^{-1} D_l\(K\^A{}_{C}\al^C_i\)\al^i_D N^l 
+\red{\uline{N^{-1} \>^3\Ga^A_{EC}K\^E{}_{D} N^C }}+\cr
&-\blue{N^{-1} \>^3\Ga^l_{ik} K\^A{}_{B}\al^B_l\al^i_D N^k }
 -N^{-1}D_D D^AN -N^{-1}D_D\(K\^A{}_{B} N^B\)  +N^{-1}D_D\(K\^A{}_{B}\al^B_i\) N^i+\cr
 &+\red{N^{-1}\>^3\Ga^A_{ED} D^EN} + \green {N^{-1}\>^3\Ga^A_{ED}K\^E{}_{C} N^C}
-\green{N^{-1} \>^3\Ga^A_{CD} K\^C{}_{B} N^B}  +\blue{N^{-1}\al^B_l  \>^3\Ga^l_{ik}\al^k_D K\^A{}_{B}N^i}+\cr
& + K\^A{}_E K\^E{}_{D} +  N^{-1}K\^E{}_{D} D_E N^A- \red{\uline{N^{-1}K\^E{}_{D}{}^3\Ga^A_{EC} N^C}} - \red{N^{-1}\>^3\Ga^A_{ED}D^EN}=\cr
 =&
N^{-1} d_0K\^A{}_{D}  -\uline{N^{-1} N^l D_lK\^A{}_{D} }
-N^{-1}D_D D^AN -\red{N^{-1}D_DK\^A{}_{B} N^B} -\uline{N^{-1}K\^A{}_{B} D_DN^B}  +\cr
& +\red{N^{-1}D_DK\^A{}_{B} N^B}+ K\^A{}_E K\^E{}_{D} +  \uline{N^{-1}K\^E{}_{D} D_E N^A} =\cr
=&
N^{-1}\( \de_0K\^A{}_{D}  -D_D D^AN  \)  + K\^A{}_E K\^E{}_{D} \cr
}
\fn$$

$$
\eqalign{
\ubR^{A}{}_{B0D}=&
 d_\mu\( \ubom^A{}_{BC}\al^C_k\) \ube^\mu_0  \al^k_D - d_k \(\ubom^A{}_{B0}\obe^0_\mu + \ubom^A{}_{BC}\obe^C_\mu \) \ube^\mu_0  \al^k_D+ \ubom^A{}_{e0}\ubom^e{}_{BD} - \ubom^A{}_{eD}\ubom^e{}_{B0}=\cr
 =&
 N^{-1}\Big(d_0\( \ubom^A{}_{BC}\al^C_k\) \al^k_D -{d_D \(\ubom^A{}_{B0}N \) }  - d_i\( \ubom^A{}_{BC}\al^C_k\) N^i  \al^k_D 
 - d_k \(\ubom^A{}_{BC} N^C \)\al^k_D + \cr
 &+d_k \(\ubom^A{}_{BC}\al^C_i \) N^i  \al^k_D+ D^AN K_{BD} +{\ubom^A{}_{C0}\ubom^C{}_{BD}}
  - K\^A{}_{D}D_B N -{\ubom^C{}_{B0}\ubom^A{}_{CD}}\Big)=\cr
 =&
 N^{-1}\Big(d_0 \>^3\Ga^A_{BD} - \uline{d_D \(NK\^A{}_B +  D_B N^A\)} +\red{d_D\({}^3\Ga^A_{BC} N^C \)}  - d_i\( \ubom^A{}_{BC}\al^C_k\) N^i  \al^k_D +\cr
 &- \red{d_D \({}^3\Ga^A_{BC} N^C \)} +d_k \(\ubom^A{}_{BC}\al^C_i \) N^i  \al^k_D+ D^AN K_{BD}  + \cr
 &+ \uline{ \(NK\^A{}_C +  D_C N^A  \)\ubom^C{}_{BD}}
 -{   {}^3\Ga^A_{CE} N^E \ubom^C{}_{BD}}
  - K\^A{}_{D}D_B N +\cr
  &- \uline{\( NK\^C{}_B +  D_B N^C  \)\ubom^A{}_{CD}}
  +{  {}^3\Ga^C_{BE} N^E \ubom^A{}_{CD}}\Big)=\cr
 =&
 N^{-1}\Big(d_0 \>^3\Ga^A_{BD} - {D_D \(NK\^A{}_B +  D_B N^A\)} 
 - \uline{d_C\>^3\Ga^A_{BD}N^C} +  \>^3\Ga^A_{BC}\al^C_k N^E d_E \al^k_D  +\uline{d_D \>^3\Ga^A_{BC}  N^C}+\cr
 &  
    + \>^3\Ga^A_{BC}d_D\al^C_i  N^i 
  + D^AN K_{BD}   -\uline{   {}^3\Ga^A_{EC}  \>^3\Ga^E_{BD} N^C}
  - K\^A{}_{D}D_B N   +\uline{  {}^3\Ga^E_{BC}  \>^3\Ga^A_{ED}N^C}\Big)=\cr
=&
 N^{-1}\Big(d_0 \>^3\Ga^A_{BD} - D_D NK\^A{}_B -  ND_DK\^A{}_B 
 - {D_D   D_B N^A} 
 - \>^3R^A{}_{BCD}N^C -  \red{\>^3\Ga^A_{BC}d_D\al^C_k N^k} +\cr
 &  
    +\red{ \>^3\Ga^A_{BC}d_D\al^C_i  N^i }
  + D^AN K_{BD}   
  - K\^A{}_{D}D_B N  \Big)=\cr
 =&
 N^{-1}\Big(-D^A \(NK_{BD}\) + D_B \(NK_{D}\^A\) + \green{D_D \(NK\^A{}_{}{B}\)}
 -\> {}^3R^A{}_{(BD)E} N^E + \frac[1/2][D_{B}, D_{D}] N^A+ \red{D_{D} D_{B} N^A}+\cr
& - \green{D_D \(NK\^A{}_B \)}
 - \red{D_D   D_B N^A} 
 - \>^3R^A{}_{BCD}N^C  
  + D^AN K_{BD}   
  - K\^A{}_{D}D_B N  \Big)=\cr
}$$
$$
\eqalign{
 =&
 N^{-1}\Big(-\red{D^A NK_{BD}} - ND^AK_{BD} + \green{D_B NK_{D}\^A} +  ND_BK_{D}\^A
 -\> \frac[1/2]{}^3R^A{}_{BDE} N^E+\cr
&  -\> \uline{\frac[1/2]{}^3R^A{}_{DBE} N^E} - \uline{\frac[1/2]{}^3R^A{}_{EDB}N^E}
 - \>^3R^A{}_{BCD}N^C  
  + \red{D^AN K_{BD} }  
  - \green{K\^A{}_{D}D_B N}  \Big)=\cr
=&
 N^{-1}\Big(- ND^AK_{BD}  +  ND_BK_{D}\^A
 -\> \uline{\frac[1/2]{}^3R^A{}_{BDE} N^E}  +\> \uline{\frac[1/2]{}^3R^A{}_{BED} N^E}
 - \>^3R^A{}_{BCD}N^C  
    \Big)=\cr
=&
 N^{-1}\Big(- ND^AK_{BD}  +  ND_BK_{D}\^A
+\red{{}^3R^A{}_{BED} N^E}
 - \red{\>^3R^A{}_{BED}N^E  }
    \Big)=\cr
=&
 \(  D_BK_{D}\^A- D^AK_{BD}  \)=
\ga^{AC} \(  D_BK_{CD}- D_CK_{BD}  \)= 
2\ga^{AC} D_{[B}K_{C]D}   \cr
}
\fn$$

$$
\eqalign{
\ubR^{A}{}_{BCD}=&
 d_C \(\ubom^A{}_{BE}\al^E_j \)   \al^j_D- d_D \(\ubom^A{}_{BE}\al_j^E\) \ube^j_C   + \ubom^A{}_{eC}\ubom^e{}_{BD} - \ubom^A{}_{eD}\ubom^e{}_{BC}=\cr
 =&
 d_C \ubom^A{}_{BD}- d_D \ubom^A{}_{BC}   + \ubom^A{}_{EC}\ubom^E{}_{BD} - \ubom^A{}_{ED}\ubom^E{}_{BC}
 + \ubom^A{}_{0C}\ubom^0{}_{BD} - \ubom^A{}_{0D}\ubom^0{}_{BC}=\cr
  =&
\>^3R^A{}_{BCD} + K\^A{}_{C} K_{BD} - K\^A{}_{D} K_{BC}= \>^3R^A{}_{BCD} + 2K\^A{}_{[C} K_{D]B} \cr
}
\fn$$

$$% \( d_\mu \ubom^a{}_{b\nu} - d_\nu \ubom^a{}_{b\mu} + \ubom^a{}_{e\mu}\ubom^e{}_{b\nu} - \ubom^a{}_{e\nu}\ubom^e{}_{b\mu}\) \ube^\mu_c  \ube^\nu_d
\eqalign{
\ubR^{A}{}_{0CD}=&
 d_C\( \ubom^A{}_{0E} \al_j^E\)   \al^j_D- d_D\( \ubom^A{}_{0E} \al_j^E\) \al^j_C  
 + \>^3\Ga^A_{EC}\ubom^E{}_{0D} - \>^3\Ga^A_{ED}\ubom^E{}_{0C}=\cr
 =&
 D_C\( \ubom^A{}_{0E} \al_j^E\)   \al^j_D- D_D\( \ubom^A{}_{0E} \al_j^E\) \al^j_C 
 + \>^3\Ga^l_{jk}\al^k_C \ubom^A{}_{0E} \al_l^E   \al^j_D- \>^3\Ga^l_{jk}\al^k_D\ubom^A{}_{0E} \al_l^E \al^j_C  
=\cr
 =&
 D_C K\^A{}_{D} - D_D K\^A{}_{C} 
 + \red{\>^3\Ga^l_{jk}\al^k_C \ubom^A{}_{0E} \al_l^E   \al^j_D}- \red{\>^3\Ga^l_{jk}\al^j_D\ubom^A{}_{0E} \al_l^E \al^k_C }
=
 2D_{[C} K\^A{}_{D]}  
\cr
}
\fn$$

}

We can hence define the Ricci tensor
$$
\ubR_{bd}= \ubR^a{}_{bad}\equiv R_{\mu\nu} \>\ube^\mu_b\ube^\nu_d
\fn$$
\CNote{
$$
\eqalign{
\ubR_{00}=& \ubR^A{}_{0A0}= - N^{-1}  \( \de_0K   -D_A D^A N \) - K^{EA}K_{E A} \cr
\ubR_{0B}=&  \ubR^C{}_{0CB}=  D_{C} \(K_{B}\^C - \de_{B}^C K\)\cr
\ubR_{A0}=& \ubR^C{}_{AC0}=   D_{C} \(K\^C{}_{A} -\de^C_{A}K\)\cr
\ubR_{AB}=& \ubR^0{}_{A0B}+ \ubR^C{}_{ACB}=
N^{-1}  \( \ga_{AE}\de_0K\^E{}_{B}   -D_B D_A N \) 
+{}^3R_{AB}  + K K_{AB}	\cr
}
$$
}
and the Ricci scalar
$$
\eqalign{
R= &\eta^{ab} \ubR_{ab}= \frac[2/N] \( \de_0K -D_A D^AN    \)+{}^3R  + K^2+K_{AB} K^{AB}\cr
}
\fn$$

Let now $\xi=\xi^\mu\>\del_\mu=\obxi^a \>e_a$ be a spacetime vector. We can define the covariant derivatives
$$
\na_\la \xi^\mu= d_\la \xi^\mu + \Ga^\mu_{\nu\la} \xi^\nu\>\>,
\qquad\qquad
\na_a \obxi^b= \ube_a^\al d_\al \obxi^b + \ubom^b{}_{ca} \obxi^c
\fl{nablaxi}$$
The second is just another expression for the tetrad component of the first one.

\CNote{
In fact, we have
$$
\eqalign{
\ube^\la_a \>\na_\la \obxi^\mu \>\obe^b_\mu=& \ube^\la_a \>\( d_\la \xi^\mu + \Ga^\mu_{\nu\la} \xi^\nu \) \>\obe^b_\mu
=  \ube^\la_a d_\la\( \xi^\mu \obe^b_\mu\) +   \( \obe^b_\mu \Ga^\mu_{\nu\la} \ube^\nu_c \ube^\la_a   - \ube^\la_a \ube_c^\mu d_\la  \obe^b_\mu \)\>\obxi^c =\cr
=&  \ube^\la_a d_\la \obxi^b +   \obe^b_\mu\(  \Ga^\mu_{\nu\la} \ube^\nu_c   +  d_\la \ube_c^\mu   \)\ube^\la_a\>\obxi^c 
=\ube^\la_a d_\la \obxi^b +  \ubom^b{}_{ca} \>\obxi^c \equiv \na_a \obxi^b\cr
}
\fn$$
}

\CNote{
For the vector $\xi=\vec n$ (i.e.~$\obxi^0=1$, $\obxi^A=0$) we get
$$
\matrix{
\na_0 \obxi^0=	0				\hfill&\hskip 2cm &
\na_0 \obxi^B=	N^{-1}D^A N		\hfill\cr
\na_A \obxi^0=	0				\hfill&\hskip 2cm &
\na_A \obxi^B=	K\^B{}_A			\hfill\cr
}
\fl{NablaNormal}$$
For the vector $\xi=\vec N$ (i.e.~$\obxi^0=0$, $\obxi^A=N^A$) we get
$$
\matrix{
\na_0 \obxi^0=	N^{-1} N^C D_C N	\hfill&\hskip 2cm &
\na_0 \obxi^B=	N^{-1} d_0 N^B  +  K\^B{}_C N^C	\hfill\cr
\na_A \obxi^0=	N^C K_{AC}		\hfill&\hskip 2cm &
\na_A \obxi^B=	D_A N^B				\hfill\cr
}
\fn$$
}

The second derivatives will be defined as
$$
\na_a\na_b \obxi^c= \ube_a^\al d_\al  \na_b \obxi^c + \ubom^c{}_{da} \na_b\obxi^d - \ubom^d{}_{ba} \na_d\obxi^c
\fn$$

\CNote{
Let us compute as an example only the second derivatives that will be used below.

For the vector $\xi=\vec n$ (i.e.~$\obxi^0=1$, $\obxi^A=0$) we get
$$
\eqalign{
%\na_A\na_0 \obxi^C=& D_A  \(N^{-1} D^C N\)  -  K\^C{}_E K\^E{}_A\cr
\na_0\na_A \obxi^C=&  N^{-1} \de_0 K\^C{}_A -N^{-2} D_A N D^C N\cr
\na_A\na_B \obxi^0=& K_{EB} K\^E{}_A	\cr
}
\fl{SecondDerivative}$$

For the vector $\xi=\vec N$ (i.e.~$\obxi^0=0$, $\obxi^A=N^A$) we get
$$
\eqalign{
%\na_A\na_0 \obxi^C=& D_A \( N^{-1} \de_0 N^C  +  K\^C{}_E N^E \)  + N^{-1}  K\^C{}_{A} N^E D_E N- K\^E{}_{A} D_E N^C \cr
\na_0\na_C \obxi^A=& D_C \( N^{-1} \de_0 N^A  +  K\^A{}_E N^E \)  + N^{-1}  K\^A{}_{C} N^E D_E N- K\^E{}_{C} D_E N^A+\cr
&+N^ED_E K\^A{}_C - N^E D\^A K_{CE} \cr
\na_A\na_B \obxi^0=&D_A   \(N^EK_{EB}\)  + K_{EA} D_BN^E - N^{-1} K_{AB} N^E D_E N\cr
}
\fn$$

We stress however that all second derivatives can be computed and we suggest the reader to do it as an exercise.
}

The Lie derivative of spin coefficients will be defined as
$$
\Lie_\xi \ubom^a{}_{bc}\equiv \obe^a_\al \Lie_\xi \Ga^\al_{\be\la} \> \ube_b^\be \ube_c^\la=\obxi^ d \ubR^a{}_{cdb} + \na_b\na_c \obxi^a
\fn$$
Notice that this is to be understood just as a {\it notation}, not as the true Lie derivative of the spin connection. The spin connection is in fact a gauge--natural 
object and one cannot define its Lie derivative along spacetime vector fields, but just along gauge transformations; see \ref{Book}.
This is instead related to the so-called Kosman lift (see \ref{OurKosman}) implicitly introduced by Kosman in order to define Lie derivative of spinors (see \ref{Kosman}).

For future convenience let us compute the Lie derivatives that will be used hereafter.

For $\xi=\vec n$ (i.e.~$\obxi^0=1$, $\obxi^A=0$) we obtain:
$$
\eqalign{
\Lie_{\vec n} \ubom^0{}_{BC}=& N^{-1}  \(\de_0K_{BC}   -D_B D_C N \) \cr
\Lie_{\vec n} \ubom^A{}_{0C}=&  N^{-1}   \de_0K\^A{}_{C}  -  N^{-2} D_C N D^A N \cr
}
\fn$$

For $\xi=N^A e_A$ (i.e.~$\obxi^0=0$, $\obxi^A=N^A$) we obtain:
$$
\eqalign{
\Lie_{\vec N} \ubom^0{}_{BC}=& D_E K_{BC} N^E + D_C   N^EK_{EB} + K_{EC} D_BN^E - N^{-1} K_{BC} N^E D_E N\cr
\Lie_{\vec N} \ubom^A{}_{0C}=& \Lie_{\vec N}K_C{}\^A +  D_C \( N^{-1} \de_0 N^A   \)
 + N^{-1}  K\^A{}_{C} N^E D_E N\cr
}
\fn$$

By using the identity 
$$
\na_b\na_c (f \obxi^a)= \na_b\na_c f \obxi^a +\na_b f   \na_c \obxi^a+\na_c f   \na_b \obxi^a +  f  \na_b \na_c \obxi^a
\fn$$
setting $\xi=e_0$, $f=N$ and $\vec m= N\vec n$ (i.e.~$\obxi^0=N$, $\obxi^A=0$) we get
$$
\Lie_{\vec m}  \ubom^a{}_{bc}= N  \Lie_{\vec n}  \ubom^a{}_{bc} + \na_b\na_c N \obxi^a +\na_b N   \na_c \obxi^a+\na_c N   \na_b \obxi^a
\fl{LiemOm}$$

\CNote{
Let us here compute for later convenience the first derivatives 
$$
\cases{
&\na_0 N= \ube_0^0 d_0 N + \ube_0^i d_i N= N^{-1} \de_0 N\cr
&\na_A N= \al_A^i d_i N = d_A N= D_A N\cr
}
\fl{NaN}$$
and the second derivatives of the lapse function
$$
\na_a \na_b N= \ube_a^\al d_\al  \na_b N -\ubom^c{}_{ba}\na_c N
\fn$$
that can be expanded as
$$
\cases{
&\na_0\na_0 N=\uline{N^{-1} d_0  \na_0 N} -\uline{N^{-1} N^i d_i  \na_0 N} -\ubom^C{}_{00}\na_C N= N^{-1} \de_0  \(N^{-1} \de_0 N\)
  -N^{-1}D^C ND_C N\cr
&\na_0\na_B N= \uline{N^{-1} d_0  \na_B N}-\uline{N^{-1} N^k d_k  \na_B N} -\ubom^0{}_{B0}\na_0 N\ -\uline{\ubom^C{}_{B0}\na_C N}=\cr
&\hskip1.1cm= \de_0\(N^{-1}  D_B N\)  -K\^C{}_B D_C N  \cr
&\na_A\na_0 N=  d_A  \na_0 N -\ubom^C{}_{0A}\na_C N = D_A  \(N^{-1} \de_0 N\) -K\^C{}_{A}D_C N \equiv \na_0\na_A N \cr
&\na_A\na_B N=  d_A  \na_B N -\ubom^0{}_{BA}\na_0 N -\ubom^C{}_{BA}\na_C N
=  D_A  D_B N -N^{-1}K_{BA} \de_0 N \cr
}
\fl{Na2N}$$
}

For future convenience let us compute the Lie derivatives that will be used hereafter.
$$
\eqalign{
\Lie_{\vec m} \ubom^0{}_{BC}=& \de_0K_{BC}   -K_{BC} N^{-1}\de_0 N  \cr
\Lie_{\vec m} \ubom^A{}_{0C}=&     \de_0K\^A{}_{C}  +   K\^A{}_CN^{-1}\de_0 N \cr
}
\fn$$

Let us remark that the vector $\vec n$ is not projectable with respect to the ADM foliation, while $\vec m$ is.
Hence the flow of $\vec m$ sends fibers into fibers, i.e.~it preserves simultaneity, while the flow of $\vec n$ does not (in general).
Let us also remark that one has $\del_0= \vec m + \vec N$.
We shall use these Lie derivatives in the applications below.

\NewSection{Covariant Conservation Laws}

In \ref{Augmented} we presented a framework based on Noether theorem to define
covariant strong conservation laws for Lagrangian systems. 
The framework here presented is practically equivalent to many frameworks independently proposed in literature (sometimes requiring extra hypotheses that are unnecessary); see \ref{Julia}, \ref{Torre}, \ref{Cavalese}, \ref{BL}, \ref{Freud}.

In this framework one doubles the fields introducing, besides the dynamical metric $g$, the so-called {\it reference metric} $\hat g$ 
and the conservation laws so obtained are interpreted as describing the {\it relative} conserved quantities 
between the reference and the dynamical fields.
This catches most of the instances of covariantization mentioned in the literature on the basis of {\it ad hoc} procedures.

The augmented covariant conserved quantities for a space section $S_t=\{x^0=t\}\subset M$ are defined as
$$
\calQ[\xi] = \Frac[1/2\ka]\int_{S_t} E[\xi] - \Frac[1/2\ka]\int_{S_t}\hat E[\xi] +\Frac[1/2\ka]\int_{\del S_t} \De[\xi] 
\fl{CQ}$$
where we set $\ka=8\pi G/c^4$ and, 
letting $u^\al_{\mu\nu}= \Ga^\al_{\mu\nu}-\de^\al_{(\mu}\Ga^\la_{\nu)\la}$, we set:
$$
 E[\xi] = \sqrt{g} \(g^{\mu\nu} \Lie_\xi u^\al_{\mu\nu} - \xi^\al R\)\> ds_\al
 \qquad\qquad
 \hat E[\xi] = \sqrt{\hat g} \(\hat g^{\mu\nu} \Lie_\xi \hat u^\al_{\mu\nu} - \xi^\al \hat R\)\> ds_\al
\fl{NoetherCurrents}$$
for the {\it Noether currents} of the field $g$ and the reference $\hat g$, while
$$
\De[\xi] = \sqrt{\hat g} \hat g^{\mu\nu}  \xi^\al (u^\be_{\mu\nu}- \hat u^\be_{\mu\nu})\> ds_{\al\be}
\fl{BoundaryCorrection}$$
denotes the {\it boundary correction} due to the augmented variational principles.

Here ``covariant'' refers to the fact that these quantities are scalars. 
However, they have to be interpreted as the mass and momentum
{\it measured by an observer} which has been fixed at the beginning.
It does not imply that mass or momentum do not depend on the observer; different observers will see different lapse and shift, different 
boundaries $\del S_t$ and different symmetry generators.
As discussed in Appendix \AppA\ one cannot expect to obtain anything better in GR.

These expressions are derived in \ref{Augmented}, to which we refer for further motivations; we wish to mention that 
these relative conserved quantities have proven to produce the expected results in many situations both standard (Schwarzschild \ref{BY}, Kerr-Newman \ref{Wald}) and non-standard (BTZ \ref{BTZ}, Kerr-AdS \ref{Kerr} and \ref{Kerr2}, Taub-Bolt \ref{Taub} and \ref{Taub2}).

Here we want to perform ADM splitting of these strong covariant conservation laws in order to investigate which extra conditions, if any, are needed to obtain standard ADM quantities.

We stress that obtaining a pseudotensor in this way is much more meaningful than what is usually done in the literature:
at least one can trace explicitly which hypotheses are required in order the expression to hold, which in turn means 
that one can keep under control which classes of preferred observers come with the pseudotensor.
We believe that these are the minimal requirements to be met in order to be allowed to use coherently non--covariant expressions in GR
at a fundamental level.

\NewSection{ADM Quantities}

Let us hereafter perform  the ADM decomposition of the augmented conserved quantities defined above.
As a first attempt we prefer to restrict to the case in which both the dynamical metric $g$ and the reference metric $\hat g$ define the same normal vector $\vec n$
at the boundary $\del S_t$ where conserved quantities are integrated.
This means the two metrics define the same lapse and shift on $\del S_t$ while the induced metrics $\ga$ and $\hat \ga$ could be different.
Notice that this condition is much less restrictive than the usual matching conditions found in literature, which impose the same metrics  on $\del S_t$.
This is a technical hypothesis and there is no fundamental reason to restrict to these cases as we here do for convenience.
More general situations may require even weaker hypotheses; further investigations will be devoted to this more general case.

We shall first establish some building blocks which will be used later to discuss the ADM splitting of the covariant conserved quantities \ShowLabel{CQ} defined in the previous Section.

Besides the foliation and the dynamical fields $g=(\ga, N, \vec N)$ we have (at the boundary)  reference fields $\hat g=(\hat \ga, N,{\vec N})$.
Let us start from the bulk parts $E[\vec n]$  and $E[\vec N]$. The expression \ShowLabel{NoetherCurrents} can be simplified to:
$$
\eqalign{
E[\xi]=&\sqrt{g} (   \eta^{ab} \Lie_\xi \ubom^c{}_{ab}  -   \eta^{cb} \Lie_\xi \ubom^e{}_{be} -\obxi^c R)\ube^0_c\>ds_0=\cr
=&N^{-1}\sqrt{g} (   \eta^{ab} \Lie_\xi \ubom^0{}_{ab}  -   \eta^{0b} \Lie_\xi \ubom^e{}_{be} -\obxi^0 R)\>dv=\cr
=&\sqrt{\ga} (     \ga^{AB} \Lie_\xi \ubom^0{}_{AB}  + \Lie_\xi \ubom^A{}_{0A}  -\obxi^0 R)\>dv\cr
}
\fl{NoetherCurrent}$$ 
where we set $d v=ds_0$ for the (coordinate) volume element on $S$.

Now setting $\xi=\vec n$ one gets:
$$
E[\vec n]=  \(K_{AB}K^{AB} -{}^3R - K^2 \)\>dv +d(\sqrt{\ga}  N^{-1} D^AN \>dv_A)
\fl{MassCurrent}$$
while for $\xi=\vec N$ one gets
$$
E[\vec N]=-2\sqrt{\ga}  \(D^AK_{CA} -D_C K \)N^C dv+d\(\( 2K_{C}\^A N^C+ N^{-1} \de_0 N^A   \) \>dv_A\)
\fl{MomentumCurrent}$$ 
Here we set $dv_A=e_A\ip dv$ and $d$ denotes formal divergence (or, equivalently, formal covariant divergence; see Appendix \AppB) which is defined to correspond to on-shell exterior differential on $S$.

The same expressions with suitable hat--objects hold for the reference currents $\hat E[{\vec n}]$  and $\hat E[{\vec N}]$.
Notice that these last currents would be associated to the vector fields $\hat{\vec n}$ and $\hat{\vec N}$, respectively, 
which in general have no {\it a priori} reason not to differ from the ones $\vec n$ and $\vec N$ computed out of the dynamical fields. 
This is why we imposed the boundary conditions $\vec n\equiv \hat{\vec n}$ and $\vec N \equiv \hat{\vec N}$.
If one wished to be completely general and two different normal vectors were to be considered, as it can be done, the reference normal unit vector $\hat{\vec n}$ would have to be 
decomposed along the frame $e_a$ associated to the dynamical metric and the corresponding currents $\hat E[\hat{\vec n}]$, $\hat E[\hat{\vec N}]$ expressed as a linear combination of the currents 
$\hat E[{\vec n}]$, $\hat E[{\vec N}]$.
This is of course feasible, but tedious and it does not seem to be really necessary for our later applications;
In fact no generality seems to be lost, at least at the classical level, since at the end the differences vanish along constraints.

\CNote{
Let us check here these splittings.
$$
\eqalign{
E[\vec n]=&\sqrt{\ga} (   \ga^{AB} \Lie_\xi \ubom^0{}_{AB}   + \Lie_\xi \ubom^A{}_{0A}  - R)\>dv=\cr
=&\sqrt{\ga} N^{-1}\Big(  \red{\de_0  K}-    \de_0\ga^{AB}K_{AB}   -\uline{D_A D^A N}    
\oplus
\red{ \de_0  K}-  N^{-1}D_A N  D^A  N \oplus \cr
&   -\red{2\de_0K} +\uline{2 D_A D^AN}    -{}^3R \> N- NK^2-NK_{AB} K^{AB}\Big)\>dv=\cr
=&\sqrt{\ga} \Big(     \uline{2K_{AB}K^{AB}}  -  N^{-2}D_A N  D^A  N + N^{-1}D_A D^AN    -{}^3R - K^2-\uline{K_{AB} K^{AB}}\Big)\>dv=\cr
=&\sqrt{\ga} \Big(    K_{AB}K^{AB}  -  \uline{N^{-2}D_A N  D^A  N} + \uline{N^{-1}D_A D^AN }   -{}^3R - K^2\Big)\>dv=\cr
=&\sqrt{\ga}   \(K_{AB}K^{AB} -{}^3R - K^2 \)\>dv +d\(\sqrt{\ga}  N^{-1} D^AN \>dv_A\)\cr
}
\fn$$
For the shift vector one has:
$$
\eqalign{
E[\vec N]=&\sqrt{\ga} (   \ga^{AB} \Lie_\xi \ubom^0{}_{AB}  + \Lie_\xi \ubom^A{}_{0A} )\>dv=\cr
=&\sqrt{\ga} ( \uline{\Lie_{\vec N}K}  +D^A\( 2K_{CA} N^C\)- 2D^AK_{CA} N^C -\red{ N^{-1} K N^C D_C N}+\cr
&+\uline{\Lie_{\vec N}K} +  D_A \( N^{-1} \de_0 N^A  \) + \red{N^{-1}  K N^C D_C N })\>dv=\cr
=&-2\sqrt{\ga}  \(D^AK_{CA} -D_C K \)N^C dv+d\(\sqrt{\ga}( 2K_{C}\^A N^C+ N^{-1} \de_0 N^A   ) \>dv_A\)\cr
}
\fn$$ 

}

The bulk terms in \ShowLabel{MassCurrent} ---namely, $H:= K_{AB}K^{AB} -{}^3R - K^2 $--- and in \ShowLabel{MomentumCurrent}  ---namely, $H_C:=-2(D^AK_{CA} -D_C K )$--- vanish on-shell (they are in fact the Hamiltonian constraints)
while the surface terms are the ADM decomposition of Komar superpotential.
This decomposition of Noether currents into a bulk term vanishing on-shell and a boundary term is the canonical splitting of Noether current (see \ref{Augmented}, \ref{Book}).

\CNote{
The $(00)$ component of fields equations is
$$
\eqalign{
\ubR_{00} -\frac[1/2] R \eta_{00}=&
 - \red{N^{-1}  \( \de_0K   -D_A D^A N \)} - \uline{K^{EA}K_{E A}} +\cr
 &+ \red{\frac[2/2N] \( \de_0K -D_A D^AN    \)}+\frac[1/2]\({}^3R  + K^2+\uline{K_{AB} K^{AB}}\)=\cr
 =&\frac[1/2]\({}^3R  + K^2-{K_{AB} K^{AB}}\)=0
}
\fn$$
and it shows that $H=0$ on-shell. 
The $(0A)$ components
$$
\eqalign{
\ubR_{0A} =&D_{C} \(K_{A}\^C - \de_{A}^C K\)= D_{C} K_{A}\^C -   D_{A}K=0
}
\fn$$
show that $H_A=0$ on-shell. 

The Komar superpotential reads as 
$$
\eqalign{
U[\xi]=& \sqrt{g} \na^\be\xi^\al ds_{\al\be}=
-2\sqrt{g} \na^{[0}\obxi^{b]}  \ube_0^0 \ube_b^i ds_{0i}=
-2\sqrt{\ga} \na^{[0}\obxi^{B]}   \al_B^i dv_{i}=\cr
=&\sqrt{\ga} \(\na^{B}\obxi^{0}  - \na^{0}\obxi^{B} \)  dv_{B}
=\sqrt{\ga} \(\ga^{BA}\na_{A}\obxi^{0}  + \na_{0}\obxi^{B} \)  dv_{B}
}
\fn$$
Setting $\xi=\vec n$ one gets
$$
U[\vec n]=\sqrt{\ga} \(\na^{B}\obxi^{0}  - \na^{0}\obxi^{B} \)  dv_{B}
=\sqrt{\ga}  N^{-1} D^BN   dv_{B}
\fl{KomarMass}$$
as above.

Setting $\xi=\vec N$ one gets
$$
\eqalign{
U[\vec N]=&\sqrt{\ga} \(  N ^{-1} d_0 N^B + \uline{K\^B{}_CN^C }  +\uline{N^CK\^B{}_{C} } \)  dv_{B} =\cr
=& \sqrt{\ga}  \( 2K\^B{}_{C} N^C+N ^{-1} d_0 N^B \)  dv_{B}
}
\fl{KomarMomentum}$$
Equations \ShowLabel{KomarMass} and \ShowLabel{KomarMomentum}
show that the surface terms in \ShowLabel{MassCurrent} and \ShowLabel{MomentumCurrent} are just the ADM splitting of the relevant Komar superpotentials.
}

Let us also remark that, in view of  
 \ShowLabel{nablaxi}, \ShowLabel{LiemOm},  and  \ShowLabel{NoetherCurrent}
we have
$$
E[\vec m]= N E[\vec n]+ \sqrt{\ga}    N^{-1}D_A N D^A N \>dv + d\(\sqrt{\ga}     D^A N \>dv_A \)
\fl{mCurrent}$$

\CNote{
In fact, we have:
$$
\eqalign{
E[\vec m]-N E[\vec n]=&
\sqrt{\ga} (     \ga^{AB} (\na_A\na_B N \obxi^0 + \frame{$\na_B N \na_A\obxi^0$} + \frame{$\na_A N \na_B\obxi^0$})  \>dv+\cr
&+\sqrt{\ga} (      \frame{$\na_A\na_0 N \obxi^A$} +  \na_0 N \na_A\obxi^A +  \na_A N \na_0\obxi^A  )\>dv=\cr
=&
\sqrt{\ga} (     \ga^{AB} (D_A  D_B N -\red{N^{-1}K_{BA} \de_0 N} ) + \red{N^{-1}\de_0 N K }+  N^{-1}D_A N D^A N  )\>dv=\cr
=&
d\(\sqrt{\ga}     D^A N \>dv_A \)+ \sqrt{\ga}    N^{-1}D_A N D^A N \>dv\cr
}
\fn$$
}

For the boundary correction \ShowLabel{BoundaryCorrection} let us first compute:
$$
\eqalign{
\sqrt{\hat g} \hat g^{\mu\nu}  \xi^\al u^\be_{\mu\nu}\> ds_{\al\be}=&
 N \sqrt{\hat\ga} \hat g^{\mu\nu} \(\obxi^{c}\ube_c^0 u^{i}_{\mu\nu}- \obxi^{c}\ube_c^i u^{0}_{\mu\nu}\)\>  ds_{0i}=\cr
=& \sqrt{\hat \ga} \(\obxi^{0} u^{i}+ \obxi^0 N^i u^{0}- N\obxi^{i} u^{0}\)\>  dv_{i}=\cr
=& \sqrt{\hat \ga} \(\obxi^{0} \(u^{i}+  N^i u^{0} \)- \obxi^{i} N u^{0}\)\al^A_i\>  dv_{A}\cr
}
\fn$$
where we set $u^{\la}:=\hat g^{\mu\nu} u^{\la}_{\mu\nu}$. 
Notice that the computation above is carried out at the boundary $\del S_t$ where the frames induced by the two metrics actually coincide under our hypotheses.
Of course one also has
$$
\eqalign{
\sqrt{\hat g} \hat g^{\mu\nu}  \xi^\al \hat u^\be_{\mu\nu}\> ds_{\al\be}=&
  \sqrt{\hat \ga} \(\obxi^{0} \(\hat u^{i}+  N^i \hat u^{0} \)- \obxi^{i} N \hat u^{0}\)\al^A_i\>  dv_{A}\cr
}
\fl{hatgxiu}$$
where we set $\hat u^{\la}=\hat g^{\mu\nu} \hat u^{\la}_{\mu\nu}$. 

%%%QUESTA C'ERA COME \Note%%%%%%%%%%%%%%%%%%%%
\CNote{
Since the two metrics induce the same frame at the boundary $\del S_t$ the matrices $\ube_a^\mu$ and $\obe^a_\mu$ representing (co)frames do not
depend on the metric and all differences between the metrics are contained in the frame metrics $\eta_{ab}$ and $\hat\eta_{ab}$. 

Let us then compute:
$$
\eqalign{
u^{0}=& \hat g^{\mu\nu} u^{0}_{\mu\nu}=\hat \eta^{ab}{\ube}_a^\mu{\ube}_b^\nu\(\Ga^0_{\mu\nu} -\de^0_\mu \Ga^\al_{\nu\be} \ube^\be_c \obe^c_\al\)=\cr
=& \hat\eta^{ab}\Ga^0_{\mu\nu} \ube_a^\mu\ube_b^\nu- \hat\eta^{ab}\ube_a^0\ube_b^\nu\Ga^\al_{\nu\be} \ube^\be_c \obe^c_\al=\cr
=& \hat\eta^{ab}\(\ube^0_c \ubom^c{}_{ab} -d_\la \ube_a^0\ube_b^\la\)- \hat\eta^{ab}\ube_a^0 \obe^c_\al \(\ube^\al_d \ubom^d{}_{bc} - d_\la \ube^\al_b \ube^\la_c\)  =\cr
=& \hat\eta^{ab}\ube^0_c \ubom^c{}_{ab} 
- \hat\eta^{ab}d_\la \ube_a^0\ube_b^\la
- \hat\eta^{ab}\ube_a^0  \ubom^c{}_{bc} 
+ \hat\eta^{ab}\ube_a^0  d_\la \ube^\la_b   =\cr
=& \hat\ga^{BC}\ube^0_0 \ubom^0{}_{BC} 
+ d_\la \ube_0^0\ube_0^\la
+\ube_0^0  \ubom^B{}_{0B} 
-\ube_0^0  d_\la \ube^\la_0   =\cr
=& \hat\ga^{BC}N^{-1} \ubom^0{}_{BC} 
+N^{-1}  \ubom^B{}_{0B} 
+ d_0 N^{-1}\ube_0^0+ d_l N^{-1}\ube_0^l
-N^{-1}  d_0 \ube^0_0 -N^{-1}  d_l \ube^l_0   =\cr
=& N^{-1}\(\hat\ga^{BC} \ubom^0{}_{BC} +  \ubom^B{}_{0B} \)
+ \red{d_0 N^{-1}N^{-1}}- d_l N^{-1} N^{-1}N^l
-\red{N^{-1}  d_0 N^{-1}} +N^{-1}  d_l \(N^{-1}N^l\)  =\cr
=& N^{-1}\(\hat\ga^{BC} \ubom^0{}_{BC} +  \ubom^B{}_{0B} \)
- \red{d_l N^{-1} N^{-1}N^l}
 +\red{N^{-1}  d_l N^{-1}N^l}+N^{-2}  d_l N^l  =\cr
=& N^{-1}\(\hat\ga^{BC} + \ga^{BC} \)K_{BC}
+N^{-2} d_l N^l  
%=2N^{-1} K +N^{-2}  d_l N^l  \cr
}
\fn$$

Accordingly, one has also
$$
\hat u^{0}= 2N^{-1} \hat\ga^{BC} \hat K_{BC}
+N^{-2} d_l N^l  
=2N^{-1} \hat K +N^{-2}  d_l N^l 
\fl{u0}
$$

%\endNote\EndBeginSection\Note
Let us also compute 
$$
\eqalign{
u^{i}=&\hat g^{\mu\nu} u^{i}_{\mu\nu}= \hat\eta^{ab}  \( \ube_a^\mu \ube_b^\nu\Ga^{i}_{\mu\nu}- \ube_a^i \obe^c_\al\ube_b^\nu  \Ga^\al_{\nu\be} \ube^\be_c\)=\cr
=& \hat\eta^{ab}  \( \ube_c^i \ubom^c{}_{ab} -d_\la \ube^i_a \ube^\la_b\)-  \hat\eta^{ab}  \ube_a^i \obe^c_\al \(\ube^\al_d\ubom^d{}_{bc} -d_\la \ube^\al_b\ube^\la_c\)=\cr
=&   \hat\eta^{ab}\ube_c^i \ubom^c{}_{ab} 
- \hat\eta^{ab}d_\la \ube^i_a \ube^\la_b
-  \hat\eta^{ab}  \ube_a^i \ubom^c{}_{bc} 
+ \hat \eta^{ab}  \ube_a^i d_\la \ube^\la_b=\cr
%=&ok\cr
=&  -\ube_c^i \ubom^c{}_{00}  + \hat\ga^{BC}\ube_c^i \ubom^c{}_{BC} 
+d_\la \ube^i_0 \ube^\la_0  - \redframe{$\hat\ga^{BC}d_k \al^i_B \al^k_C$}
+ \ube_0^i \ubom^C{}_{0C} -  \hat\ga^{BC}  \al_B^i \ubom^c{}_{Cc} 
-  \ube_0^i d_\la \ube^\la_0+  \greenframe{$\hat\ga^{BC}  \al_B^i d_k \al^k_C$}=\cr
=&  - \al^i_C\ubom^C{}_{00}  + \uline{\hat\ga^{BC}\ube_0^i \ubom^0{}_{BC}}  + \uuline{\hat\ga^{BC} \al^i_A\ubom^A{}_{BC}} 
+d_\la \ube^i_0 \ube^\la_0  
+ \uline{\ube_0^i \ubom^C{}_{0C}} -  \hat\ga^{BC}  \al^i_B\ubom^0{}_{C0} - \uuline{ \hat\ga^{BC} \al^i_B \ubom^A{}_{CA} }
-  \ube_0^i d_\la \ube^\la_0+\cr
& - \redframe{$\hat\ga^{BC} \al^i_{BC} $} -  \greenframe{$\hat\ga^{BC}  \al_B^i  \al^k_{C} d_D \al^D_k$}=\cr
=&   -N^{-1} N^i \(\hat\ga^{BC} \ubom^0{}_{BC} +  \ubom^C{}_{0C} \)
 + \hat\ga^{BC}\al^i_A\( {}^3\Ga^A_{BC} -  \de^A_B\>  {}^3\Ga^D_{CD} \)
 - \al^i_A\ubom^A{}_{00}  -  \hat\ga^{BC} \al^i_B \ubom^0{}_{C0} 
+d_\la \ube^i_0 \ube^\la_0  
-  \ube_0^i d_\la \ube^\la_0+\cr
& - \redframe{$\hat\ga^{BC} \al^i_{BC} $} -  \greenframe{$\hat\ga^{BC}  \al_B^i  \al^k_{C} d_D \al^D_k$}=\cr
=&   -N^{-1} N^i \(\hat \ga^{BC} + \ga^{BC}\)K_{BC}
 + \hat \ga^{BC}\al^i_A\> \({}^3u^A{}_{BC}\)
  - 2N^{-1}  D^i N 
+d_0 \ube^i_0 \ube^0_0  +d_l \ube^i_0 \ube^l_0  
-  \ube_0^i d_0 \ube^0_0-  \ube_0^i d_l \ube^l_0+\cr
& - \redframe{$\hat\ga^{BC} \al^i_{BC} $} -  \greenframe{$\hat\ga^{BC}  \al_B^i  \al^k_{C} d_D \al^D_k$}=\cr
}$$
$$
\eqalign{
u^{i}
=&   -N^{-1} N^i  \(\hat \ga^{BC} + \ga^{BC}\)K_{BC}
  + \hat \ga^{BC}\al^i_A\> \({}^3u^A{}_{BC}\)
    - 2N^{-1}  D^i N 
+\cr
&-\uline{N^{-1}d_0\(N^{-1} N^i\) }  +\uuline{d_l \(N^{-1}N^i\)N^{-1} N^l}+ \uline{N^{-1} N^i d_0 N^{-1}}+\cr
& -  \uuline{N^{-1} N^i  d_l \( N^{-1} N^l\)}- \redframe{$\hat\ga^{BC} \al^i_{BC} $} -  \greenframe{$\hat\ga^{BC}  \al_B^i  \al^k_{C} d_D \al^D_k$}=\cr
=&   -N^{-1} N^i  \(\hat \ga^{BC} + \ga^{BC}\)K_{BC}
  + \hat \ga^{BC}\al^i_A\> \({}^3u^A{}_{BC}\)
    - 2N^{-1}  D^i N 
+\cr
&-{N^{-2} d_0N^i }  +{ N^{-2}d_l N^i N^l}- {N^{-2} N^i   d_l N^l}
 - \redframe{$\hat\ga^{BC} \al^i_{BC} $} -  \greenframe{$\hat\ga^{BC}  \al_B^i  \al^k_{C} d_D \al^D_k$}\cr
}
\fn$$
and for later convenience
$$
\eqalign{
u^{i} + N^i u^0=&
-\red{N^{-1} N^i  \(\hat \ga^{BC} + \ga^{BC}\)K_{BC}}
  + \hat \ga^{BC}\al^i_A\> \({}^3u^A{}_{BC}\)
    - 2N^{-1}  D^i N 
+\cr
&-{N^{-2} d_0N^i }  +{ N^{-2}d_l N^i N^l}- \green{N^{-2} N^i   d_l N^l}
 - \redframe{$\hat\ga^{BC} \al^i_{BC} $} -  \greenframe{$\hat\ga^{BC}  \al_B^i  \al^k_{C} d_D \al^D_k$}
  \oplus \cr
&+ \red{N^{-1}N^i\(\hat\ga^{BC} + \ga^{BC} \)K_{BC}}
+\green{N^{-2} N^id_l N^l }=\cr
=&
  \hat \ga^{BC}\al^i_A\> \({}^3u^A{}_{BC}\)
    - 2N^{-1}  D^i N 
    -{N^{-2} d_0N^i }  +{ N^{-2}d_l N^i N^l}
+\cr
&
 - \redframe{$\hat\ga^{BC} \al^i_{BC} $} -  \greenframe{$\hat\ga^{BC}  \al_B^i  \al^k_{C} d_D \al^D_k$}
\cr
}
\fl{deltaPrime}$$

Accordingly, one has also
$$
\hat u^{i} + N^i \hat u^0=\al^i_A\> \({}^3 \hat u^A\)
    - 2N^{-1}  \hat D^i N 
    -{N^{-2} d_0N^i }  +{ N^{-2}d_l N^i N^l}
 - \redframe{$\hat\ga^{BC} \al^i_{BC} $} -  \greenframe{$\hat\ga^{BC}  \al_B^i  \al^k_{C} d_D \al^D_k$}
\fl{u+Nu}$$
}
%\endNote
%\EndSection

Setting $\xi=\vec n$ we obtain
$$
\eqalign{
\De[\vec n]=  \sqrt{\hat \ga}   \hat \ga^{BC}\> \({}^3u^A_{BC}  -{}^3\hat u^A_{BC}\)\>  dv_{A}
}
\fn$$
%%%%%%%%%%%%%%%%%

Setting $\xi=\vec N$ one gets instead
$$
\eqalign{
\De[\vec N]=& \sqrt{\hat \ga}    \(2 \hat K 
-\(\hat\ga^{BC} + \ga^{BC} \)K_{BC} \)N^{A}\>  dv_{A}\cr
}
\fn$$

Of course, being $\De[\xi]$ linear in $\obxi^a$, one has $\De[\vec m]=N\De[\vec n]$.

\CNote{
Let us here compute the above expressions.

Setting $\xi=\vec n$, we obtain
$$
\eqalign{
\De[\vec n]= & \sqrt{\hat \ga} \( u^{i}+  N^i u^{0} -  \hat u^{i}-  N^i \hat u^{0} \)\>  dv_{i}=\cr
=& \sqrt{\hat \ga} \Big(   \hat \ga^{BC}\al^i_A\> \({}^3u^A{}_{BC}\)
    - \red{2N^{-1}  D^i N }
    -\green{N^{-2} d_0N^i }  +\blue{ N^{-2}d_l N^i N^l}+\cr
%+\cr&
 &- \redframe{$\hat\ga^{BC} \al^i_{BC} $} -  \greenframe{$\hat\ga^{BC}  \al_B^i  \al^k_{C} d_D \al^D_k$}
  + \redframe{$\hat\ga^{BC} \al^i_{BC} $} +  \greenframe{$\hat\ga^{BC}  \al_B^i  \al^k_{C} d_D \al^D_k$}
+ \cr
&- \al^i_A\> \({}^3 \hat u^A\)
    + \red{2N^{-1}  D^i N} 
    +\green{N^{-2} d_0N^i }  -\blue{ N^{-2}d_l N^i N^l}
 \Big)\>  dv_{i}=
\cr
=& \sqrt{\hat \ga}   \hat \ga^{BC}\> \({}^3u^A_{BC}  -{}^3\hat u^A_{BC}\)\>  dv_{A}
%=\cr=& \sqrt{\ga} \(  \ga^{jk}\> \({}^3u^i{}_{jk}\)- 2N^{-1}  D^i N -N^{-2}d_0 N^i   +N^{-2}d_l N^i N^l \)\>  dv_{i}
}
\fn$$

For $\xi=\vec N$
$$
\eqalign{
\De[\vec N]=&-\sqrt{\hat \ga}   N \(u^{0}-\hat u^0\)N^{i}\>  dv_{i}=\cr
=& \sqrt{\hat \ga}   N \(2N^{-1} \hat K +\red{N^{-2}  d_l N^l }
-N^{-1}\(\hat\ga^{BC} + \ga^{BC} \)K_{BC}
-\red{N^{-2} d_l N^l} \)N^{i}\>  dv_{i}\cr
}
\fn$$
}

Now we are ready to compute the conserved quantity $\calQ[\xi]$ given by \ShowLabel{CQ}.

For $\xi=\vec n$ one gets
$$
\eqalign{
\calQ[\vec n]=&\Frac[1/2\ka]\int_{S_t} \sqrt{\ga}  H \>dv -\Frac[1/2\ka]\int_{S_t} \sqrt{\hat \ga}  \hat H \>dv +\cr
&+{\Frac[1/2\ka]\int_{\del S_t}  \sqrt{\ga}  N^{-1} D^AN \>dv_A}-{\Frac[1/2\ka]\int_{\del S_t}  \sqrt{\hat \ga}  N^{-1} \hat D^AN \>dv_A}+\cr
&+\Frac[1/2\ka]\int_{\del S_t} \sqrt{\hat \ga}   \hat \ga^{BC}\> \({}^3u^A_{BC}  -{}^3\hat u^A_{BC}\)\>  dv_{A}
}
\fn$$
The bulk terms vanish on-shell, while we can set
$$
C[\vec n]= \( \(\sqrt{\ga}N^{-1} D^AN    -  \sqrt{\hat \ga}N^{-1} \hat D^AN  \) 
+ \sqrt{\hat \ga}   \hat \ga^{BC}\> \({}^3u^A_{BC}  -{}^3\hat u^A_{BC}\)\)\>  dv_{A}
\fl{Qn}$$
for the {\it boundary current},
so that one has $\calQ[\vec n]=(2\ka)^{-1}\int_{\del S_t} C[\vec n]$.

For $\xi=\vec N$ one gets thence
$$
\eqalign{
\calQ[\vec N]=& \Frac[1/2\ka]\int_{S_t} \sqrt{\ga} H_C N^C \>dv -  \Frac[1/2\ka]\int_{S_t} \sqrt{\hat \ga} \hat H_C N^C \>dv+\cr
&+\Frac[1/2\ka]\int_{\del S_t}\sqrt{\ga}\( 2K_{C}\^A N^C+{N^{-1} \de_0 N^A}   \) \>dv_A
-\Frac[1/2\ka]\int_{\del S_t}\sqrt{\hat \ga}\( 2\hat K_{C}\^A N^C+{N^{-1} \de_0 N^A}   \) \>dv_A+\cr
&+\Frac[1/2\ka]\int_{\del S_t} \sqrt{\hat \ga}    \(2\hat K 
-\(\hat\ga^{BC} + \ga^{BC} \)K_{BC} \)N^{A}\>  dv_{A}
}
\fn$$
The bulk terms vanish on-shell, while we have
$$
\eqalign{
C[\vec N]=& \sqrt{\ga}\( 2K_{C}\^A N^C+{N^{-1} \de_0 N^A}   \) \>dv_A
-\sqrt{\hat \ga}\( 2\hat K_{C}\^A N^C+{N^{-1} \de_0 N^A}   \) \>dv_A+\cr
&+ \sqrt{\hat \ga}    \(2 \hat K 
-\(\hat\ga^{BC} + \ga^{BC} \)K_{BC} \)N^{A}\>  dv_{A}
}
\fl{QN}$$
for the boundary current,
so that one has $\calQ[\vec N]=(2\ka)^{-1}\int_{\del S_t} C[\vec N]$.

In view of \ShowLabel{mCurrent} we can also compute the conserved quantity
$$
\eqalign{
\calQ[\vec m]=&\Frac[1/2\ka]\int_{S_t} \sqrt{\ga}  N H \>dv -\Frac[1/2\ka]\int_{S_t} \sqrt{\hat \ga} N \hat H \>dv +\cr
&+{\Frac[1/2\ka]\int_{S_t} { \( \sqrt{\ga}  N^{-1}D_A N D^A N- \sqrt{\hat \ga}  N^{-1}\hat D_A N \hat D^A N\)} \>dv}+\cr
&+{\Frac[1/\ka]\int_{\del S_t}  	{\(\sqrt{\ga}  D^AN -\sqrt{\hat \ga}  \hat D^AN  \)} \>dv_A}
+\Frac[1/2\ka]\int_{\del S_t} \sqrt{\hat \ga}  N \hat \ga^{BC}\> \({}^3u^A_{BC}  -{}^3\hat u^A_{BC}\)\>  dv_{A}\cr
}
\fn$$

\CNote{
In fact, in view of \ShowLabel{MassCurrent}, \ShowLabel{mCurrent} and \ShowLabel{deltaPrime}, on-shell  we have:
$$
\eqalign{
2\ka\calQ[\vec m]=&{\int_{\del S_t}  \uline{\sqrt{\ga}   D^AN}\>dv_A}-{\int_{\del S_t}  	\uuline{\sqrt{\hat \ga}  D^AN} \>dv_A}+\cr
&+\int_{\del S_t} \sqrt{\hat \ga}  N \hat \ga^{BC}\> \({}^3u^A_{BC}  -{}^3\hat u^A_{BC}\)\>  dv_{A}+\cr
+&\int_{\del S_t}\uline{\sqrt{\ga}     D^A N} dv_A + \int_{S_t}\sqrt{\ga}    N^{-1}D_A N D^A N \>dv+\cr
-&\int_{\del S_t}\uuline{\sqrt{\hat \ga}   \hat  D^A N} dv_A - \int_{S_t}\sqrt{\hat \ga}    N^{-1}\hat D_A N \hat D^A N \>dv=\cr
=&{\int_{S_t} \( \sqrt{\ga}   N^{-1}D_A N D^A N -\sqrt{\hat \ga}   N^{-1}\hat D_A N \hat D^A N  \) \>dv}+\cr
&+2{\int_{\del S_t} \({\sqrt{\ga}   D^AN}- {\sqrt{\hat \ga}  \hat D^AN} \)\>dv_A}+\int_{\del S_t} \sqrt{\hat \ga}  N \hat \ga^{BC}\> \({}^3u^A_{BC}  -{}^3\hat u^A_{BC}\)\>  dv_{A}\cr
}
\fn$$
}

On shell  we can set
$$
\eqalign{
B[\vec m]=&  \( \sqrt{ \ga}  N^{-1}D_A N D^A N-  \sqrt{ \hat \ga}  N^{-1}\hat D_A N \hat D^A N\) \>dv\cr
C[\vec m]=& 2\(\sqrt{\ga}   D^AN - \sqrt{\hat \ga}   \hat D^AN\) \>dv_A
%- \sqrt{\hat \ga}  \(2D^AN-N^{-1}D_A N D^A N\) \>dv_A+\cr&
+ \sqrt{\hat \ga}  N \hat \ga^{BC}\> \({}^3u^A_{BC}  -{}^3\hat u^A_{BC}\)\>  dv_{A}\cr
}
\fl{Qm}$$
so that one has $\calQ[\vec m]=(2\ka)^{-1} \(\int_{S_t} B[\vec m]+ \int_{\del S_t} C[\vec m]\)$.

At this point one can require some extra condition in order to further simplify these expressions. 
One usually requires the metric and the reference metric to match at the boundary, so that, still at the boundary, one also has $\ga=\hat\ga$,
while of course the derivatives of the metrics (and the relative connections) are not required to match.
Under these further hypotheses one gets:
$$
\eqalign{
&\tilde C[\vec n]= \sqrt{\ga}   \ga^{BC}\({}^3u^A_{BC}  -   {}^3\hat u^A_{BC}\)\>  dv_{A}\cr
&\tilde C[\vec m]= B[\vec m]+ N  \sqrt{\ga}  \ga^{BC} \({}^3u^A_{BC}  -{}^3\hat u^A_{BC}\)\>  dv_{A}\cr
&\tilde C[\vec N]= 2\sqrt{\ga}\(\( K_{C}\^A -K\de^A_C\)
- \( \hat K_{C}\^A   -\hat K\de^A_C \) \)N^C\>dv_A\cr
}
\fl{AugmentedMatchedCQ}$$

If $M\simeq \R^4$ and  the reference metric happens to be Minkowski, if we use Cartesian coordinates for the reference metric, 
we have $\hat u^A_{BC}=0$, $\hat K_{AB}=0$
on the boundary and \ShowLabel{AugmentedMatchedCQ} give
$$
\eqalign{
&\calC[\vec n]= \Frac[1/2\ka] \int_{\del S_t} \sqrt{\ga}   \ga^{BC}\>{}^3u^A_{BC}  \>  dv_{A}=: M^n_{ADM}\cr
&\calC[\vec m]=\Frac[1/2\ka]\int_{S_t} B[\vec m]+   M^m_{ADM}
\qquad \qquad
		  M^m_{ADM}:=  \Frac[1/2\ka]\int_{\del S_t} N\sqrt{\ga}   \ga^{BC}\>{}^3u^A_{BC}  \>  dv_{A}\cr
&\calC[\vec N]= \Frac[1/\ka]\int_{\del S_t} \sqrt{\ga}\( K_{C}\^A -K\de^A_C\)N^C \>dv_A=:P_{ADM}\cr
}
\fl{ADMcq}$$
i.e.~the standard ADM conserved quantities, though obtained from augmented covariant conservation laws.

Let us stress that our hypotheses are in any case weaker than the usual asymptotical flatness. 
Here we just require that the metric $g_{\mu\nu}$ goes to the reference $\hat g_{\mu\nu}$ at space infinity, no matter how fast it goes.
%,while asymptotical flatness requires a fall off at least as fast as $r^{-1}$.
\def\sp{{\buildrel{+} \over s}}
\def\sm{{\buildrel{-} \over s}}
\def\tm{{\buildrel{-} \over t}}
\def\km{{\buildrel{-} \over k}}

In fact in the literature there are a number of slightly different notions of {\it asymptotically flat} %(or asymptotically Minkowskian) 
spacetimes. All of them 
require the dynamical metric  $g$  to match a flat reference metric $\hat g$ on the boundary $\del S_t$. 
However, they often differ on how fast this match is obtained. One defines a quantity $r$ which approaches $r\arr \infty$ on $\del S_t$ and
the components of $g-\hat g$ (and their derivatives) are required to be infinitesimal of certain order as $r\arr \infty$.
One needs to control also the derivatives since, of course, a function can be infinitesimal but with a derivative which is not.
The order of infinitesimal required strongly depends on what exactly asymptotically flatness is required for, in particular which quantities 
one wants to control.

Here we shall consider two different sets of hypotheses.
First, we shall say that a spacetime is {\it asymptotically flat} (AF) when $g-\hat g$ scale as $r^{-1}$, that first derivatives scale as $r^{-2}$, and so on.

We shall also consider {\it BoM-asymptotically flat} (BoMAF) spacetimes (see \ref{Murchadha}, \ref{Szaba}) that, to the best of our knowedge, is the prescriptions which are best suited for
conservation laws and initial value problem. Field fall-off is considered to ensure that spacetime has no even supertranslations and still there are enough initial conditions to
have a well--posed initial value problem.

Let $S$ be a space leaf, $x^i$ a global coordinate system and $\ga$ the the metric induced on the space leaf. 
Let also also define $r^2= \de_{ij} x^i x^j$. The spacetime metric $g$  is said to be {\it BoM-asymptotically flat} (BoMAF) if  (for $r\gg 0$) one has
$$
\ga_{ij} (x)= \de_{ij} + \frac[1/r] \sp{}_{ij}(\frac[x/r])+ h_{ij}(x) 
\fl{BM3metric}$$
where the function $\sp_{ij}$ is an even smooth function on $S^2$ (i.e.~$\sp{}_{ij}(x)= \sp{}_{ij}(-x)$) and $h_{ij}$ falls off like $r^{-1-\ep}$ (for some $\ep>0$),
the derivatives $\del_k h_{ij}$ like  $r^{-2-\ep}$ and so on.

\CNote{An even (odd) function is a function $f(x)$ such that
$$
\cases{
&f(x)= f(\la x) \quad \forall \la\in \R^+\cr
&f(x)= f(-x) \qquad (f(x)= -f(-x))\cr
}
\fn$$
Even functions will be denoted by $\sp(x)$ while odd functions will be denoted by $\sm(x)$.
Derivatives of even (odd) functions are easily shown to be odd (even).
}

Moreover, the conjugate momenta $\pi^{ij}(x):= \sqrt{\ga}\( K^{ij}- K \ga^{ij}\)$ are tensor densities such that
$$
\pi^{ij}(x)=\frac[1/r^2] \tm{}^{ij}(\frac[x/r])+ k^{ij}(x) 
\fl{BM3momentum}$$ 
where the function $\tm{}^{ij}$ is an odd smooth function on $S^2$ and $k^{ij}(x)$ falls off like $r^{-2-\ep}$ (for some $\ep>0$),
the derivatives $\del_m k^{ij}$ like  $r^{-3-\ep}$ and so on.
The lapse and shift can be recasted as
$$
N(x)= \hat N+  \km(\frac[x/r]) + n(x)
\qquad
N^i(x)= \hat N^i+ \km{}^i(\frac[x/r]) + n^i(x)
\qquad
(\hat N, \hat N^i \in \R)
\fl{BM3LapseShift}$$
where $\km$ and $\km{}^i$ are odd smooth functions on $S^2$ and $(n, n^i)$ fall off like $r^{-\ep}$ (for some $\ep>0$),
the derivatives $(\del_k n, \del_k n^i)$ like  $r^{-1-\ep}$ and so on.

These conditions \ShowLabel{BM3metric},  \ShowLabel{BM3momentum}, and \ShowLabel{BM3LapseShift}
will be hereafter called {\it BoM-asymptotic flatness}.
Let us stress that BoMAF conditions are stricter than AF conditions as far as the $3$-metric is concerned (in view of \ShowLabel{MetricADMDec}, BoMAF dictates falling off as $r^{-1}$ but with specific parity) 
but they are weaker about lapse and shift (e.g.~BoMAF prescribes $N^i\sim r^{-\ep}$ while AF prescribes $N^i\sim r^{-1}$)

Both these definitions of asymptotically flatness are coordinate dependent; in the literature there are also intrinsic definitions (see, e.g.,\ref{Geroch}).
Of course the intrinsic definition is a better notion in GR and should be preferred with respect to coordinate definitions as the one we adopted here. 
However, we stress that here we are discussing ADM integrals which are obtained by pseudotensors. Precisely, we are obtaining pseudotensors by a (partial) gauge fixing of the coordinate freedom. As usual when one performs a partial gauge fixing, a preferred class of transformations parametrizing the residual gauge freedom
is automatically selected and, in view of this breaking of covariance, physical quantities cannot be expected to be manifestly generally covariant.
In this context we {\it need} to use coordinate dependent notion of asymptotically flatness, since the intrinsic one would not respect the partial gauge fixing we have been doing.

In the following Section we shall see that the same result holds under various set of hypotheses starting from different expressions.

\NewSection{Pseudotensors}

Reference fields are not very popular in literature so that one often tries to avoid them.
Despite it is well--known, also in Newtonian Physics, that absolute energies cannot be endowed with a meaning and that only relative energies are 
well-defined physical quantities, people like absolute prescriptions in GR.  
One way to get rid of the reference metric is to choose the observers for which reference contributions vanish.
To this purpose, let us consider the following quantity
$$
\calQ'[\xi] = \Frac[1/2\ka]\int_{S_t}E[\xi] + \Frac[1/2\ka]\int_{\del S_t}\De'[\xi] 
\fl{PseudoTensor}$$
with
$$
 E[\xi] = \sqrt{g} \(g^{\mu\nu} \Lie_\xi u^\al_{\mu\nu} - \xi^\al R\)\> ds_\al
 \qquad\qquad
\De'[\xi] =\sqrt{g} g^{\mu\nu}  \xi^\al u^\be_{\mu\nu}\> ds_{\al\be}
\fn$$
One can fix the reference background so that  $\hat E[\xi]=0$, %$\hat R^\al{}_{\be\mu\nu}=0$, the foliation so that  $\hat{\Dal} \xi^\al=0$ 
and the coordinates in such a way that one has $\hat u^\la_{\mu\nu}=0$ at the boundary.
For instance one can fix Minkowski metric as a reference and asymptotically Cartesian coordinates in which the reference metric is constant and its Christoffell symbols vanish.
Under these assumptions this is equivalent to augmented conserved quantities.
However, choosing coordinates corresponds to a partial gauge fixing which breaks general covariance. In fact, $u^\be_{\mu\nu}$ 
is not a tensor and hence the boundary correction is not a scalar. The reference field was originally introduced exactly to covariantize
this quantity; see \ref{Cavalese}.

The first term $E[\xi]$ splits as above while the second term $\De'[\xi] $ splits as \ShowLabel{hatgxiu}, though without hats,
The quantity $\calQ'[\xi] $ has then a bulk term which vanishes on-shell, plus a boundary current $C'[\xi]$ which receives a contribution from the
Noether current $ E[\xi] $ and a contribution from the boundary correnction $\De'[\xi] $.
Accordingly, for $\xi=\vec n$ one gets
$$
\calQ'[\vec n]=M^n_{ADM} +\Frac[1/2\ka]\int_{\del S_t} C'[\vec n]
\fl{Qprimen}$$
where we defined the {\it extra current}
$$
C'[\vec n]=- \sqrt{\ga}N^{-1} \(       D^A N 
   + {N^{-1} d_0N^A }  
   -{ N^{-1}N^Bd_B N^A } \)\>  dv_{A}
\fn$$

\CNote{
In fact, in view of equations \ShowLabel{MassCurrent} and \ShowLabel{u+Nu}, one has
$\calQ'[\vec n]=(2\ka)^{-1}\int_{\del S_t} Q'[\vec n]$ and:
$$
\eqalign{
Q'[\vec n] =& \sqrt{\ga} \uline{ N^{-1} D^AN} \>dv_A
+ \sqrt{\ga} \(\> \({}^3 u^A\)
    - \uline{2N^{-1}  D^A N  }
    -{N^{-2} d_0N^A }  +{ N^{-2}N^Bd_B N^A } \)\>  dv_{A}+\cr
&- \sqrt{\ga} \(\frame{$ N^{-2} N^i N^ld_l\al^A_i$}
 + \frame{$\ga^{BC} \al^i_{BC} \al^A_i$} +  \frame{$\ga^{AC}   \al^k_{C} d_D \al^D_k$} \)\>  dv_{A}=\cr
=& 
 \sqrt{\ga} \({}^3 u^A\)dv_{A}
-\sqrt{\ga} \(
     N^{-1}  D^A N 
   + {N^{-2} d_0N^A }  -{ N^{-2}N^Bd_B N^A } \)\>  dv_{A}\cr
}
\fn$$
where the framed terms vanish when one uses adapted parametrizations ($\al^i_A=\de^i_A$).
}

Under AF assumptions, since the metric approaches Minkowski at the boundary, the shift $N^i=g^{0i}\sim r^{-1}$, $N-1\sim  r^{-1}$, and  one has $\sqrt{\ga} dv_A\sim r^2$. 
Accordingly, terms like $\sqrt{\ga} N^{-2}N^Bd_B N^A \sim r^{-1}$ and the corresponding integral vanishes at the boundary.
Under the same assumptions,  $\sqrt{\ga}N^{-1} D^A N \sim 1$ and $\sqrt{\ga}N^{-2}d_0N^A \sim 1$ and the corresponding integral in general does not vanish.
Of course in particular situations one can obtain better behaviours, e.g.~when the shift is time independent.

In the case of AF solutions the quadratic term in the shift does not contribute and one obtains
$$
\calQ'_{AF}[\vec n]=M^n_{ADM} -\Frac[1/2\ka]\int_{\del S_t} \sqrt{\ga}N^{-1} \(       D^A N 
   + {N^{-1} d_0N^A }  \)\>  dv_{A}
\fl{QAFn}$$

Under BoMAF assumptions, analysis needs more details.
In this case one has to consider lapse and shift fall-off \ShowLabel{BM3LapseShift} together with the matching condition $g=\hat g$ at the boundary and the Minkowski limit for $\hat g$.
Then one has $\hat N=1$, $\hat N^i=0$ (as well as $\km(x)=0$ and $\km{}^i(x)=0$ identically).
Then one has  $\sqrt{\ga} N^{-2}N^Bd_B N^A \sim r^{1-\ep}$.
Similarly, one has $\sqrt{\ga}N^{-1} D^A N \sim  r^{1-\ep}$ and $\sqrt{\ga}N^{-2}d_0N^A \sim r^{1-\ep}$.
In other words, under BoMAF assumptions all terms contribute (and singularly diverge!) and the current is not reduced: $\calQ'_{BoMAF}[\vec n]
\equiv \calQ'[\vec n]$ given by \ShowLabel{Qprimen}.

%In fact, the extra currents $C'[\vec n]$ vanish only if very severe hypotheses on the asymptotics are required. 
%In particular, asymptotic flatness is not enough. In fact $\sqrt{\ga}\sim r^2$ so, for the integrals to converge in the limit $r\arr \infty$, one needs terms going to zero faster than $r^{-2}$, i.e.~at least quadratic in the lapse and shift. Terms like $N^{A} d_C N^C\sim r^{-3}$ will not contribute  if the metric is asymptotically flat.
%On the contrary, terms like $\de_0 N^A\sim r^{-2}$  will in general contribute while terms as $D^AN \sim r^{-3/2}$ can even cause divergencies.

For $\xi=\vec N$ one gets
$$
\calQ'[\vec N]=P_{ADM} +\Frac[1/2\ka]\int_{\del S_t} C'[\vec N]
\fl{QprimeN}$$
where we defined the extra current by
$$
C'[\vec N]=\sqrt{\ga}   N^{-1}\(  \de_0 N^A  - N^{A} d_l N^l\)\>  dv_{A}
\fn$$

\CNote{
In fact, in view of equations \ShowLabel{MomentumCurrent} and \ShowLabel{u0}, one has
$\calQ'[\vec N]=(2\ka)^{-1}\int_{\del S_t} Q'[\vec N]$ and:
$$
\eqalign{
Q'[\vec N] =&\( 2K_{C}\^A N^C+ N^{-1} \de_0 N^A   \) \>dv_A
-\sqrt{\ga} N^{i}  \(2 K +N^{-1}  d_l N^l\)\>  dv_{i}=\cr
=&2\sqrt{\ga}\( K_{C}\^A  - K \de^A_C\) N^{C}\>dv_A
+\sqrt{\ga}   N^{-1}\(  \de_0 N^A  - N^{A} d_l N^l\)\>  dv_{A}\cr
}
\fn$$
}

In the case of AF solutions the quadratic term in the shift does not contribute and one obtains
$$
\calQ'_{AF}[\vec N]=P_{ADM} +\Frac[1/2\ka]\int_{\del S_t} \sqrt{\ga}   N^{-1}  \de_0 N^A \>  dv_{A}
\fl{QAFN}$$
with the integrand going as $\sqrt{\ga}   N^{-1}  \de_0 N^A \sim 1$.

Again under BoMAF assumptions one has the extra current unchanged $\calQ'_{BoMAF}[\vec N]\equiv \calQ'[\vec N]$ given by \ShowLabel{QprimeN}.
In general both integrand terms diverge. 

If the extra current does not vanish there are corrections which are necessary to obtain reasonable results. 
We shall discuss this on a specific example below.
We stress that standard expressions for ADM quantities are checked on specific simple examples in which one already knows which mass and momentun is reasonable to expect. Only examples can validate or reject prescriptions for conserved quantities. We shall discuss some example and interpretation below.

Less severe (and known in the literature: see \ref{Sinicco} and references quoted therein) hypotheses are required if one considers the
conserved quantity
$$
\calQ'[\vec m]=M^m_{ADM}+\Frac[1/2\ka]\int_{\del S_t}C'[\vec m]
\fl{Qprimem}$$
where we defined the extra current as
$$
C[\vec m]=\sqrt{\ga}N^{-1}  \( d_0N^A   -{ N^Bd_B N^A } \)\>  dv_{A}
\fn$$

\CNote{
In fact, one has:
$$
\eqalign{
2\ka\calQ'[\vec m] =& \int_{S_t} \sqrt{\ga} N D_A(  N^{-1} D^AN )\>dv
+\int_{\del S_t} \sqrt{\ga}N \({}^3 u^A\)dv_{A}+\cr
&-\int_{\del S_t} \sqrt{\ga} \(
      {2D^A N }
   + {N^{-1} d_0N^A }  -{ N^{-1}N^Bd_B N^A } \)\>  dv_{A}+\cr
  & +\int_{\del S_t}\sqrt{\ga}    {D^A N} dv_A + \int_{S_t}\sqrt{\ga}    N^{-1}D_A N D^A N \>dv=\cr
=& \int_{S_t}   \red{d(  \sqrt{\ga}D^AN \>dv_A)} - \int_{S_t} \sqrt{\ga} \green{N^{-1} D_A N   D^AN} \>dv
+\int_{\del S_t} \sqrt{\ga}N \({}^3 u^A\)dv_{A}+\cr
&-\int_{\del S_t} \sqrt{\ga} \(
      \red{2D^A N }
   + {N^{-1} d_0N^A }  -{ N^{-1}N^Bd_B N^A } \)\>  dv_{A}+\cr
  & +\int_{\del S_t}\sqrt{\ga}   \red{D^A N} dv_A + \int_{S_t}\sqrt{\ga}    \green{N^{-1}D_A N D^A N} \>dv=\cr%
=&\int_{\del S_t} \sqrt{\ga}N \({}^3 u^A\)dv_{A}-\int_{\del S_t} \sqrt{\ga} \(
   {N^{-1} d_0N^A }  -{ N^{-1}N^Bd_B N^A } \)\>  dv_{A}\cr%
}
\fn$$
}

In the case of AF solutions the quadratic term in the shift does not contribute and one obtains
$$
\calQ'_{AF}[\vec m]=M^m_{ADM} +\Frac[1/2\ka]\int_{\del S_t}\sqrt{\ga}N^{-1}  d_0N^A  \>  dv_{A}
\fl{QAFm}$$

The extra current  $C'[\vec m]$ is now controlled by the shift only, since it does not depend on the derivatives of the lapse, contrarily to $C'[\vec n]$.
However, it  may be non-zero even for asymptotically flat metrics owing to the term $d_0N^A$.

Again under BoMAF assuptions the extra current is generic: $\calQ'_{BoMAF}[\vec m]\equiv \calQ'[\vec m]$ as given by \ShowLabel{Qprimem}.

We can now easily obtain again the result of \ref{Sinicco} by computing the boundary Hamiltonian $\calH=\calQ'[\del_0]$:
$$
\calH=\calQ'[\vec m]+ \calQ'[\vec N]= 
\Frac[1/2\ka]\int_{\del S_t} H_{RT}+ \Frac[1/2\ka]\int_{\del S_t} C'[\vec m] +  C'[\vec N]
\fl{Hamiltonian}$$
where we defined the {\it Regge-Teitelboim boundary Hamiltonian}
$$
H_{RT}=\sqrt{\ga}N \({}^3 u^A\)dv_{A}+2\sqrt{\ga}\( K_{C}\^A  - K \de^A_C\) N^{C}\>dv_A
\fl{RHamiltonian}$$
and the {\it extra current}
$$
C'[\del_0]:= C'[\vec m]+ C'[\vec N]=\sqrt{\ga} N^{-1}\(N^Bd_B N^A   - N^{A} d_B N^B\)\>  dv_{A}
\fn$$

\CNote{
In fact, one has:
$$
\eqalign{
2\ka\calH=&\int_{\del S_t} \sqrt{\ga}N \({}^3 u^A\)dv_{A}-\int_{\del S_t} \sqrt{\ga} N^{-1}\(
   \red{ \de_0N^A }  -{ N^Bd_B N^A } \)\>  dv_{A} +\cr
   &+2\int_{\del S_t}\sqrt{\ga}\( K_{C}\^A  - K \de^A_C\) N^{C}\>dv_A
+\int_{\del S_t} \sqrt{\ga}   N^{-1}\( \red{ \de_0 N^A}  - N^{A} d_l N^l\)\>  dv_{A}=\cr
=&\int_{\del S_t} \sqrt{\ga}N \({}^3 u^A\)dv_{A}+2\int_{\del S_t}\sqrt{\ga}\( K_{C}\^A  - K \de^A_C\) N^{C}\>dv_A+\cr
   &+\int_{\del S_t} \sqrt{\ga} N^{-1}\(N^Bd_B N^A   - N^{A} d_B N^B\)\>  dv_{A}\cr}
\fn$$
}

The full Hamiltonian receives a (vanishing on-shell) bulk contribution from the Hamiltonian constraints, 
a contribution from the Regge-Teitelboim boundary Hamiltonian (see \ref{RT})
as well as a boundary contribution from the extra current  $C'[\del_0]$ (see \ref{Sinicco}).
The extra current can be controlled only by spatial derivatives of the shift.
This extra term vanishes for AF solutions,
but if the metric does not meet the fall off prescription for asymptotically flatness, as in the case of BoMAF, these corrections still guarantee the correct results.
We shall see this below in an example.

To summarize, we have shown how the standard ADM mass and momentum can be obtained from Noether theorem, 
associated to $\vec n$ (or $\vec m$) and $\vec N$, respectively, provided some extra hypotheses are imposed;
one can cancel the extra terms using a matched reference or using the pseudotensor \ShowLabel{PseudoTensor} when
hypotheses (stricter than asymptotical flatness) are imposed. 
The Hamiltonian is a particularly lucky case since in that case asymptotical flatness is sufficient for ADM mass and momentum as well.

Before considering examples,
we shall show how asymptotically flatness becomes in fact sufficient if one  requires $\del_0$ to be a Killing vector.
The Killing equations for $\xi=\del_0\equiv \vec m+\vec N$ are hence
$$
\cases{
&\na^0 \obxi^0 + \na^0 \obxi^0=0
\quad\then   d_0 N=0
 \cr
&\na^0 \obxi^i + \na^i \obxi^0=0
\quad\then  D^i N + N^k K_k{}^i  - \(D^i N + K\^i{}_k N^k\)- N^{-1} \de_0 N^i=0
\quad\then d_0 N^i=0
\cr
&\na^j \obxi^i + \na^i \obxi^j=0
\quad\then D^{(j} N^{i)}+ NK^{ij}=0
\quad\then d_0 \ga_{ij}=0
\cr
}
\fn$$

\CNote{
For $\xi=\del_0$ we have
$$
\cases{
&\na^0 \obxi^0=-N^{-1} \(\de_0 N + N^kD_k N\)= -N^{-1} d_0 N\cr
&\na^j \obxi^0=D^j N + N^k K_k\^j\cr
&\na^0 \obxi^i=- \(D^i N + K\^i{}_k N^k\)- N^{-1} \de_0 N^i \cr
&\na^j \obxi^i= D^j N^i+ NK^{ij}\cr
}
\fn$$ 
}

If Killing equations hold true, then the time--derivatives of the lapse and the shift vanish and AF is enough to control
the extra currents in the mass and in the momentum.

Of course in case of BoMAF hypotheses, extra control on the terms quadratic in the shift is needed.

We stress, however, that we did not use all Killing equations to obtain control; hence also here the vector $\xi=\del_0$ does not need to be necessarily Killing, while
this is certainly a sufficient condition.

Moreover, with specific solutions, one could also do better than this. For example, for a specific solutions and foliations the shift may happen to fall off faster than prescribed, which in particular allows to obtain control on all these quantities.

\NewSection{Examples}

As long as different prescriptions for conserved quantities give the same result they of course enforce one another and the choice is a matter of taste which may depend on easy calculability or on how strictly they implement fundamental principles.
In the previous Section we presented a number of sets of hypotheses which ensure, in different situations, that standard expressions for ADM are recovered from
covariant conservation laws. However, one should not be too fond on standard expressions, expecially outside the scope in which they have been derived.

When different prescriptions provide different results one can really judge which one, if any, is still valid and which one is not.
Here we shall consider the Schwarzschild solution in different coordinates and apply all the above prescriptions and compare them.
Let us remark here that Schwarzschild is one of the relatively few cases in which the Newtonian limit is quite well--understood 
and one can make a comparison with classic definitions.
Hereafter we shall use Maple Tensor package (see \ref{Maple}) for explicit calculations.

The Schwarzschild metric in pseudo-Cartesian coordinates $(t, x, y, z)$ reads as
$$
g=-f^2(r) dt^2 + dx^2+ dy^2+ dz^2 +\Frac[2m/ f^2(r)r^3] (xdx+ydy+zdz)^2
\fl{Sm}$$
where we set $f^2(r)=1-\frac[2m/r]$ and $r^2=x^2+y^2+z^2$. 
We are here considering the spacetime region with $r>2m$; the ADM foliation is given by the projection on coordinate time $t$ and fibers are $S_t=\{x^0=t\}$.
The boundary region where we shall perform all integrations is the space infinity $r\arr \infty$.
We have the normal unit vector $\vec n= f^{-1}(r) \del_0$; which means lapse $N=f(r)$ and zero shift $\vec N=0$.
The extrinsic curvature is $K_{ij}=\frac[1/2N]\del_0\ga_{ij}\equiv0$.

The corresponding standard ADM mass and momentum \ShowLabel{ADMcq} are
$$
\eqalign{
&M^n_{ADM}= m N^{-1} \Limit m\cr
&M^m_{ADM}=  m\Limit m\cr
&P_{ADM}\equiv 0\cr
}
\fn$$

The corrected ADM prescription \ShowLabel{Qprimen}, \ShowLabel{Qprimem}, \ShowLabel{QprimeN} are:
$$
\eqalign{
&\calQ'[\vec n]=  m  N^{-1}- \frac[1/4] m  N^{-1} \Limit \frac[3/4] m\cr
&\calQ'[\vec m]= m \Limit m\cr
&\calQ'[\vec N]= P_{ADM} \equiv0
}
\fl{SchwQprime}$$ 

The Hamiltonian \ShowLabel{Hamiltonian} gives:
$$
\calH=M^m_{ADM} + P_{ADM}= m \Limit  m
\fn$$
since the shift is zero.

The reference metric can be chosen to be the limit of \ShowLabel{Sm} for $m\arr m_0$, i.e.
$$
\hat g=-f^2_0(r) dt^2 + dx^2+ dy^2+ dz^2 +\Frac[2m_0/ f^2_0(r)r^3] (xdx+ydy+zdz)^2
\fn$$
where we set $f^2_0(r)=1-\frac[2m_0/r]$. This is of course Minkowski metric when $m_0=0$.
The two metrics match at $r\arr \infty$. They define the same asymptotic lapse and shift.
The augmented conserved quantities \ShowLabel{CQ} 
 are
$$
\eqalign{
&\calQ[\vec n]\Limit  \frac[3/4]\(m-m_0\)\cr
\calQ[\del_0]\equiv &\calQ[\vec m] \Limit  m-  m_0\cr
&\calQ[\vec N] \equiv0
}
\fn$$

All this is quite standard;  we are in quasi-Cartesian coordinates and with an asymptotically flat solution.
The shift is zero and $\del_0$ is a Killing vector. All hypotheses made above are met and all prescriptions presented above agree.
Notice that one has $\hat E[\vec m]=0$, while one has $\hat E[\vec n]=- \hat D_A\(\sqrt{\hat \ga}  N^{-1} \hat D^A N\)$ 
which accounts for the fact that  the pseudotensor for $\vec m$ behaves better than the pseudotensor for $\vec n$.
In any event the augmented conserved quantity produces the correct result.

\CNote{
The bulk contribution for $\xi=\vec n$ of the reference metric is:
$$
\eqalign{
\hat E[\vec n]=&\sqrt{\hat \ga}\(\hat\ga^{AB}\Lie_{\vec n} \hat{\ubom}^0{}_{AB}+ \Lie_{\vec n} \hat{\ubom}^A{}_{0A} - \hat R\) dv=\cr
=&\sqrt{\hat \ga}\(N^{-1}\( \hat\ga^{AB}\de_0 \hat K_{AB} -\hat D_A\hat D\^A N\)+ N^{-1} \de_0 \hat K -N^{-2} \hat D_AN \hat D^AN \) dv=\cr
=&\sqrt{\hat \ga}\(N^{-1}\( \de_0 \hat K+ 2N \hat K_{AB} \hat K^{AB} -\hat D_A\hat D\^A N\)+ N^{-1} \de_0 \hat K -N^{-2} \hat D_AN \hat D^AN \) dv=\cr
=&\sqrt{\hat \ga}\(N^{-1}\(2 \de_0 \hat K+ 2N \hat K_{AB} \hat K^{AB}\) -\hat D_A\(N^{-1}\hat D\^A N\) -2N^{-2} \hat D_AN \hat D^AN \) dv\cr
}
\fn$$
Considering  that $\hat K_{AB}=0$ and that, even under the milder assumption of BoMAF, the term $-2N^{-2} \hat D_AN \hat D^AN$ falls off like $ r^{-2-\ep}$
one has that $\hat E[\vec n]=- \hat D_A\(\sqrt{\hat \ga}  N^{-1} \hat D^A N\)$.
This is, in view of Stokes theorem, a boundary contribution and it was exactly responsible for the deviation found in $\calQ'[\vec n]$; see \ShowLabel{SchwQprime}.

For $\xi=\vec m$ one has
$$
\eqalign{
\hat E[\vec m]=&\sqrt{\hat \ga}\(\hat\ga^{AB}\Lie_{\vec m} \hat{\ubom}^0{}_{AB}+ \Lie_{\vec m} \hat{\ubom}^A{}_{0A} - \hat R\) dv=\cr
=&\sqrt{\hat \ga}\( \hat\ga^{AB}\de_0 \hat K_{AB} -\red{N^{-1} \de_0 N \hat K} + \de_0 \hat K + \red{N^{-1}\de_0 N \hat K}\) dv=\cr
=&2\sqrt{\hat \ga}\( \de_0 \hat K+N \hat K_{AB} \hat K^{AB}  \) dv= 0\cr
}
\fn$$
where we used the fact that the extrinsic curvature is vanishing.
}

Let us now check what happens when we consider different coordinates.
In view of the general covariance principle (moreover, in its weaker formulation about {\it passive} changes of coordinates)
one would expect reasonable physical quantities to be unaffected, provided the coordinate change is globally defined in the same integration domain (as it will be in all cases considered hereafter).
The Schwarzschild metric in standard (quasi-polar) coordinates $(t, r, \te, \phi)$ reads as
$$
g=-f^2(r) dt^2 + f^{-2}(r)  dr^2 + r^2 d\Om^2
\fl{Smp}$$
with $f^2(r)=1-\frac[2m/r]$. 
The relation with previous quasi-Cartesian coordinates reads as
$$
\cases{
&t=t\cr
&x= r\sin(\te)\cos(\phi)\cr
&y= r\sin(\te)\sin(\phi)\cr
&z= r\cos(\te)\cr
}
\fn$$
We are here considering again the spacetime region with $r>2m$; the ADM foliation is given as above since we did not change coordinate time.
We have the normal unit vector $\vec n= f^{-1}(r) \del_0$; which means lapse $N=f(r)$ and zero shift $\vec N=0$.
The extrinsic curvature (being a tensor) is again zero.
The corresponding standard ADM mass and momentum \ShowLabel{ADMcq} are
$$
\eqalign{
&M^n_{ADM}= -r N\Limit -\infty\cr
&M^m_{ADM}=   -r f^2(r)\Limit -\infty\cr
&P_{ADM}\equiv 0\cr
}
\fn$$
Since these are not tensor objects it is no surprise that their values are different from the corresponding quantities in quasi-Cartesian coordinates.

The corrected ADM prescription \ShowLabel{Qprimen}, \ShowLabel{Qprimem}, \ShowLabel{QprimeN} are:
$$
\eqalign{
&\calQ'[\vec n]= -rN - \frac[m/4N]  \Limit -\infty\cr
&\calQ'[\vec m]= M^m_{ADM} \Limit -\infty\cr
&\calQ'[\vec N]= P_{ADM} \equiv0
}
\fn$$ 
In fact the contribution of the reference metric (also for $m_0=0$) in polar coordinates is definitely non--zero; 
on the contrary, it scales as $r\Limit \infty$, and, in view of its relation with augmented conserved quantities, one does not expect $\calQ'[\xi]$
to be meaningful or finite in these coordinates.

The Hamiltonian \ShowLabel{Hamiltonian} gives:
$$
\calH=M^m_{ADM} \Limit -\infty
\fn$$
since the shift is zero.
If one keeps considering the reference contributions then one obtains the expected finite result for the Regge--Teitelboim Hamiltonian also in these coordinates.

The reference metric can be chosen to be the limit of \ShowLabel{Smp} for $m\arr m_0$, i.e.
$$
\hat g=- f^2_0(r)dt^2 + f_0^{-2}(r)dr^2 + r^2 d\Om^2
\fn$$
where we set $f_0(r)=\sqrt{1-\frac[2m_0/r]}$. This is of course Minkowski metric when $m_0=0$.
The two metrics match at $r=\infty$. This reference induces the same lapse and shift at the boundary $\del S_t$.
The augmented conserved quantities \ShowLabel{CQ} are
$$
\eqalign{
&\calQ[\vec n]\Limit  \frac[3/4]\(m-m_0\)\cr
&\calQ[\vec m] \Limit m-m_0\cr
&\calQ[\vec N] \equiv0
}
\fn$$

Here the corrected conserved quantity $\calE$ and $\calC[\vec n]$  still provide the expected result despite standard ADM expressions fail since their hypotheses
are not satisfied.
In these coordinates the reference contribution cannot be neglected since $\hat u^\al_{\mu\nu}$ is not zero; in fact it diverges on the boundary. 
This explains why $M^n_{ADM}$, $M^m_{ADM}$, $\calQ'[\vec n]$ and $\calQ'[\vec m]$ diverge.
Since $\del_0$ is a Killing vector and the shift is zero, one can expect $\calQ'[\vec m]= M^m_{ADM}$.
Finally, covariance accounts for the fact that augmented conservd quantities are the same as in quasi-Cartesian coordinates. 

We can now change again coordinates and choose
$$
\cases{
&\tau=t+ \sqrt{2m}\int \Frac[\sqrt{r}dr/r-2m] \cr
&x= r\sin(\te)\cos(\phi)\cr
}
\qquad\qquad
\cases{
&y= r\sin(\te)\sin(\phi)\cr
&z= r\cos(\te)\cr
}
\fn$$
In these new coordinates $(\tau, x, y, z)$ the Schwarzschild metric reads as
$$
g= -d\tau^2 + \de_{ij} \( dx^i \pm\sqrt{\frac[2m/r^3]} x^id \tau\)\(dx^j \pm\sqrt{\frac[2m/r^3]} x^j d \tau\)
\fl{SchwarzschildMetricLaMaitre}$$
where $r^2=(x^1)^2+(x^2)^2+(x^3)^2$.  
Since we changed coordinate time we are now using a different foliation $\tau=\hbox{constant}$.
In this case we have $N^2=1$, $\ga_{ij}=\de_{ij}$ and $N^i=\pm\sqrt{\frac[2m/r^3]} x^i$.
This shift does not meet the fall-off prescription usually required for AF, while they meet BoMAF prescriptions though with some extra parity conditions.
Notice that the shift vector is still time-independent.
The extrinsic curvature is now 
$$
K_{ij}= \mp\sqrt{\Frac[2m/r^3]}\( \de_{ij}-\Frac[3/2] \Frac[x_i x_j/r^2]\)
\fn$$
Of course, $K$ is a space tensor and does not depend on coordinates, but now that we are changing ADM  foliation it refers to other space manifolds in $M$.

Since the space metric is $\ga_{ij}=\de_{ij}$ and $N=1$, the quantity $^3u^k_{ij}$ vanishes and we find
$$
\cases{
&M^n_{ADM}=M^m_{ADM}=0\cr
&P_{ADM}\Limit 2m\cr
}
\fn$$
Again no much surprise that we get different results since these are conserved quantities measured by different observers.

When extra currents are taken into account one gets for the mass
$$
\calQ'[\vec n]=\calQ'[\vec m]= \int_{\del S_t} \sqrt{\ga} N^id_i N^k dv_k\Limit -\Frac[ m/4]
\fn$$
while for the shift
$$
\calQ'[\vec N]= P_{ADM}- \int_{\del S_t} \sqrt{\ga} N^kd_i N^i dv_k\Limit 2m -\Frac[ 3m/4]= \Frac[ 5m/4]
\fn$$

This recovers the Hamiltonian found in \ref{Sinicco}
$$
\calH=-\Frac[ m/4] +\Frac[ 5m/4]= m \not= M^m_{ADM}+P_{ADM}
\fn$$ 
which is correct. Notice that the extra currents are essential for getting the correct result; without extra currents the Regge-Teitelboim boundary Hamiltonian
would produce wrong result $M^m_{ADM}+P_{ADM}=2m$, according to the fact that the metric expression \ShowLabel{SchwarzschildMetricLaMaitre} is not AF.

The reference can be chosen as
$$
\hat g= -d\tau^2 + \de_{ij} \( dx^i \pm\sqrt{\frac[2m_0/r^3]} x^id \tau\)\(dx^j \pm\sqrt{\frac[2m_0/r^3]} x^j d \tau\)
\fl{referenceLaMaitre}$$
This defines the same lapse and shift at space infinity, and for $m_0=0$ it reduces to Minkowski.

Augmented conserved quantities associated to $\vec n=\vec m$ and $\vec N$ and the reference \ShowLabel{referenceLaMaitre} for $m_0=0$ are:
$$
\cases{
&\calQ[\vec n]=\calQ[\vec m]\Limit -\Frac[m/4]\cr
&\calQ[\vec N]\Limit \Frac[5m/4]\cr
}
\fn$$
Again augmented conserved quantities reproduce the correct mass, momentum and Hamiltonian, also in this case when the solution is not
asymptotically flat. 
The augmented conserved quantity associated to $\del_\tau=\vec m$ coincides by definition with the Hamiltonian.

Here we are outside the scope of the definition of ADM quantities. In the chosen coordinates the solution is not manifestly AF.
If we accept as a coherence check that $\calH= M^m_{ADM}+ P_{ADM}$, the standard ADM quantities define the Regge-Teitelboim Hamiltonian density
which is manifestly affected by an anomalous factor problem. 
On the contrary, the corrected ADM quantities ($\calQ'[\vec n]$ and $\calQ'[\vec N]$) define the expected Hamiltonian with no anomalous factor.
Moreover the corrected ADM quantities coincide with the value of the augmented conserved quantities which once again provide the correct result in a covariant way that, in the form \ShowLabel{CQ}, is even independent on the foliation machinery.

\NewSection{Conclusions and Perspectives}

The ADM mass and momentum of Schwarzschild spacetime are obtained as Noether quantities associated to particular symmetries.
We found that ADM mass is associated to the normal unit vector $\vec n$ of the foliation, while ADM momentun is associated to the shift $\vec N$.
This is true in pseudo-Cartesian coordinates, but it remains true also in pseudo-spherical coordinates provided one uses a matched reference to 
suitably compensate the infinities. Notice that the prescriptions based on pseudotensor here fail. 

{The relation of ADM pseudotensor with covariant conservation laws allows us to trace exactly which hypotheses are needed to reduce the covariant quantities
to the ADM pseudotensor. These hypotheses include also the definition of preferred observer (the one associated with the frame $e_a$ used as symmetry generators). ADM conserved quantities emerged as the covariant quantities measured by those preferred observers, accounting explicitly for non-covariance.}

{The extra hypotheses required  emerge from the need of cancelling terms which one wish to get rid of. We pointed out that different terms can be cancelled
using different techiques: 
by restricting symmetry generators to Killing vectors 
or by cancelling contributions from dynamical fields with contributions from reference field (which in this case are required to match)
or by restricting asymptotic behaviours.}

{All these techniques are legitimate in particular contexts. 
However, we stress that when one discusses conserved quantities  from a fundamental viewpoint
one cannot forget that a generic solution in GR has no Killing vector to be used and different asymptotics provide different equally physically legitimate sectors of the theory.
Of course Killing vectors or asymptotic behaviours can be used in specific situations, 
though matchings, when viable, provides in our opinion a preferred, generic, covariant tool.}

{Future investigations will be devoted to consider if the standard treatment inspired to SR and energy-momentum tensors can be reproduced for ADM quantities
and one can take advantage from the techniques developed for covariant conservation laws (for example to discuss the relation between 
energy-momentum tensors and Hilbert stress tensors; see Appendix \AppA).}

\NewAppendix{\AppA}{Conservation Laws in Special Relativity}

In Special Relativity (SR) the standard treatment of conserved quantities is based on structures which do survive when one tries the extension to GR.
The same simple notion that $4$-momentum is a Lorentz vector cannot be extended in any way to GR.
This can be easily seen by tracing the procedure of SR using the language to be used in GR; in other words we can try and regard SR as a special solution of GR
with the Minkowski metric $g=\eta$ which being flat is trivially a solution of Einstein equations.
We stress that this is equivalent to treat Minkowski spacetime as a Lorentzian manifold (which happens to be isometric to $\R^4$) instead 
of using its affine structure.
Accordingly, one should systematically mind the difference between points in $M$ and its tangent vectors.

For simplicity, we shall consider a Klein-Gordon (KG) matter field; this is simpler than the general case but it is sufficient to simply make most of our points.
KG scalar field is a section of the configuration bundle $M\times \R$ with coordinates $(x^\mu, \phi)$.
The KG Lagrangian is
$$
L_{m}= \frac[\sqrt{g}/2]\( \phi_\mu\phi_\nu g^{\mu\nu} -m^2\phi^2\)
\fn$$
where $ \phi_\mu=\na_\mu\phi$ denotes first derivatives of KG field.
The momentum densities are
$$
p_{\mu\nu}=\Frac[\del L_m/\del g^{\mu\nu}]=\frac[1/2]\(\sqrt{g} \phi_\mu\phi_\nu - L_m g_{\mu\nu} \)
\qquad
p^\mu =\Frac[\del L_m/\del \phi_\mu]=\sqrt{g} \phi\^\mu 
\qquad
p =\Frac[\del L_m/\del \phi]= -\sqrt{g} m^2 \phi
\fn$$

The variation of the matter Lagrangian reads as
$$
\de L_m= \frac[\sqrt{g}/2] H_{\mu\nu} \de g^{\mu\nu} +E\de \phi + \na_\mu\( p^\mu\de \phi\)
\fn$$
where we defined the KG tensor density
$$
E=p-\na_\mu p^\mu=-\sqrt{g}\(\Dal \phi+m^2\phi\)
\fn$$
which represents field equations $E=0$ for the KG field (thus, by definition, it vanishes on-shell);
the Hilbert stress tensor $H$ is defined by $p_{\mu\nu}= \frac[\sqrt{g}/2] H_{\mu\nu}$.
It is defined to be a symmetric tensor.
When one couples to gravity in GR, matter acts as a source of gravitational field through $H_{\mu\nu}$.
In literature this is often called the {\it energy-momentum tensor} while we shall use that name for a different quantity
coming from Noether theorem which only in special cases (one of which is in fact KG matter field) coincides with the Hilbert stress tensor.

One can prove the {\it covariance identity} which holds true for any spacetime vector field $\xi$
$$
\na_\mu\(\xi^\mu L_m\)=\frac[\sqrt{g}/2] H_{\mu\nu} \Lie_\xi g^{\mu\nu} +p\Lie_\xi \phi  +p^\mu \Lie_\xi \phi_\mu 
\fl{CovId}$$
which can be easily proved by simply expanding Lie derivatives
$$
\Lie_\xi g^{\mu\nu}= -\(\na^\mu\xi^\nu +\na^\nu\xi^\mu\)
\qquad
\Lie_\xi \phi= \xi^\al \phi_\al
\qquad
\Lie_\xi \phi_\mu= \xi^\al \na_\mu \phi_{\al} + \na_\mu\xi^\al \phi_\al
\fn$$

The covariance identity can be recasted as 
$\na_\mu \calE^\mu[\xi]=-\frac[\sqrt{g}/2] H_{\mu\nu} \Lie_\xi g^{\mu\nu} -E\Lie_\xi \phi $
by a suitable covariant integration by parts 
where we set 
$$
\calE[\xi]=\calE^\mu[\xi] \>ds_\mu=\(p^\mu \Lie_\xi\phi -\xi^\mu L_m\) \>ds_\mu= \sqrt{g}T^\mu_\al \xi^\al\>ds_\mu
\fn$$
for the Noether current.
The tensor $T^\mu_\al$ is more appropriately called {\it energy momentun tensor}; we shall comment later about its relation with $H_{\mu\nu}$.

In a general SR situation this Noether current is not conserved; in fact one has $E=0$ on-shell  but $H\not=0$, so that
one has 
$$
d \calE[\xi]= \frac[\sqrt{g}/2] H_{\mu\nu} \Lie_\xi g^{\mu\nu}
\fl{ConservationID}$$
This extra term on the right hand side is the direct consequence of the fact that Minkowski metric in SR does not obey field equations and it is instead imposed as a freezed structure on spacetime. 
In GR one couples with gravity, Hilbert Lagrangian contributes by a further term which, together with matter contribution, factorize Einstein-matter field equations and they together vanish on-shell. 
This cancellation cannot be obtained in SR.
However, if one restricts to consider Killing vectors for the metric (i.e.~$\Lie_\xi g=0$) then the right hand side vanishes and Noether current is conserved.
 In this case one can define the conserved quantities
$$
Q[\xi]=\int_{S_t}  \calE[\xi]=\int_{S_t} T^0_\al\xi^\al dv
\fn$$ 
which are conserved since the Noether current is conserved (this time not because of field equations though thanks to Killing equation). 
In Cartesian coordinates,  for $\xi=\del_0$ one defines the energy $P_0=Q[\del_0]$, while for 
$\xi=\del_i$ one defines the momentum $P_i= Q[\del_i]$.
Of course, there are infinitely many Cartesian coordinate systems. Lorentz transformations change Cartesian coordinates the quantities $P_\mu=(P_0, P_i)$ transform as a
Lorentz vector.
However, let us stress that $P_	\mu$ are integral quantities which are not associated to any particular point of Minkowski space. It is only using the affine structure of Minkowski space that one can define a free covector $\vec P=P_\mu dx^\mu$; this is not a vector field (the components are numbers not functions) nor
a tangent vector to some point of Minkowski. It is only in view of integrability of parallel transport on Minkowski (which holds true iff the metric is flat)
that one can define {\it constant vector fields} to be identified with free vectors. We stress that on a generic Lorentzian manifold there is nothing like that.
 
The properties of energy-momentum tensor are obtained by considering the identity \ShowLabel{ConservationID}. Expanding both sides one gets
$$
\cases{
&\na_\mu T^\mu_\al=0\cr
&T^\mu_\al=H^\mu_\al\quad\then
\cases{
&T_{[\mu\al]}=0\cr
&T_{(\mu\al)}=H_{\mu\al}\cr
}\cr
}
\fn$$
In this way one can prove that the energy-momentum tensor $T$ is conserved, it is symmetric and it coincides with the Hilbert stress tensor.

For more general matter fields (e.g.~the electromagnetic field) the Noether current is in the more general  form
$$
\calE^\mu = \sqrt{g}\(T^\mu_\al \xi^\al + T^{\mu\be}_\al \na_\be\xi^\al \)
\fl{firstOrderConservation}$$
The conservation law implies then on-shell
$$
\cases{
& T^{(\mu\be)}_\al=0 \cr
& T^\mu_\al+ \na_\nu T^{\mu\nu}_\al= H^\mu_\al \cr
& \na_\mu T^\mu_\al = 0\cr
}
\fn$$
where we used the fact that in Minkowski spacetime covariant derivatives commute.
Again we have conservation, but the energy-momentum tensor and the Hilbert stress tensor differ considerably. 
The second item, however, provides the relation between these two tensors in terms of the higher order tensor $T^{\mu\nu}_\al$
(which is skew in $[\mu\nu]$).

One can integrate  \ShowLabel{firstOrderConservation} by parts
$$
\calE^\mu = \sqrt{g}(T^\mu_\al- \na_\be T^{\mu\be}_\al )\xi^\al+ \na_\be(\sqrt{g}T^{\mu\be}_\al \xi^\al)
= \sqrt{g} H^\mu_\al \xi^\al+ \na_\be(\sqrt{g}T^{\mu\be}_\al \xi^\al)
\fn$$
which in view of antisymmetry defines the superpotential (which is vanishing for KG Lagrangian)
$$
U=\frac[\sqrt{g}/2]T^{\mu\be}_\al \xi^\al ds_{\mu\nu}
\fn$$
The superpotential defines conserved quantities as boundary quantities and controls the relation between the energy-momentum tensor and the Hilbert stress tensor;
in particular it tells us how to build a symmetric tensor $H_{\mu\nu}$ out of the energy-momentum tensor $T_{\mu\nu}$ which is not symmetric in general.

To summarize, one cannot expect to be able to define anything like a $4$-momentum tensor, nor to control its transformation rules with respect to Lorentz transformations (which are not even defined in GR). Of course, in the case of AF solutions, one can repeat the SR argument at infinity where one can define
asymptotic Cartesian coordinates, asymptotic Lorentz transformations, asymptotic Killing vectors and asymptotic free Lorentz covariant $4$-momentum covector.
On the other hand, in GR general covariance considerably improves conservation laws which hold in general without any need of requiring Killing vectors.

\NewAppendix{\AppB}{Frequently Made Comments about Conservation Laws in GR}

 \ni {\bf Covariant conservation laws are not conservation laws}
 
This argument goes back to Einstein who also gave a physical explanation of non-covariance in term of equivalence principle and
interaction energy; see e.g.\ \ref{Weinberg} page 126. 

When the argument is expressed as a general statement ({\it Covariant conservation laws {\rm cannot} be conservation laws}) 
it is wrong and based on a wrong understanding of what one should mean by ``covariant conservation laws''.

In fact, it is obvious that in order of having a covariant conservation law the current $\calE$ must be an object for which covariant derivative is defined, i.e. a
tensor or tensor density object. Applying the argument to pseudotensors is based on a wrong definition of covariant derivative.

If $\calE^\mu $ is a tensor then of course covariant conservation law is not a continuity equation (due to the connection term which acts as a source).
However, if the current is a tensor density of weight 1, i.e.~$\calE=\sqrt{g} \calE^\mu ds_\mu$, then the covariant derivative must be modified by a further term originated by the density character
$$
\na_\nu\( \sqrt{g} \calE^\mu \) = d_\nu\( \sqrt{g} \calE^\mu \)  + \Ga^\mu_{\la\nu}\sqrt{g} \calE^\la - \Ga^\la_{\la\nu} \sqrt{g} \calE^\mu
\fn$$
This last terms is needed to make $\na_\nu\( \sqrt{g} \calE^\mu \)$ a tensor density.
The covariant conservation laws is thence
$$
\na_\mu\( \sqrt{g} \calE^\mu \) = d_\mu\( \sqrt{g} \calE^\mu \)  + \Ga^\mu_{\la\mu}\sqrt{g} \calE^\la - \Ga^\la_{\la\nu} \sqrt{g} \calE^\mu
\equiv
 d_\mu\( \sqrt{g} \calE^\mu \)  
\fn$$
exactly a continuity equation (we are assuming a torsionless connection).

In other words, this argument instead of ruling out covariant conservation laws, does in fact show how one should require currents to be vector densities,
i.e.~$(m-1)$-forms.
By the way $(m-1)$-forms is what Noether theorem naturally produces as currents.

\ms\ni {\bf Conserved quantities must be expressed by pseudotensors since they are not observer invariants}

Of course, energy is not observer invariant nor covariant; also in Special Relativity it is the zero component of a $4$-vector, namely the $4$-momentum vector.
In Appendix $\AppA$ we summarized the standard treatment of (Lorentz) covariance of the $4$-momentum in SR, stressing the issues which will turn 
meaningless when passing to GR. The very notion of a Lorentz free vector is tricky to be extendend to GR.

In any event, aiming to a covariant prescription for conserved quantity does not mean to obtain an invariant (or covariant) quantity.
Covariant quantities do in fact depend on the observer since different observers evaluate them along different vector fields and integrate them along different
spatial submanifolds; see \ref{Kerr2}. Moreover, in view of augmented variational prescriptions (see \ref{Augmented}) different observers may choose different backgrounds (see \ref{Kerr} and \ref{Kerr2}) and different control variables (see \ref{Nester}). 

We stress once again that integrating generic pseudotensor objects does explicitly depend on coordinates and hence it directly  contradicts the general covariance
principle (not only {\it manifest} general covariance good looking rule). Of course this does not completely rule out pseudotensor, but requires that when pseudotensors are used an explicit fixing of coordinate gauge and a physical motivation for it should be provided.
Just in the same way as in electrodynamics one can use the wave equation, but just after defining Lorentz gauge and showing that one can perform Lorentz gauge fixing in the generic situation.

\ms\ni {\bf Killing vectors are necessary to define covariant conservation laws}

They are not. Komar superpotential was originally defined for a timelike Killing vector but it defined a strong conservation laws for any spacetime vector field;
that is $\calE= 2\na_\mu \(\sqrt{g}\na^\al \xi^\mu\) \> ds_\al$ is always covariantly conserved; see \ref{Book}.
This is quite strightforward to be proved directly and indipendently of any theoretical approach.

Augmented conserved quantities provide a whole set of examples of strong conservation laws which can be applied when no Killing vector exists.
The Hilbert Lagrangian is generally covariant, it is hence natural to expect a conservation law attached to any symmetry generator, i.e.~to any spacetime vector field.

The ``myth'' of Killing vectors was originated by a continuous deformation of arguments used in SR, where the metric structure is fixed and hence it must be preserved.
This is exacly what GR left behind of SR and why GR is a more fundamental framework than SR.

Of course GR expressions can be often simplified if a Killing vector is assumed, but in view of the general covariance principle one cannot rely on it at a fundamental
level. Just as for pseudotensors it can be used in special situations but not in the generic case when there is no Killing vector at all.

\ 

\ms\ni {\bf Matching with the reference is the same as asymptotical flatness}

{
This is not the case. Asymptotical flatness is a stronger request than the matching with Minkowski.
Asymptotic flatness  requires particular fall off constraints (which by the way are coordinate dependent) while matching 
simply requires that the metric becomes the reference metric at the boundary, with no particular fall-off required.
}

\NewAppendix{\AppC}{Different routes from covariant conservation laws to ADM}

{
Let us here summarize the relation between the different quantities defined above.
}

$${\AbstractStyle
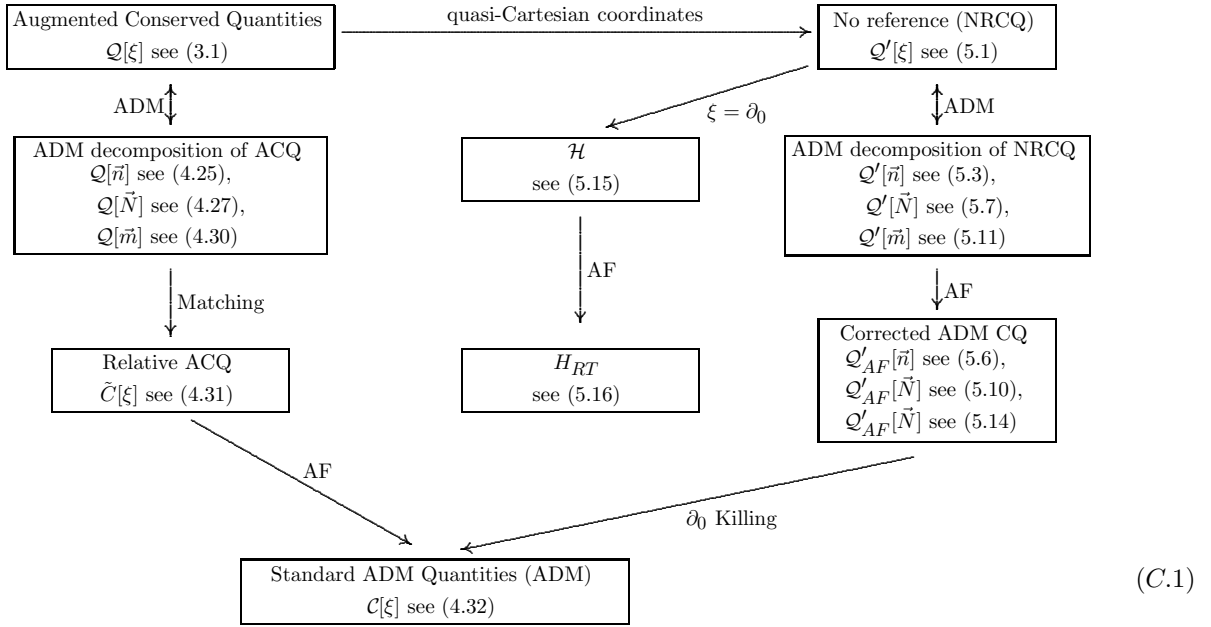
\begindc{\commdiag}[1]
\obj(0,210)[ACQ]{\Box to {4.2cm}{Augmented Conserved Quantities}{$\calQ[\xi]$ see \ShowLabel{CQ}}}
\obj(0,150)[ACQ2]{\Box to {4cm}{ADM decomposition of ACQ}{\vbox{
	\hbox{$\calQ[\vec n]$ see \ShowLabel{Qn},}
	\hbox{ $\calQ[\vec N]$ see \ShowLabel{QN},}
	\hbox{ $\calQ[\vec m]$ see \ShowLabel{Qm}}
}}}
\obj(0,80)[MACQ]{\Box to {3cm}{Relative ACQ}{$\tilde C[\xi]$ see \ShowLabel{AugmentedMatchedCQ}}}
\obj(100,0)[ADM]{\Box to {5cm}{Standard ADM Quantities (ADM)}{$\calC[\xi]$ see \ShowLabel{ADMcq}}}
\obj(290,210)[NBG]{\Box to {3cm}{No reference (NRCQ)}{$\calQ'[\xi]$ see \ShowLabel{PseudoTensor}}} 
\obj(290,150)[NBG2]{\Box to {3.9cm}{ADM decomposition of NRCQ}{\vbox{
	\hbox{$\calQ'[\vec n]$ see \ShowLabel{Qprimen},}
	\hbox{ $\calQ'[\vec N]$ see \ShowLabel{QprimeN},}
	\hbox{$\calQ'[\vec m]$ see \ShowLabel{Qprimem}}
}}} 
\obj(290,80)[SADM]{\Box to {3cm}{Corrected ADM CQ}{\vbox{
	\hbox{$\calQ'_{AF}[\vec n]$ see \ShowLabel{QAFn},}
	\hbox{$\calQ'_{AF}[\vec N]$ see \ShowLabel{QAFN},}
	\hbox{$\calQ'_{AF}[\vec N]$ see \ShowLabel{QAFm}}
}}} 
\obj(155,160)[Ham]{\Box to {3cm}{$\calH$}{see \ShowLabel{Hamiltonian}}} 
\obj(155,80)[HamRT]{\Box to {3cm}{$H_{RT}$}{see \ShowLabel{RHamiltonian}}} 
\mor(155,150)(155,90){AF}
\mor(250,200)(155,170){$\xi=\del_0$}
\mor(0,200)(0,165){}\mor(0,165)(0,200){ADM}
\mor(0,130)(0,85){Matching}
\mor(0,68)(100, 10){AF}
\mor{ACQ}{NBG}{quasi-Cartesian coordinates}
\mor(290,200)(290,165){ADM}\mor(290,165)(290,200){}
\mor(290,130)(290,95){AF}
\mor(290,50)(100,12){$\del_0$ Killing}
%\mor{M1}{M2}{}[\atleft, \solidline] \mor(40,33)(110,33){}[\atleft, \solidline]
\enddc}
\fn$$
\ms
{
Let us just remark that, for example, $\calQ[\vec n]$ depends both on the metric and the reference.
However, if one simply forgets about the reference contributions to $\calQ[\vec n]$ 
(as in the spirit of pseudotensors where quasi-Cartesian coordinates are chosen in order to kill the contributions from the reference) then $\calQ'[\vec n]$ is not obtained.
In fact, in computing $\calQ'[\vec n]$ we first performed a number of cancellations between metric and reference terms and then forgot about the reference.
Forgetting about the reference at first leaves some metric contributions uncanceled. In other words, we could say that
making the ADM decomposition and forgetting about the reference are two ``non-commuting operations''.}

%%%%%%%%%%%%%%%%%%%%%%%%%%%%%%%%%%%%%%%%%%%%%%%%%%%
\Acknowledgements

This work is partially supported by MIUR: PRIN 2005 on {\it Leggi di conservazione e termodinamica in meccanica dei continui e teorie di campo}.  
We also acknowledge the contribution of INFN (Iniziativa Specifica NA12) and the local research founds of Dipartimento di Matematica of Torino University.

\ 
 
\ShowBiblio

\end